\documentclass{article}

\usepackage[numbers]{natbib}
\usepackage{arxiv}

\usepackage[utf8]{inputenc} 
\usepackage[T1]{fontenc}    
\usepackage{hyperref}       
\usepackage{url}            
\usepackage{booktabs}       
\usepackage{amsfonts}       
\usepackage{nicefrac}       
\usepackage{microtype}      
\usepackage{lipsum}		
\usepackage{graphicx}
\usepackage{doi}
\usepackage{mathptmx}
\usepackage{mathptmx}
 \usepackage{amsmath}
 \usepackage{tabularx}
 \usepackage{amssymb}

\title{Pseudo-2D RANS: a LiDAR-driven mid-fidelity model for simulations of wind farm flows}

\author{ {Stefano Letizia}\\
	Wind Fluids and Experiments (WindFluX) Laboratory\\ Mechanical Engineering Department\\ The University of Texas at
Dallas\\ 800 W Campbell Rd, 75080 Richardson, TX, USA\\
	\And
	{Giacomo Valerio Iungo} \\
	Wind Fluids and Experiments (WindFluX) Laboratory\\ Mechanical Engineering Department\\ The University of Texas at
Dallas\\ 800 W Campbell Rd, 75080 Richardson, TX, USA\\
	\texttt{valerio.iungo@utdallas.edu} \\
}

\renewcommand{\shorttitle}{Pseudo-2D RANS: a LiDAR-driven mid-fidelity model for simulations of wind farm flows}

\hypersetup{
pdftitle={A template for the arxiv style},
pdfsubject={q-bio.NC, q-bio.QM},
pdfauthor={David S.~Hippocampus, Elias D.~Striatum},
pdfkeywords={First keyword, Second keyword, More},
}

\begin{document}
\maketitle

\begin{abstract}
	Next-generation models of wind farm flows are increasingly needed to assist the design, operation, and performance diagnostic of modern wind power plants. Accuracy in the descriptions of the wind farm aerodynamics, including the effects of atmospheric stability, coalescing wakes, and the pressure field induced by the turbine rotors, and low computational costs are necessary attributes for such tools. The Pseudo-2D RANS model is formulated to provide an efficient solution of the Navier-Stokes equations governing wind-farm flows installed in flat terrain and offshore. The turbulence closure and actuator disk model are calibrated based on wind LiDAR measurements of wind turbine wakes collected under different operative and atmospheric conditions. A shallow-water formulation is implemented to achieve a converged solution for the velocity and pressure fields across a farm with computational costs comparable to those of mid-fidelity engineering wake models. The theoretical foundations and numerical scheme of the Pseudo-2D RANS model are provided, together with a detailed description of the verification and validation processes. The model is assessed against a large dataset of power production for an onshore wind farm located in North Texas showing a normalized mean absolute error of 5.6\% on the 10-minute-averaged active power and 3\% on the clustered wind farm efficiency, which represent 8\% and 24\%, respectively, improvements with respect to the best-performing engineering wake model tested in this work.
\end{abstract}


\section{Introduction}\label{sec:Intro}

A deeper understanding of the physical processes governing the operation of modern wind farms is instrumental to enable reduction of the levelized cost of energy (LCOE) and increase of wind energy penetration in the global power system \cite{SanzRodrigo2017}. Specifically, many mechanisms governing the aerodynamics of wind turbines immersed in the atmospheric turbulent boundary layer remain still elusive nowadays \cite{Veers2019}. By extracting kinetic energy from the incoming flow, wind turbines generate wakes, which can persist for several kilometers downstream \cite{Christiansen2006,Platis2018,Lundquist2019}, and cause power losses and enhanced fatigue loads for downstream wind turbines \cite{Porte-agel2019}. Wind industry stakeholders identify turbine wakes as one of the major causes for energy losses for onshore power plants in North America \cite{Bailey2014}. Quantitative studies of wake effects based on the analysis of supervisory control and data acquisition (SCADA) data showed energy losses between 2-4\% for onshore sites \cite{El-Asha2017} and 10\% to 20\% of the annual energy production (AEP) for offshore wind farms \cite{Barthelmie2010}. Nonetheless, a recent analysis of data covering ten years of operations for the Lillgrund offshore wind farm revealed a reduction in power capture as high as 28\% of the nominal capacity \cite{Sebastiani2020}. 

The prediction of wake impact on wind farm performance is complicated by the influence of the atmospheric stability on wake recovery \cite{Magnusson1994,Iungo2014,CarbajoFuertes2018,Zhan2020}, which results in significantly reduced power losses for high turbulence intensity, convective atmospheric conditions compared to low turbulence intensity, stable atmospheric conditions \cite{BarthelmieWE2010,Hansen2012,El-Asha2017}. An additional hurdle for predicting wake-induced power losses is represented by the coalescence of multiple wakes, which creates a spatially heterogeneous and highly turbulent flow within the wind farm boundary layer. Furthermore, the pressure field created by the thrust of the rotors induces blockage \cite{MeyerForsting2017,Wu2017,Ebenhoch2017,Bleeg2018,Branlard2020,Schneemann2021} and speedups \cite{Nishino2012,McTavish2014,MeyerForsting2016}, which can significantly impact wind turbine performance, yet generally not modeled in engineering approaches due to the excessive computational burden that a coupled solution of the continuity and momentum equation entails when including the pressure field.

During the past four decades, several wake and wind farm models have been proposed, spanning a breadth of approaches and levels of sophistication. Wake models can be broadly classified into two categories: analytical models and computational fluid mechanics (CFD) methods \cite{Porte-agel2019}. The simplest analytical models assume a top-hat velocity profile and calculate wake width and velocity deficit based on mass \cite{Jensen1983} or momentum \cite{Frandsen2006} conservation. Other analytical wake models assume self-similarity of the wake velocity profiles, in analogy to the shear-flow theory \cite{Schlichting1979,Abramovich1963}, to predict maximum velocity deficit and wake width as a function of the downstream distance from the turbine rotor \cite{Lissaman1979,Voutsinas1990,Larsen1988}. The self-similarity assumption is also at the basis of the Gaussian wake model \cite{Bastankhah2014} and the subsequent evolutions, such as the Ishihara model \cite{Ishihara2018}, the elliptic-Gaussian model \cite{Xie2015}, the Gaussian plume dispersion model \cite{Cheng2018}, the double-Gaussian model \cite{Schreiber2020}, and the super-Gaussian models \cite{Shapiro2019,Blondel2020}. It is noteworthy that the self-similarity hypothesis is strictly valid only if the wake velocity deficit is significantly smaller than the freestream velocity, which is typically true only in the very far-wake \citep{Pope2000}.

The extremely low computational cost and the simple implementation contributed to the popularity of engineering wake models, which are still included in several state-of-the-art tools for wind farm design and modeling \cite{Crasto2012,Mortensen2016,Gebraad2016}. The main drawbacks of most analytical wake models are the need for tuning parameters regulating the wake recovery and the shape of the velocity profiles \citep{Zhan2020WES}. Furthermore, engineering wake models typically require the use of superposition principles in the case of multiple wakes, which introduce a further inaccuracy \cite{Gunn2016,Zong2020}. Nevertheless, if properly calibrated, these models exhibit a mean absolute error in the prediction of wake power losses smaller than 15\% and, unexpectedly, comparable to that of more complex CFD models \cite{Sebastiani2020,Barthelmie2006,Barthelmie2009,Moriarty2014,Archer2018}.

CFD models, especially Reynolds-Averaged Navier-Stokes (RANS) and Large Eddy Simulations (LES) have been increasingly employed to simulate wind farm flows by solving the governing equations for turbulent flows \cite{Santoni2018,Santoni2020}. Specifically, LES have provided insightful information on the turbine wake flow physics (see reviews by \cite{Mehta2014,Breton2017}); however, their high computational cost ($\sim 10^3-10^4$ CPU hours \cite{Porte-agel2019}, depending on the Reynolds number\cite{Sanderse2011}) represent a serious hindrance for their utilization in computationally intensive tasks, such as online control, real-time diagnostic applications, and layout optimization. Furthermore, the sub-grid model needs to be carefully selected to avoid inconsistent results near the ground \cite{Bou-Zeid2005}, for stratified flows \cite{Wan2011} and in the proximity of the turbines \cite{Porte-agel2019}.

Although very sensitive to the type of turbulent closure adopted \cite{Rethore2009,Breton2017}, RANS models have computational costs generally three orders of magnitude smaller than LES \cite{VanDerLaan2015}. The time required for a RANS simulation is intrinsically dependent on the simplifying assumption applied to the Navier-Stokes equations and the solver implemented. Linearization of the advection term is a commonly adopted approximation for early RANS models \cite{Taylor1980,Sforza1981,Crespo1985} and received renewed interest more recently thanks to the achieved low computational costs. Modern linearized models include FUGA \cite{Ott2011}, ORFEUS \cite{Ebenhoch2017}, the curled wake model \cite{Martinez-Tossas2019}. Another popular class of models is the parabolic RANS, where, by leveraging to the boundary-layer approximation, computational costs are reduced by solving the RANS equations from the inlet moving in the downstream direction by neglecting the elliptic nature of subsonic flows. The classic models of Liu \cite{Liu1983}, Ainslie \cite{Ainslie1988}, and UPMWAKE \cite{Crespo1985} belong to this category. More recently, a parabolic axisymmetric solver for turbine arrays was optimally calibrated based on high-fidelity simulations \cite{Iungo2015,Iungo2017,Santhanagopalan2018}. The commercial tool WakeBlaster \cite{Bradstock2020} is also a parabolic 3D wind farm model that can solve the flow of a medium-size wind farm in a few seconds. Furthermore, the curled wake model in his fully linearized version \cite{Martinez-Tossas2019} and the newest release \cite{Martinez-Tossas2020} use a parabolic solution to reduce the computational time. The main drawback of parabolic models is the breakdown of the boundary layer approximation in the near wake \cite{Pope2000} and the inability to simulate the effects of pressure (namely blockage and speedups) on the wind field \cite{Bleeg2018}. 

A more detailed flow characterization can be achieved through elliptic RANS models, which solve the full set of RANS equations at the cost of a significantly higher computational requirement. Crespo and Hend\'andez\cite{Crespo1991} formulated an elliptic implementation of UPMWAKE, whose results showed only slight differences with respect to the original parabolic version. Masson et al.\cite{Masson1997} formulated a steady laminar axisymmetric RANS solver based on the actuator disk theory, subsequently enhanced through the inclusion of a turbulence model\cite{Leclerc1999} and a 3D formulation \cite{Ammara2002}. 
In an attempt to reduce the dimensionality and, thus, computational burden, several authors resolved the wind farm flow at hub height in a purely 2D fashion \cite{Soleimanzadeh2014,Annoni2015,King2016,Adcock2018}, which however can result in an excessive curvature of the streamlines due to the unphysical vertical confinement \cite{King2017}.

A crucial aspect of a RANS model is the modeling of the turbulent Reynolds stresses, aimed at capturing the complex role of atmospheric turbulence, blade- and wake-generated turbulence, and wake dynamics, such as wake meandering \cite{Larsen2008,Keck2012}. The vast majority of the RANS models adopt the linear turbulent eddy viscosity hypothesis \cite{Boussinesq1897}, with just a few examples of Reynolds Stress transport \cite{Gomez-Elvira2005,Cabezon2011} and non-linear eddy viscosity closures \cite{VanderLaan2013}. It is worth mentioning that the eddy viscosity approach has been disproved based on theoretical arguments and experimental evidence for flows undergoing rapid distortion \cite{Pope2000}, and particularly for the flow near wind turbines \cite{Rethore2009}. Nevertheless, the eddy viscosity can provide a useful characterization of mean quantities for simple shear flows \cite{Pope2000}.

Algebraic closures for turbine wakes specify the value of the turbulent eddy viscosity or mixing length a priori \cite{King2017}, based on the Monin-Obukhov similarity theory \citep{Liu1983,Ammara2002,Ott2011,Ebenhoch2017,Martinez-Tossas2019}, wake characteristics \cite{Sforza1981} or both \cite{Ainslie1988,Lange2003,Bradstock2020}. A fairly complex algebraic closure is implemented in the dynamic wake meandering and includes effects of atmospheric stability \cite{Keck2014a}, shear, and added turbulence \cite{Keck2015}. Some authors pursued a more data-driven approach, by calibrating the turbulence model against LES \cite{Iungo2017} or experimental data collected with light detection and ranging (LiDAR) systems \cite{Iungo2018,Adcock2018}. The standard two-equation models (namely $\kappa-\epsilon$ and $\kappa-\omega$), despite their well-documented accuracy in the solution of other classic fluid mechanics problems, are known to be over-diffusive in the near wake \cite{Cabezon2011,Prospathopoulos2011}, where the rapid flow changes, high velocity, and pressure gradients undermine the fundamental assumptions of the eddy viscosity model \cite{Rethore2009}. This has spurred significant research work in the development of advanced turbulence closure for turbine wake flow \cite{ElKasmi2008,VanDerLaan2015,VanDerLaan2015b,Hennen2017}.

The foregoing appraisal of the capabilities and limitations of the state-of-the-art turbine wake models inspired the formulation of a novel mid-fidelity model for wind farm flows: the Pseudo-2D RANS (P2D-RANS hereafter). This tool is conceived to provide a comprehensive description of the time-averaged wind farm aerodynamics, including multiple wake interactions and the effects of atmospheric stability and pressure field while keeping the computational costs low enough to simulate a medium-size farm in a few CPU seconds. With these aims, three main novelties are introduced: $i)$ the 3D flow equations are reduced to a 2D mathematical problem by adopting the shallow-water model with tailor-made corrections for near-wake vertical fluxes and dispersive terms; $ii)$ the partially-parabolized Navier Stokes (PPNS) equations are solved by retaining the full elliptic pressure field yet ensuring numerical efficiency of the iterative marching scheme of the numerical solver; $iii)$ the turbulence closure and actuator disc model are optimally calibrated based on LiDAR data clustered based on incoming wind speed and turbulence intensity \citep{Zhan2020,Zhan2020WES}. We will show in the following that the P2D-RANS is also capable of accurately simulating highly challenging wind conditions where the pressure field induced by the turbine rotors can modify the incoming wind leading to local speedups, and, in turn, increased power capture, which has been confirmed experimentally. Accuracy of the P2D-RANS is assessed for the case study of an onshore wind farm located in North Texas including 25 2.3 MW turbines. The current implementation of the code can simulate the steady-state operation of the investigated wind farm in around 80 s on an i5 single-core laptop computer.


The remainder of this manuscript is organized as follows: the first two sections describe the conceptual (Sec. \ref{sec:Conceptual}) and numerical (Sec. \ref{sec:Computational}) modeling of the depth-averaged, partially-parabolized Navier-Stokes equations. Sec.  \ref{sec:Calibration} provides a description of the experimental site and the dataset used for the tuning of the LiDAR-driven actuator disk and turbulence closure, as well as a detailed outline of the calibration procedure. The novel approach is then verified (Sec. \ref{sec:Verification}), validated against real power data, and compared to other models of wind turbine wakes (Sec.  \ref{sec:Validation}). Finally, concluding remarks are discussed in Sec. \ref{sec:Conclusions}.

\section{Conceptual model}\label{sec:Conceptual}
\begin{figure}[b!]
\centerline{\includegraphics[width=0.65\textwidth]{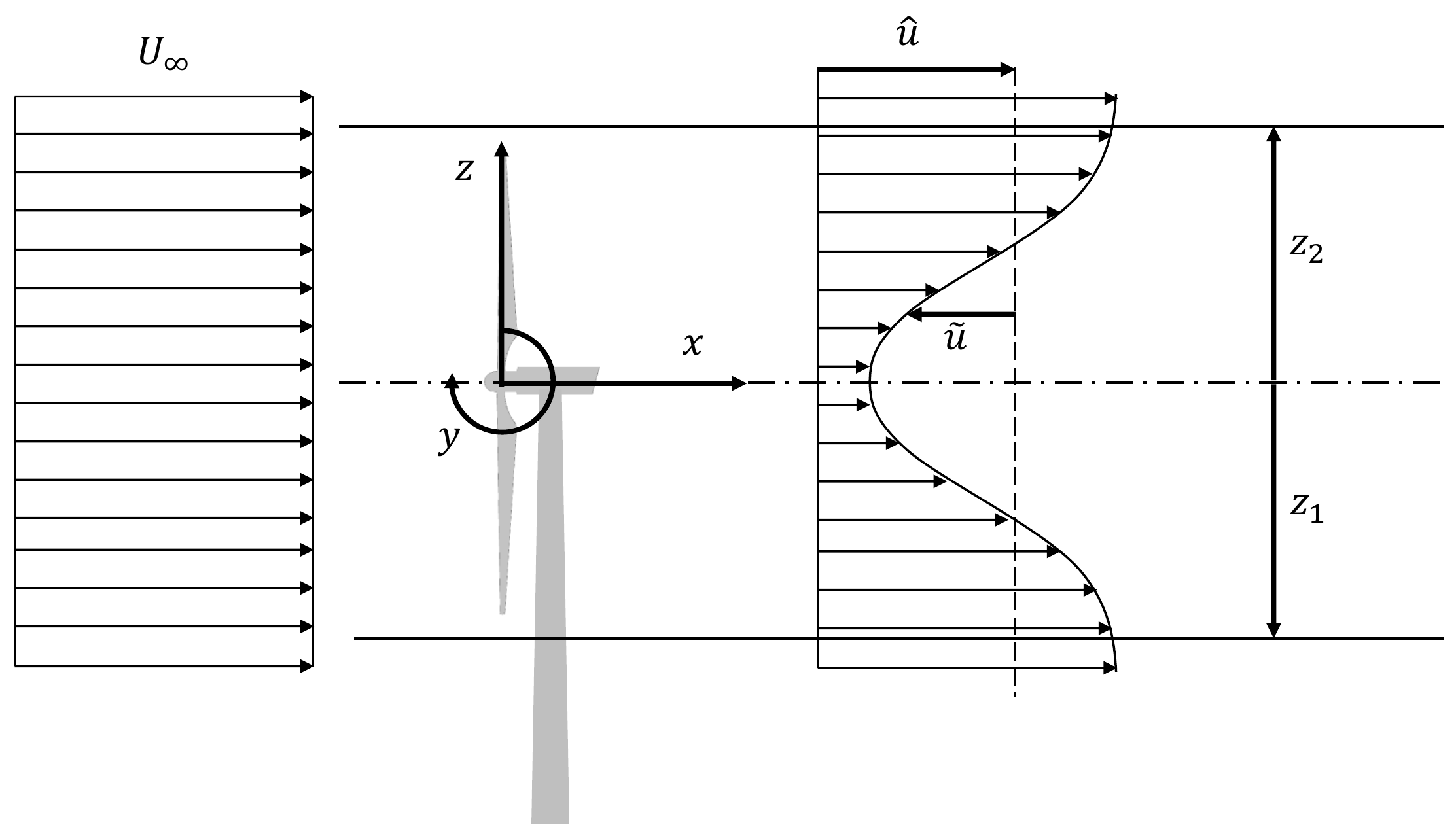}}
\caption{Schematic of the reference system of the P2D-RANS and the shallow-water decomposition of velocity.\label{fig:Turbine_sketch}}
\end{figure}

Under steady conditions, the flow around wind turbines can be described through the RANS equations for incompressible turbulent flows:
\begin{equation}\label{eq:RANS}
\begin{cases}
\frac{\partial \overline{u}_j}{\partial x_j}=0\\
\overline{u}_j\frac{\partial \overline{u}_i}{\partial x_j}=-\frac{\partial \overline{p}}{\partial x_i}+\nu  \frac{\partial^2 \overline{u}_i}{\partial x_j \partial x_j}- \frac{\partial \overline{u'_i u'_j}}{\partial x_j} + \overline{f}_i,
\end{cases}
\end{equation}
where $\overline{\mathbf{u}}$, $\overline{p}$, and $\overline{\mathbf{f}}$ are the Reynolds-averaged velocity vector, pressure, and body force vector, respectively ($\overline{f}_1$ representing the turbine rotor thrust), and $\nu$ is the molecular air viscosity. Within the present notation, repeated indices imply summation and bold quantities indicate vectorial quantities. The $x_1$-axis (or $x$) is aligned with the turbine axis and points downstream, $x_3$ (or $z$) is the vertical direction, and $x_2$ (or $y$) is the traversal direction defined according to a right-handed reference system (Fig. \ref{fig:Turbine_sketch}). The velocity components in the $(x,~y,~z)$ directions are $(u,~v,~w)$, respectively. Throughout the paper, all the physical quantities have been made non-dimensional through the turbine diameter, $D$ (length), incoming wind speed, $U_\infty$ (velocities), undisturbed dynamic pressure $\rho U_\infty^2$, (pressure), and the purely dimensional terms $U_\infty D$ (viscosity) and $\rho U_\infty^2/D$ (forces per unit volume).

The Boussinesq hypothesis allows to express the deviatoric part of the Reynolds stress tensor to be proportional to the strain rate tensor through the turbulent eddy-viscosity, $\nu_T$, as follows\cite{Boussinesq1897}:
\begin{equation}\label{eq:EV}
    \overline{u'_i u'_j}= \frac{1}{3}\delta_{ij} \overline{u'_i u'_i}
    -\nu_T \left (\frac{\partial \overline{u}_j}{\partial x_i}+\frac{\partial \overline{u}_i}{\partial x_j} \right),
\end{equation}
where $\delta_{ij}$ is the Kronecker delta. By incorporating the first term on the RHS of Eq. (\ref{eq:EV}) into a modified pressure and neglecting the molecular viscosity, we obtain:
\begin{equation}\label{eq:RANS_EV}
\begin{cases}
\frac{\partial \overline{u}_j}{\partial x_j}=0\\
\overline{u}_j\frac{\partial \overline{u}_i}{\partial x_j}=-\frac{\partial \overline{p}}{\partial x_i}+\nu_T  \frac{\partial^2 \overline{u}_i}{\partial x_j \partial x_j} + \overline{f}_i.
\end{cases}
\end{equation}

For the P2D-RANS, the dimensionality of the problem is reduced from 3D to 2D by adopting a depth-averaged (or shallow water) approximation  of the RANS equations, which requires the horizontal length and velocity scales of the mean flow to be significantly greater than their respective vertical quantities \citep{Cea2007}. For wind farm flows, this assumption may not be satisfied in presence of significant vertical shear (which is at the moment not considered and can eventually be added a posteriori through a superposition model \cite{Leclerc1999,Larsen2008}) or complex terrain, whose modeling is beyond the scope of this work.

We define a generic Reynolds-averaged scalar, $\overline{\xi}$, as the sum of a vertically-averaged value, $\hat{\xi}$, and a $z$-varying (or dispersive) component, $\tilde{\xi}$, as sketched in Fig. \ref{fig:Turbine_sketch}. The depth-averaged parameter is defined as:
\begin{equation}\label{eq:DepthAverage}
    \hat{\xi}(x,y)=
\frac{1}{z_2-z_1}\int_{z_1}^{z_2} \overline{\xi}(x,y,z) dz 
\end{equation}
where $z_1$ and $z_2$ are the vertical coordinates of the bottom and top boundaries, respectively, of the integration domain. In this work, these boundaries are selected as $z_2=-z_1= D/2$ to include the entire rotor layer, which is convenient to study wind turbines with uniform hub height and rotor diameter on flat terrain. 

Application of the operator defined in Eq. (\ref{eq:DepthAverage}) to both sides of Eqs. (\ref{eq:RANS_EV}) in conservative form, yields:
\begin{equation}\label{eq:RANS_SW}
\begin{cases}
\frac{\partial \hat{u}}{\partial x}+\frac{\partial \hat{v}}{\partial y}+\frac{\overline{w}_2-\overline{w}_1}{z_2-z_1} =0
\\
\frac{\partial \hat{u}} {\partial x}\hat{u}+\frac{\partial \hat{u}}{\partial y}\hat{v}+\frac{\overline{u}_2\overline{w}_2-\overline{u}_1\overline{w}_1-\hat{u}(\overline{w}_2-\overline{w}_1)}{z_2-z_1}
=-\frac{\partial \hat{p}}{\partial x}+
\nu_T \left( \frac{\partial^2{\hat{u}}}{\partial x^2}
+\frac{\partial^2{\hat{u}}}{\partial y^2}\right)\\
+\hat{f_x} + \frac{\nu_T}{z_2-z_1} \left(\frac{\partial{\overline{u}}}{\partial z}\bigg|_2-\frac{\partial{\overline{u}}}{\partial z}\bigg|_1 \right)
-\frac{\partial\widehat{\tilde{u}\tilde{u}}}{\partial x}-\frac{\partial \widehat{\tilde{u}\tilde{v}}}{ \partial y}
\\
\frac{\partial \hat{v}} {\partial x}\hat{u}+\frac{\partial \hat{v}}{\partial y}\hat{v}+\frac{\overline{v}_2\overline{w}_2-\overline{v}_1\overline{w}_1-\hat{v}(\overline{w}_2-\overline{w}_1)}{z_2-z_1}
=-\frac{\partial \hat{p}}{\partial y}+
\nu_T \left( \frac{\partial^2{\hat{v}}}{\partial x^2}+ \frac{\partial^2{\hat{v}}}{\partial y^2}\right)\\
+
\hat{f_y} +
\frac{\nu_T}{z_2-z_1}  \left(\frac{\partial{\overline{v}}}{\partial z}\bigg|_2-\frac{\partial{\overline{v}}}{\partial z}\bigg|_1 \right)
-\frac{\partial\widehat{\tilde{u}\tilde{v}}}{\partial x}-\frac{\partial \widehat{\tilde{v}\tilde{v}}}{ \partial y},\\
\end{cases}
\end{equation}
where the subscript 1 and 2 refer to 3D quantities evaluated at $z_1$ and $z_2$, respectively. Eqs. (\ref{eq:RANS_SW}) include depth-averaged ($\hat{.}$), 3D ($\overline{.}$), and dispersive ($\tilde{.}$) terms. While the first are explicitly solved by the P2D-RANS model, the last two contributions need to be modeled as a function of the depth-averaged field. By neglecting the wake swirling motion, which is only present in the near wake \cite{Iungo2013} and has negligible effects on the streamwise velocity and power \cite{Meyers2010}, and assuming an axisymmetric wake velocity field, which can be recovered for cases with no yaw misalignment of the turbine rotor by subtracting the incoming vertical profile of the incoming streamwise velocity \cite{Medici2006,Chamorro2009}, it is possible to calculate these fluxes following the approach proposed by Boersma et al. \cite{Boersma2018}. 
Specifically, this allows defining linear operators that map the depth-averaged field $(\hat{u},\hat{v})$ onto the corresponding axisymmetric streamwise velocity field $(\overline{u}_x,\overline{u}_r)$, as follows (see Appendix I for details):
\begin{equation}
\label{eq:3D_corr}
\begin{cases}
\overline{u}_x=M_u^{-1}(\hat{u})\\
\overline{u}_r=M_v^{-1}(\hat{v}).\\
\end{cases}
\end{equation}
Once the axisymmetric velocity components corresponding to a given provisional depth-averaged field are known, the vertical fluxes and the dispersive stresses in Eqs. (\ref{eq:RANS_SW}) can be numerically evaluated and fed as source terms into the model in an iterative way, as it will be detailed in Sec. \ref{sec:Computational}. In the case of side-wise multiple wake interaction, which causes a breakdown of the assumption of axisymmetric flow, an appropriate superposition of the single-wake vertical and dispersive fluxes is performed, as it will be described in Sec.  \ref{sec:Verification}.

\section{Computational model}\label{sec:Computational}
The depth-averaged RANS equations (\ref{eq:RANS_EV}) are solved in terms of velocity components using a marching scheme approach in the streamwise direction, $x$, while the pressure field is calculated iteratively due to its inherently elliptic nature. Problems where the streamwise diffusion is negligible but the pressure gradient is not, are referred to as Partially-Parabolized Navier-Stokes (PPNS) \cite{Pletcher2012}, and provide advantages in terms of computational cost and data usage compared to fully elliptic solvers \cite{Pratap1975}. An essential feature of PPNS algorithms is the implementation of a pressure correction necessary to calculate the pressure field and enforce continuity to guarantee a divergence-free velocity field. 

\begin{figure*}
\centerline{\includegraphics[width=1\textwidth, trim=0 0 0 0,clip]{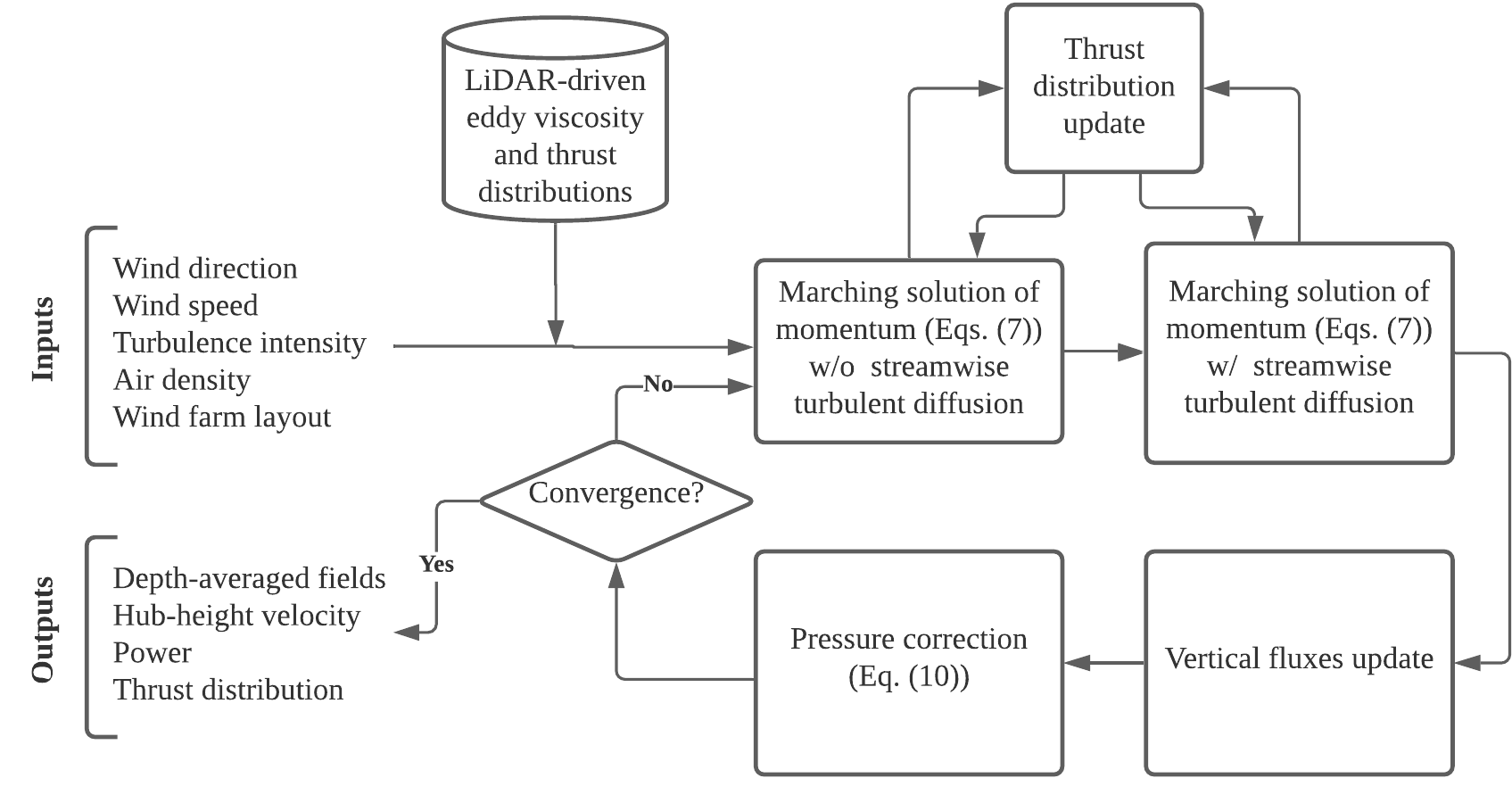}}
 \caption{Workflow of the P2D-RANS.}\label{fig:P2D_sketch}
\end{figure*}

The PPNS scheme adopted for the P2D-RANS is inspired by previous works for ducted flows \cite{Briley1974,Pratap1975,Moore1979,Dodge1977}, with the main difference that no velocity correction is enforced throughout the sweeps. The global pressure field is indeed corrected at the end of each parabolic sweep \cite{Chilukuri1980}. The provisional velocity field does not satisfy mass conservation until the pressure correction achieves convergence (see Fig. \ref{fig:P2D_sketch}).

\begin{figure*}
\centerline{\includegraphics[width=0.35\textwidth, trim=0 0 0 0,clip]{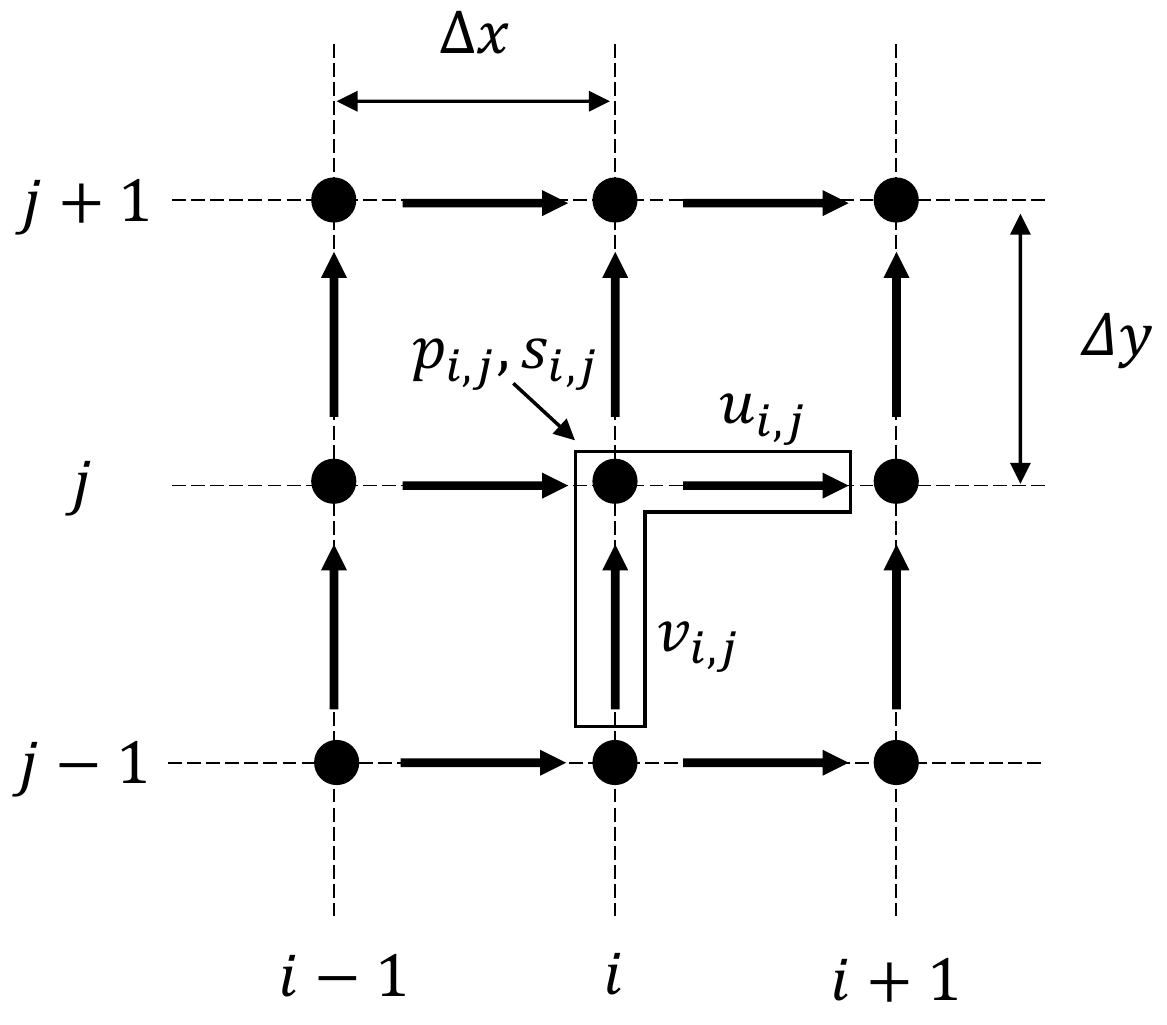}}
 \caption{Numerical grid of the P2D-RANS.}\label{fig:P2D_grid}
\end{figure*}

The numerical scheme adopted for the momentum equations is identical to that documented in \citep{Chilukuri1980}, except for the inclusion of the streamwise turbulent diffusion terms  
in Eqs. (\ref{eq:RANS_SW}), which are retained to enhance accuracy and numerical stability. The scheme is fully implicit and the non-linear terms are treated via the lagging coefficient technique \cite{Pletcher2012}. The uniform grid is staggered so that each velocity node is placed between two pressure nodes (Fig. \ref{fig:P2D_grid}). The momentum equations in their discretized form are:
\begin{equation}\label{eq:uv_discretized}
    \begin{cases}
         \hat{u}_{i,j} \frac{\hat{u}_{i+1,j}-\hat{u}_{i,j}}{\Delta x}+\hat{v}_{i+1,j} \frac{\hat{u}_{i+1,j}-\hat{u}_{i+1,j-1}}{2\Delta y}+\hat{v}_{i+1,j+1} \frac{\hat{u}_{i+1,j+1}-\hat{u}_{i+1,j}}{2\Delta y}=\\ 
         =\frac{\hat{p}_{i+1,j}-\hat{p}_{i+2,j}}{\Delta x}+\nu_T \frac{\hat{u}_{i+2,j}-2\hat{u}_{i+1,j}+\hat{u}_{i,j}}{\Delta x^2}+\nu_T \frac{\hat{u}_{i+1,j+1}-2\hat{u}_{i+1,j}+\hat{u}_{i+1,j-1}}{\Delta y^2}+ \hat{f}_{x,i+1,j}\\
         \frac{\hat{u}_{i,j}+\hat{u}_{i,j-1}}{2} \frac{\hat{v}_{i+1,j}-\hat{v}_{i,j}}{\Delta x}+\hat{v}_{i,j} \frac{\hat{v}_{i+1,j+1}-\hat{v}_{i+1,j-1}}{2\Delta y}=\\
         =\frac{\hat{p}_{i+1,j-1}-\hat{p}_{i+1,j}}{\Delta y}+\nu_T\frac{\hat{v}_{i+2,j}-2\hat{v}_{i+1,j}+\hat{v}_{i,j}}{\Delta x^2}+\nu_T \frac{\hat{v}_{i+1,j+1}-2\hat{v}_{i+1,j}+\hat{v}_{i+1,j-1}}{\Delta y^2}+ \hat{f}_{y,i+1,j},
    \end{cases}
\end{equation}
where the indices $i$ and $j$ refer to the $i^\text{th}$ streamwise and $j^\text{th}$ spanwise locations, respectively. It is noteworthy that the 3D correction terms are tacitly included in the volumetric forces $\hat{f}$. The elliptic terms $\hat{u}_{i+2,j},\hat{v}_{i+2,j},\hat{p}_{i+2,j}$ are evaluated from the previous sweep. As indicated in Fig. \ref{fig:P2D_sketch}, Eqs. (\ref{eq:uv_discretized}) are solved parabolically marching from the inlet to the outlet and twice for each sweep, first without streamwise diffusion and then with the streamwise diffusion activated.

The pressure correction is derived from first principles in Appendix B. The final form of the Poisson equation for the calculation of the corrective pressure field, $p^c$, that is applied at the end of each sweep to correct the provisional velocity is:
 \begin{equation}\label{eq:Press_correction1}
 \begin{array}{l}
    \frac{\partial^2 \hat{p}^c}{\partial x^2}+\frac{\partial^2 \hat{p}^c}{\partial y^2}
     =\hat{u}\frac{\partial \nabla}{\partial x}+\hat{v}\frac{\partial \nabla}{\partial y}-\nu_T\left( \frac{\partial^2 \nabla}{\partial x^2}+\frac{\partial^2 \nabla}{\partial y^2}\right)
 \end{array}
 \end{equation}
 or, more concisely:
   \begin{equation}\label{eq:Press_correction2}
 \begin{array}{l}
    \nabla^2 \hat{p}^c
     =\frac{D\nabla}{Dt}-\nu_T\nabla^2\nabla, 
 \end{array}
 \end{equation}
where $\nabla$ is the divergence of the provisional velocity field, including the correction for vertical mass fluxes. From a numerical standpoint, the pressure correction is discretized as:
 \begin{equation}\label{eq:p_discretized}
    \frac{\hat{p}^c_{i+1,j}-2\hat{p}^c_{i,j}+\hat{p}^c_{i-1,j}}{\Delta x^2}+\frac{\hat{p}^c_{i,j+1}-2\hat{p}^c_{i,j}+\hat{p}^c_{i,j-1}}{\Delta y^2}=s_{i,j}
 \end{equation}
 where:
   \begin{equation}
   \begin{array}{l}
         s_{i,j}=\hat{u}_{i-1,j} \frac{\nabla_{i,j}-\nabla_{i-1,j}}{2\Delta x}+\hat{u}_{i,j} \frac{\nabla_{i+1,j}-\nabla_{i,j}}{2\Delta x}+\hat{v}_{i,j} \frac{\nabla_{i,j}-\nabla_{i,j-1}}{2\Delta y}
         \\+\hat{v}_{i,j+1} \frac{\nabla_{i,j+1}-\nabla_{i,j}}{2\Delta y}
         -\nu_T\left(\frac{\nabla_{i+1,j}-2\nabla_{i,j}+\nabla_{i-1,j}}{\Delta x^2}+\frac{\nabla_{i,j+1}-2\nabla_{i,j}+\nabla_{i,j-1}}{\Delta y^2}\right)
     \end{array}
 \end{equation}
 and:
  \begin{equation}
    \nabla_{i,j}=\frac{\hat{u}_{i,j}-\hat{u}_{i-1,j}}{\Delta x}+\frac{\hat{v}_{i,j+1}-\hat{v}_{i,j}}{\Delta y} + \frac{\overline{w}_{2,i,j}-\overline{w}_{1,i,j}}{z_2-z_1},
 \end{equation}
where the last term represents the vertical mass flux obtained as explained in Sec.  \ref{sec:Conceptual}. 
 
The Poisson equation for the pressure is solved through the $backslash$ Matlab solver for sparse linear systems \cite{HaglundElGaidi2018}. To speed up the convergence, the pressure correction is applied alternatively to all the cells and in blocks of 2$\times$2 cells \cite{Murthy1990}. Convergence is considered to be achieved when the mass residual, defined as the integral of the absolute value of the mass sources and sinks normalized by the incoming mass flow rate is below 1\% \citep{Chilukuri1980}. 

Dirichlet boundary conditions on velocity ($\hat{u}=1, \hat{v}=0$) and pressure ($\hat{p}=0$) are prescribed at the inlet and outlet, respectively. Non-homogeneous Neumann conditions are applied to the side boundaries by imposing the spanwise gradients of the velocity components to be equal to their respective values at the nearest $y$ locations, which minimizes the back-propagation of boundary effects in the inner domain  \cite{Iungo2018}. Finally, homogeneous Neumann boundary condition is applied to the remaining boundaries for pressure, which is consistent with the Navier-Stokes equations \cite{Pletcher2012}. The spatial resolution of the grid is selected as $\Delta x=0.125$ and $\Delta y=0.05$, while the domain extends 15$D$ away from the turbines in all directions. These parameters are selected after a sensitivity analysis to ensure a maximum relative error on the streamwise component smaller than 1$\%$ in every point of the domain for both single-turbine and full wind farm simulations.

\section{Calibration}\label{sec:Calibration}
\subsection{Experimental Dataset}
\begin{figure*}
\centerline{\includegraphics[width=\textwidth]{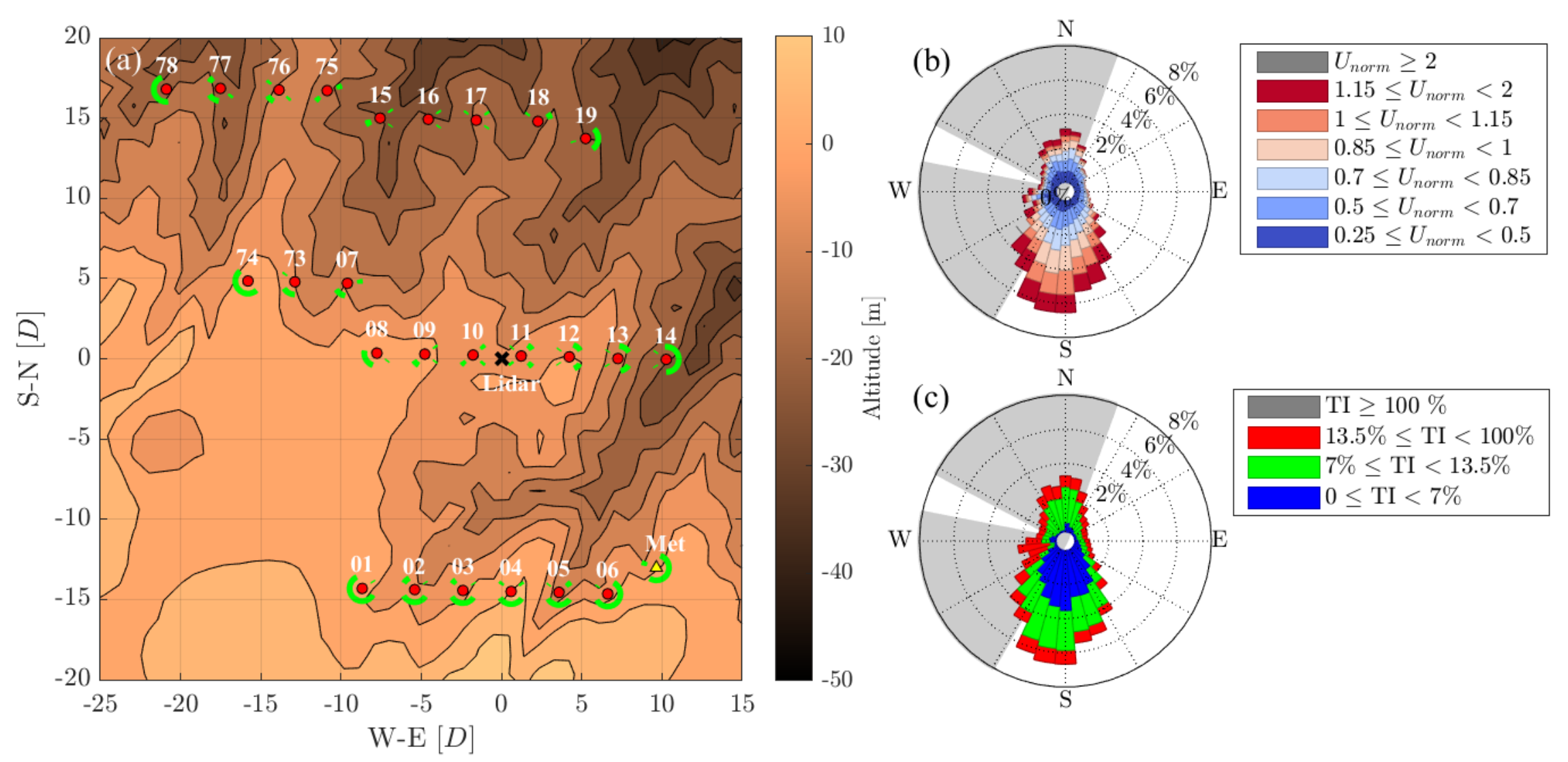}}
\caption{Characterization of the wind farm under investigation: (a) topographic map of the site; (b) directional histogram of the normalized wind speed at hub height (Eq. (\ref{eq:U_norm})); (c) directional histogram of the incoming turbulence intensity at hub height. The green sectors in (a) represent the unwaked wind sectors, while the grey ones in (b) and (c) correspond to wind sectors where the met mast is likely affected by turbine wakes.}\label{fig:Panhandle_map_v2}
\end{figure*}

To enhance accuracy in predictions of the intra-wind-farm velocity field and wind turbine power capture, the P2D-RANS has been developed by encompassing a fully data-driven actuator disk model and turbulence closure, both calibrated through wind LiDAR measurements of individual wakes generated by utility-scale wind turbines. The onshore wind farm under examination is located in North Texas and includes 25 identical 2.3 MW turbines with a rotor diameter of $D=108$ m and a hub height of $H=80$ m (Fig. \ref{fig:Panhandle_map_v2}). A WindCube 200S pulsed Doppler wind LiDAR equipped with a rotating scanning head was deployed from August 2015 to March 2017 (see Fig. \ref{fig:Panhandle_map_v2}). This section provides a concise overview of the field campaign, since a detailed description of the site, experimental strategy, and LiDAR data processing can be found in previous related publications \cite{El-Asha2017,Zhan2020,Maulik2021}. 

A meteorological (met) mast is located at the South-East corner of the farm and recorded for the whole duration of the field campaign mean and standard deviation of horizontal wind speed and direction at 36-m, 60-m, and 80-m heights, barometric pressure at 2-m and 75-m heights, and temperature at 3-m and 75-m heights. Furthermore, SCADA data in the form of mean and standard deviation of hub-height wind speed, active power, and other operational parameters were acquired.

The wind LiDAR was set up with a range gate of 50 m and an accumulation time of 500 ms. The scanning wind LiDAR performed narrow Plan Position Indicator (PPI) scans with low elevation angles (typically $3^{\circ}$), azimuthal range of $20^{\circ}$, and rotation speed of the scanning head of $2^{\circ}$ s$^{-1}$, leading to a typical scanning time for a single PPI scan of 10 s. PPI scans were performed for wind directions included within the sector 145$^\circ$-235$^\circ$, when the wakes of turbines 1-6 are advected towards the LiDAR location. Overall, 16765 quality-controlled PPI scans are available for the present study. 

Individual LiDAR samples undergo a quality control process, including data rejection based on the LiDAR signal intensity, which is performed based on the dynamic filtering \cite{Beck2017}. After realigning the wake according to the 10-minute-averaged estimated wind direction, the horizontal equivalent velocity field, $u_\text{eq}$, is made non-dimensional through the 10-minute-averaged incoming vertical wind profile, according to the procedure outlined in Zhan et al. \cite{Zhan2020}. Finally, the LiDAR measurements are reported in an axisymmetric-equivalent reference frame $(x,r)$, where the radial position, $r$, represents the distance from the expected wake center \citep{Zhan2020}. The re-aligned and non-dimensional horizontal-equivalent wake velocity data are clustered in bins based on operative conditions (i.e. aerodynamic rotor thrust) and incoming wind turbulence intensity. The selection of the bin edges is guided by a preliminary characterization of the wind farm performance as a function of the inflow conditions, which is the object of the following subsection.

\subsection{Wind turbine and wind farm performance}
The operation of a pitch-regulated turbine can be characterized as a function of the hub-height density-corrected wind speed \cite{IEC61400_12_1}, as follows:
\begin{equation}\label{eq:U_norm}
    U_\text{norm}=U_\text{hub}\left(\frac{\rho}{\rho_0}\right)^{1/3},
\end{equation}
where $U_\text{hub}$ is the 10-minute-averaged horizontal wind speed at hub height, $\rho$ is the measured air density, and $\rho_0=1.225$ kg m$^{-3}$ is the reference air density. Wake turbulent mixing and recovery can be conveniently investigated through the turbulence intensity, TI, defined as the ratio between the standard deviation of the hub-height wind speed over the mean of the same quantity. Specifically, Zhan et al \cite{Zhan2020} identified through an atmospheric stability analysis values of $7\%$ and $13.5\%$ as thresholds of the undisturbed turbulence intensity bounding stable (TI$_\infty<7\%$), near-neutral (TI$_\infty\in[7,13.5)\%$), and convective (unstable) (TI$_\infty\geq13.5\%$) conditions for the site under investigation. In general, the subscript $\infty$ refers to quantities recorded only by the SCADA data of turbines unaffected by wakes and, thus, representative of the incoming wind field. In this work, wind sectors affected by wakes are defined according to the 61400-12, Annex B of the IEC standards \cite{IEC61400_12_2}.

Wind speed at hub height, $U_\text{hub}$, is retrieved based on the measurements of the nacelle anemometer, which undergoes quality control and correction processes. Firstly, occurrences of unrealistic velocity and turbulence intensity, extreme veer, yaw misalignment, idling sensors, and zero or negative active power are rejected (see Table \ref{tab:Quality_check}). Then the nacelle transfer function is calculated based on met-tower normalized wind speed collected at 80-m height and the nacelle anemometer of Turbine 06 (namely the closest turbine to the met tower) only for unwaked wind sectors, in compliance with the International IEC Standards \cite{IEC61400_12_2}. Such transfer function cannot be applied directly to the whole park since biases of up to 5\% between different SCADA anemometers are present. The mentioned biases are estimated by comparing the raw individual power curves \cite{Sebastiani2020}. Therefore, all the wind speed data are re-scaled by a multiplicative factor using the data of Turbine 06 as a reference before the correction for the nacelle flow distortion.
\begin{table}[]
\centering
\caption{Breakdown of the quality check on met and SCADA data.}\label{tab:Quality_check}
\begin{tabular}{lll}
 & Formulation   & Rejection rate {[}\%{]} \\
 \hline
Veer                  & $\frac{d\theta_w}{dz}>0.25 ^\circ$ m$^{-1}$ & 16.3                    \\
Idling wind direction & $\frac{d \theta_w}{dt}=0$                  & 2.1                     \\
Yaw misalignment      & $|\theta_\text{yaw}-\theta_w|>30 ^\circ$   & 1.6                     \\
Idling power          & $\frac{d P_\text{norm}}{dt}=0$             & 0.04                    \\
No power              & $P_\text{norm}\leq0$                        & 9.9                     \\
Zero velocity         & $U_\text{norm}=0$                          & 0.02                    \\
Off-range TI          & TI$\leq0$ or TI$>100 \%$                    & 1.1\\                  
\end{tabular}
\end{table}

The experimental turbine power curve is calculated by leveraging the data from all the turbines following the IEC Standards \cite{IEC61400_12_2}. To flag power curtailments associated with off-design operations, an initial nominal power curve is generated based on the power of reference Turbine 06 only and the met-tower normalized wind speed and only for unwaked sectors. Power reads exhibiting a deficit larger than 300 kW from the power estimated through the nominal power curve ($\sim 1\%$ of the cases) are then rejected \citep{Hamilton2020}. The final power curves are generated using all the available SCADA data (August 2015 to April 2017) and differentiated as a function of TI \cite{Teng2020}. It is noteworthy that power in region III can exceed the nameplate value by virtue of the power-boost mode, which enables the generator to produce additional power under favorable operative conditions.
\begin{figure*}
\centerline{\includegraphics[width=0.9\textwidth]{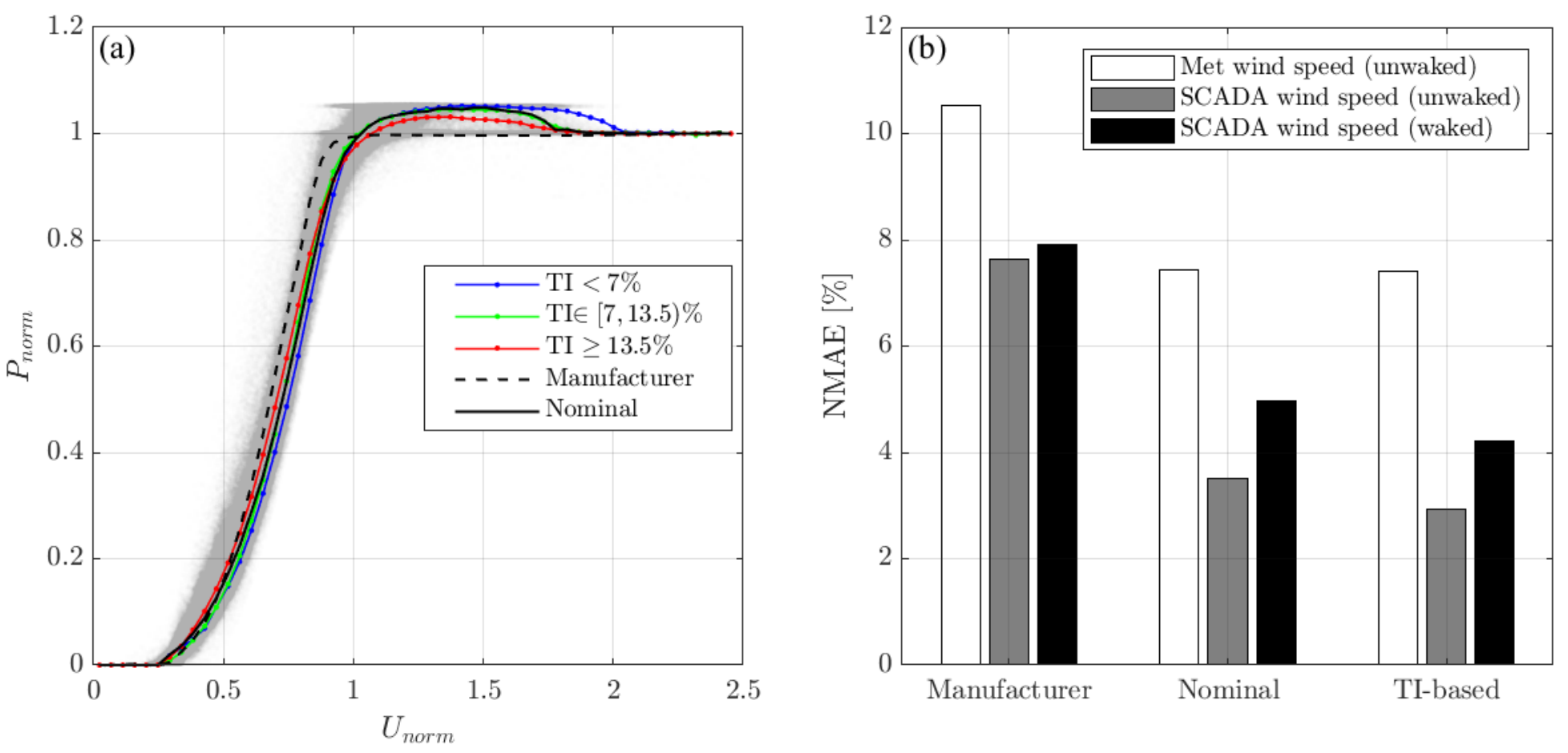}}
 \caption{(a) Power curves for different stability classes. The grey dots represent quality-controlled 10-minute-averaged data. (b) Normalized Mean Absolute Error (NMAE) of the SCADA 10-minute-averaged power and the power obtained through several power curves and using different inputs as in incoming velocity.}\label{fig:Power_curve}
\end{figure*}
The TI-based normalized power ($P_\text{norm}$) curves are reported in Fig. \ref{fig:Power_curve}(a). The accuracy of this method is assessed quantitatively by evaluating the error between the measured power and the power obtained from the different power curves and using either met or SCADA (corrected) hub-height wind speeds as input variables. The error metric used here is the normalized mean absolute error \cite{Moriarty2014}, which is defined as:
\begin{equation}
    \text{NMAE}(x,x_\text{ref})=\frac{\sum_i|x_i-x_{i,\text{ref}}|}{\sum_i x_{i,\text{ref}}},
\end{equation}
where $x$ and $x_\text{ref}$ represent the modeled and reference value of the target quantity, respectively.
This error analysis indicates that the best agreement with the SCADA data is achieved by using the TI-based power curves and the corrected SCADA wind speed as input. Slightly worse accuracy is obtained by using the nominal power curve. The highest error associated with the met tower wind speed is due to the presence of spatial heterogeneity of the flow over the farm \cite{Hamilton2020}. Further, higher uncertainty is observed for the power output of waked turbines, likely due to the  more heterogeneous velocity distribution over the rotor area not captured by the nacelle anemometer. 

\begin{figure*}
\centerline{\includegraphics[width=\textwidth]{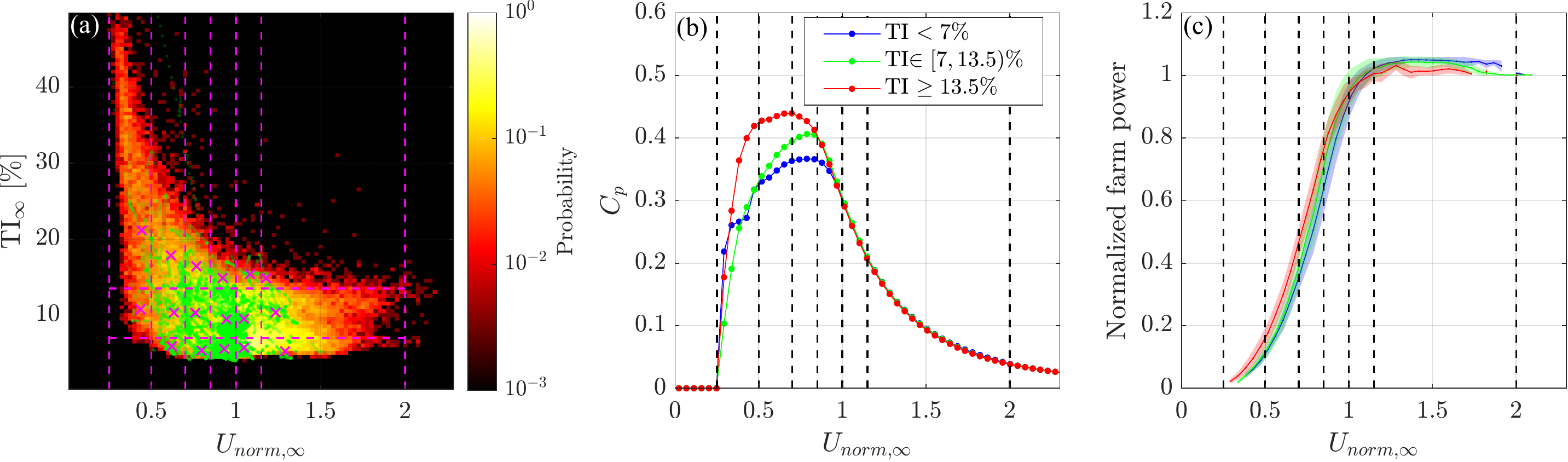}}
\caption{Selection of the bin edges: (a) two-dimensional pdf of $U_{\text{norm},\infty}$ and TI$_\infty$ (green dots correspond to available LiDAR scans, while crosses indicate the bin centroids); (b) average $c_p$ experimental curves; (c) overall normalized farm power as a function of $U_{\text{norm},\infty}$, where the shaded areas represent standard deviation. The dashed lines separate the different bins in $U_{\text{norm},\infty}$.}\label{fig:Cluster_stats}
 \end{figure*}

Similar procedures are used to evaluate the power coefficient, $c_p$, and total farm power curve, which guide the selection of the bin edges of the undisturbed normalized hub-height velocity, $U_{\text{norm},\infty}$, used for the clustering of the LiDAR data. The bin selection is summarized in Fig. \ref{fig:Cluster_stats}. Specifically, the heat map in Fig. \ref{fig:Cluster_stats}(a) shows the occurrence of a $[U_{\text{norm},\infty},$ TI$_\infty]$ pairs based on the all available SCADA data and averaged over the set of unwaked turbines. 
The green dots refer to the SCADA data paired with the wake LiDAR scans, which can be seen to cover satisfactorily most of the typical wind conditions, except for the $U_{\text{norm},\infty}<0.5$, TI$_\infty<13.5\%$ region. For this study, bin edges in $U_{\text{norm},\infty}$ are selected as $[0.25, 0.5, 0.7, 0.85, 1, 1.15, 2]$ (see Fig. \ref{fig:Cluster_stats}(b)), and are purposely refined in the proximity of the turbine rated power (i.e. $U_{\text{norm},\infty}=1$), where a stronger variation of $c_p$ and, thus, thrust coefficient, are expected. Finally, as highlighted by the farm power curve calculated including all the turbines regardless of their waked/unwaked state (Fig. \ref{fig:Cluster_stats}(c)), for $U_{\text{norm},\infty}>1.15$, the average total farm power is practically equal to the nominal capacity of the wind plant. The numerical modeling of such cases is irrelevant for the assessment of the proposed wind farm model in terms of power capture, and further bins for very high $U_{\text{norm},\infty}$ are thus not defined. Table \ref{tab:Cluster_points} reports the number of raw and quality-controlled experimental LiDAR samples available for each cluster.

\begin{table}
\caption{Number (in millions) and percentage (in brackets) of quality-controlled LiDAR samples available for each cluster.}\label{tab:Cluster_points}
\centering
\begin{tabular}{lccc} 
& TI$_\infty <7\%$ & TI$_\infty \in [7,13.5)\%$ & TI$_\infty\geq13.5\%$\\
\hline
$U_{\text{norm},\infty}\in[0,0.25)$ & 0    & 0.04 (0.3)  &       0.24 (1.8)\\
$U_{\text{norm},\infty}\in[0.5,0.7)$ &0.29 (2.2) &0.88 (6.6) &0.52 (3.9) \\
$U_{\text{norm},\infty}\in[0.7,0.85)$ &0.77 (5.8) &1.32 (9.9) &0.42 (3.1) \\
$U_{\text{norm},\infty}\in[0.85,1)$ &1.77 (20.7) &2.87 (21.5) &0.3 (2.2)\\
$U_{\text{norm},\infty}\in[1,1.15)$ &0.58 (4.4) &1.46 (11.0) &0.16 (1.2) \\
$U_{\text{norm},\infty}\in[1.15,2)$ &0.21 (1.6) &0.48 (3.6) &0.03 (0.2) \\
\end{tabular}
\end{table}

\subsection{Data-driven calibration of the RANS model}
\begin{figure*}[b!]
\centerline{\includegraphics[width=\textwidth]{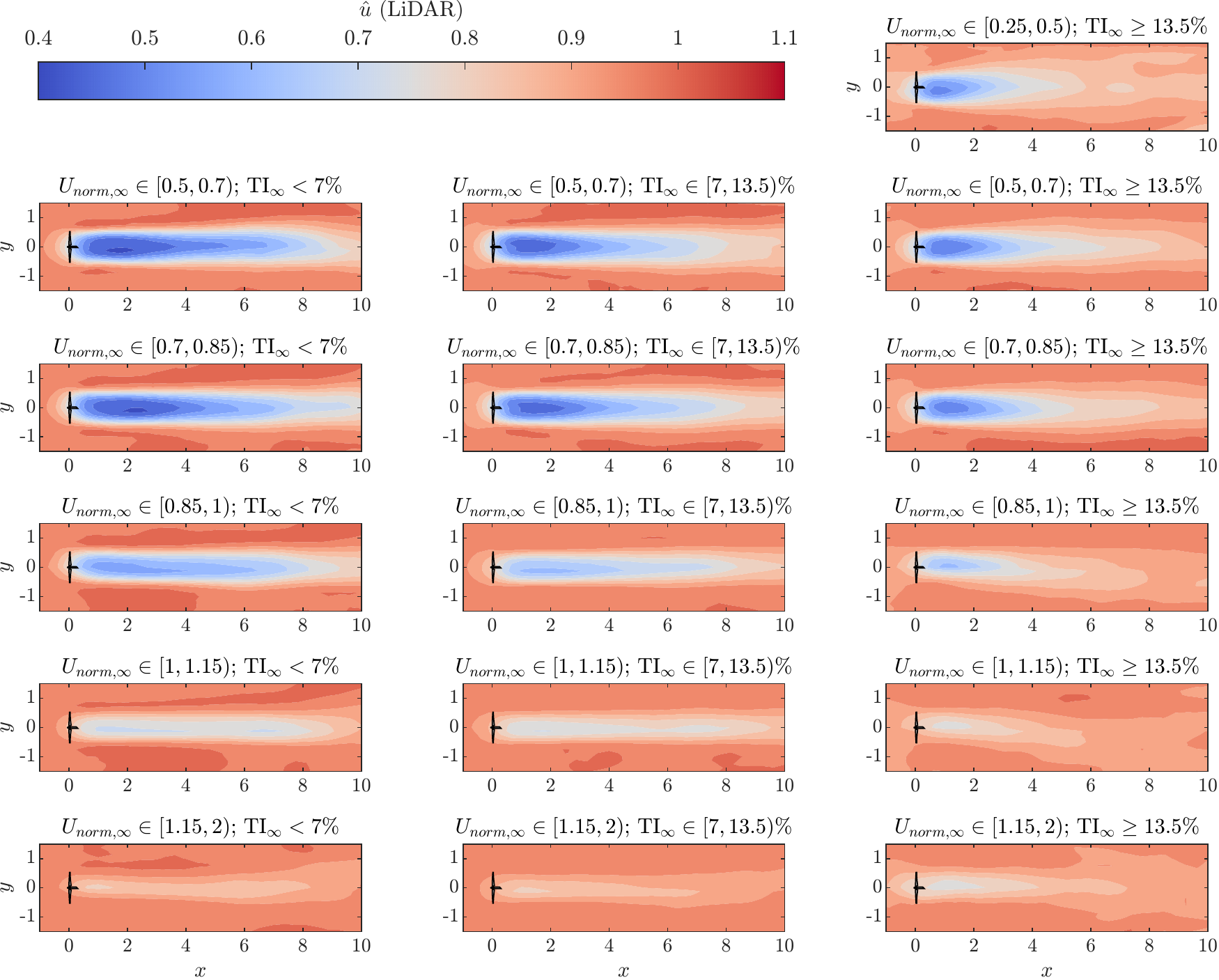}}
 \caption{Mean streamwise, depth-averaged velocity fields ($\hat{u}$) calculated from the LiDAR data for the various bins defined based on hub-height wind speed, $U_{\text{norm},\infty}$, and, turbulence intensity, $TI_\infty$.}\label{fig:Cal_u_LiDAR}
 \end{figure*}

The mean streamwise velocity for each cluster is reconstructed using the LiDAR Statistical Barnes Objective Analysis (LiSBOA) tool \cite{Letizia2021,Letizia2021b} on a Cartesian grid with fundamental half-wavelengths $\Delta n_{0,x}=1.5 D,\Delta n_{0,r}= 0.5 D$ and region with local data spacing $\Delta \tilde{d}>0.25$ are rejected (refer to \cite{Letizia2021} for more theoretical details on LiSBOA). Data voids are filled through a bi-harmonic interpolation algorithm \cite{inpaintnans}. The velocity fields in axisymmetric coordinate are finally depth-averaged to build the reference dataset of $\hat{u}(x,y,U_{\text{norm},\infty},\text{TI}_\infty)$, which are reported in Fig. \ref{fig:Cal_u_LiDAR}. Results are not shown for two bins, which fail to satisfy the statistical constraints imposed by the LiSBOA due to the limited availability of data points. The used methodology for the characterization of the wake velocity field captures satisfactorily the reduced velocity deficit occurring for cases with lower $c_t$ (i.e. for $U_{\text{norm},\infty} \gtrapprox 0.85$), as well as the faster wake recovery occurring with increasing turbulence intensity. The quantification of the wake variability with incoming wind speed and turbulence intensity is instrumental for the calibration of the actuator disk and the turbulence closure model of the P2D-RANS and, more in general, for accurate modeling of the wind farm flow.
 
The tuning of the turbulent eddy-viscosity and axial force distribution are performed following the methodology introduced by Iungo et al. \cite{Iungo2018}, but adopting depth-averaged velocity fields and the P2D-RANS model instead of the axisymmetric flow and RANS model of the original publication. The present calibration procedure includes the following steps:
\begin{itemize}
     \item the near wake length, $x_\text{NW}$, namely the extension of the region where the pressure gradients are relevant, is estimated as the streamwise location where the  minimum rotor-averaged velocity occurs \cite{Vermeulen1980};
     \item a constant turbulent eddy-viscosity is estimated by minimizing the mean absolute percentage error (MAPE) between the average streamwise wake velocity predicted with the P2D-RANS and that obtained from the LiDAR data, where:
     \begin{equation}
         \text{MAPE}=\bigg\langle\frac{|\hat{u}_\text{P2D-RANS}(x,|y|<1)-\hat{u}_\text{LiDAR}(x,|y|<1)|}{\hat{u}_\text{LiDAR}(x,|y|<1)}\bigg\rangle,
     \end{equation}
     with $\langle \rangle$ indicating space average. In this case, the modified P2D-RANS solver does not include the rotor region, instead the $\hat{u}$ velocity profile calculated from the LiDAR data at $x_\text{NW}$ is injected as inlet boundary condition, while assuming constant pressure to calculate $\hat{v}$. The optimization algorithm used to estimate turbulent eddy-viscosity is described in Iungo et al.\cite{Iungo2018};
     \item once the optimal $\nu_T$ is known, the axial thrust force on the turbine rotor is estimated by correcting iteratively the streamwise forcing, $\hat{f}_x$, in the full version of the P2D-RANS model to match the experimental velocity profile at $x_\text{NW}$ as:
     \begin{equation}  
     \begin{array}{l}
     \int_{-\infty}^{+\infty} \hat{f_x}^{n+1}(x,y)~dx=\int_{-\infty}^{+\infty} \hat{f_x}^{n}(x,y)~dx
     +\hat{u}_\text{P2D-RANS}(x_{NW},y)^2-\hat{u}_\text{LiDAR}(x_\text{NW},y)^2,
     \end{array}
     \end{equation}
     where the index $n$ is the iteration counter. This method represents a simplified version of the technique based on the local momentum budget \cite{Iungo2018}. 
     It is noteworthy that the integrals in the $x$-direction are included only to compensate for the streamwise smoothing of the forces applied in the P2D-RANS to prevent numerical instabilities (further details on the formulation of the aerodynamic loads are provided in Appendix \ref{app:Loads}). The algorithm stops when variation in the above-defined MAPE lower than 0.1\% are attained (generally with less than 5-6 iterations).
\end{itemize}
 
\begin{figure*}[b]
\centerline{\includegraphics[width=\textwidth]{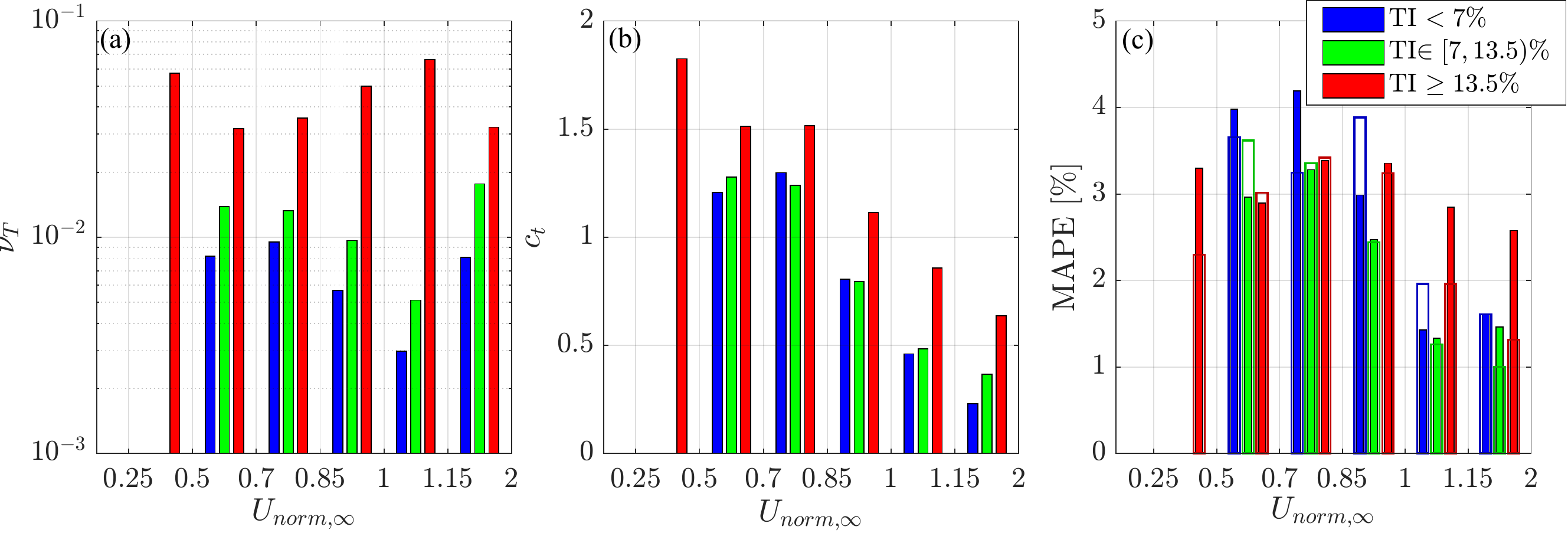}}
 \caption{Calibration of the P2D-RANS: (a) optimal turbulent eddy-viscosity; (b) optimal thrust coefficient, $c_t$; (c) mean absolute percentage error (MAPE) (empty bars represent the MAPE for the far wake velocity field after the optimization of $\nu_T$ only, and the full ones represent the final MAPE after the estimation of the axial load).}\label{fig:Cal_EV_Ct}
 \end{figure*}

\begin{figure*}[b!]
\centerline{\includegraphics[width=\textwidth]{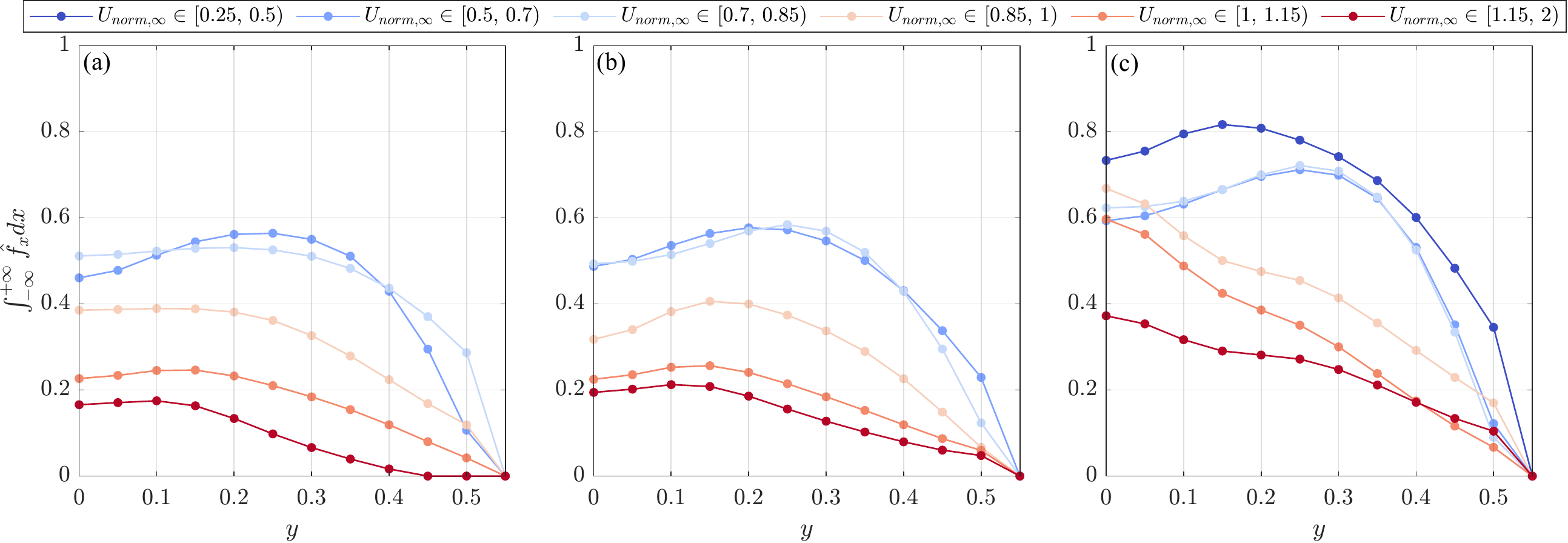}}
 \caption{Optimal axial forcing distribution after calibration: (a) TI$_\infty<7\%$; (b) TI$_\infty$ $\in[7,13.5)\%$; (c) TI$_\infty\geq13.5\%$.}\label{fig:Cal_loads}
 \end{figure*}

The results of the calibration procedure are summarized in Figs. \ref{fig:Cal_EV_Ct} and \ref{fig:Cal_loads}.
The behavior of the optimal turbulent eddy-viscosity (Fig. \ref{fig:Cal_EV_Ct}(a)) lends interesting insight into the wake recovery process. The most consistent pattern is the increase of $\nu_T$ as a function of TI$_\infty$, which is expected since TI$_\infty$ is linked to the magnitude of the turbulent momentum fluxes in the wake. The turbulent eddy viscosity increases by nearly tenfold in convective conditions compared to stable conditions, which confirms the necessity of including effects of atmospheric stability in wind farm models. Furthermore, the turbulent eddy-viscosity for low and moderate TI$_\infty$ decreases as a function of $U_{\text{norm},\infty}$, except for the last bin $U_{\text{norm},\infty}\in [1.15,2)$. Our interpretation is that the $\nu_T$ optimizer identifies effects related to the wake-added turbulence, which is indeed stronger for higher $c_t$ values (i.e. low $U_{\text{norm},\infty}$) conditions due to the enhanced shear-generated turbulence. The proportionality of the turbulent eddy viscosity to the velocity deficit, $\Delta u$, (i.e. $c_t$) is predicted by the shear flow theory, which produces the well-known scaling $\nu_T \sim r_w \Delta u$, where $r_w$ is the wake radius, for a self-similar axisymmetric wake \cite{Sforza1981,Ainslie1988,Lange2003}. 

It is noteworthy that $\nu_T$ for convective conditions increases between $U_{\text{norm},\infty}=0.5$ to $1.15$. This trend, which is reversed compared to what is observed for low to moderate incoming TI$_\infty$ conditions, is likely due to a different wake recovery mechanism occurring for unstable atmospheric conditions, where the incoming turbulence significantly contributes to the momentum diffusion in the wake, especially through the flow dynamics connected with wake meandering \cite{Larsen2008}. Furthermore, the inverse proportionality between $\nu_T$ and $c_t$ observed in convective conditions may be due to the damping of the largest incoming coherent structures operated by the rotor already documented by Chamorro et al. \cite{Chamorro2012}, which is expected to be more severe for high $c_t$. A more thorough investigation of this interesting behavior will be the object of future work. Finally, the reduction of $\nu_T$ observed for the last bin $U_{\text{norm},\infty}\in[1.15,2)$ is ascribed to the very shallow wake deficit, which makes the definition of an eddy viscosity quite elusive.

Fig. \ref{fig:Cal_EV_Ct}(b) and \ref{fig:Cal_loads} show $c_t$ and the spanwise distribution of depth-averaged thrust force. The thrust coefficient is practically constant for $0.25 \leq U_{\text{norm},\infty}<0.7$ (i.e. region II of the power curve), then it decreases due to the blade pitching and tip-speed ratio reduction operated by the controller to limit the power output to the nameplate capacity. The only available thrust-force profile at very low $U_{\text{norm},\infty}$ corresponds to convective conditions, which shows an overall higher aerodynamic force. The force distribution is less sensitive to the atmospheric stability, although slightly higher values are estimated in high TI conditions \citep{Iungo2018}.

\begin{figure*}
\centerline{\includegraphics[width=\textwidth]{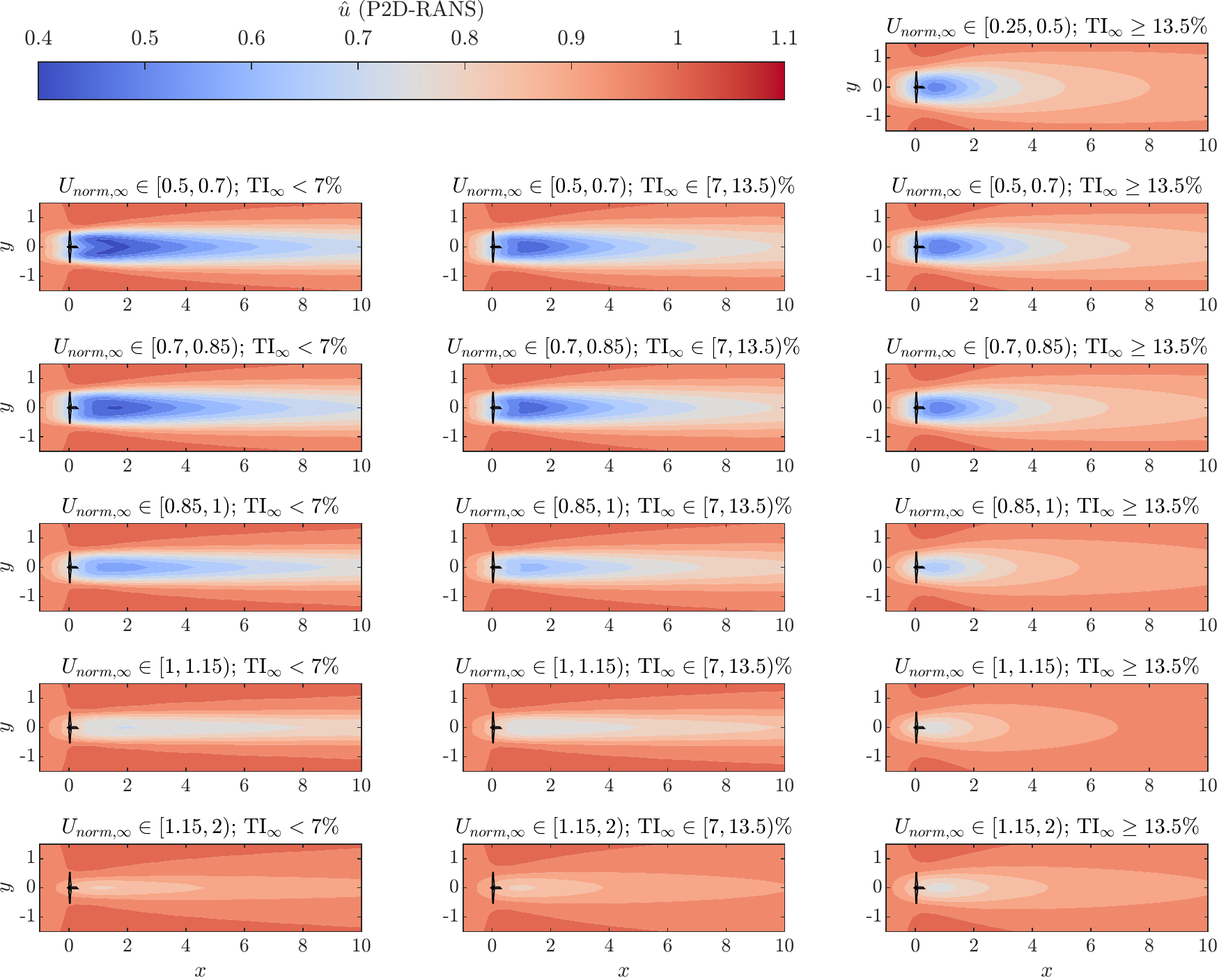}}
 \caption{Mean streamwise, depth-averaged velocity fields ($\hat{u}$) of the P2D-RANS after calibration.}\label{fig:Cal_u_P2D}
 \end{figure*}

\begin{figure*}
\centerline{\includegraphics[width=\textwidth]{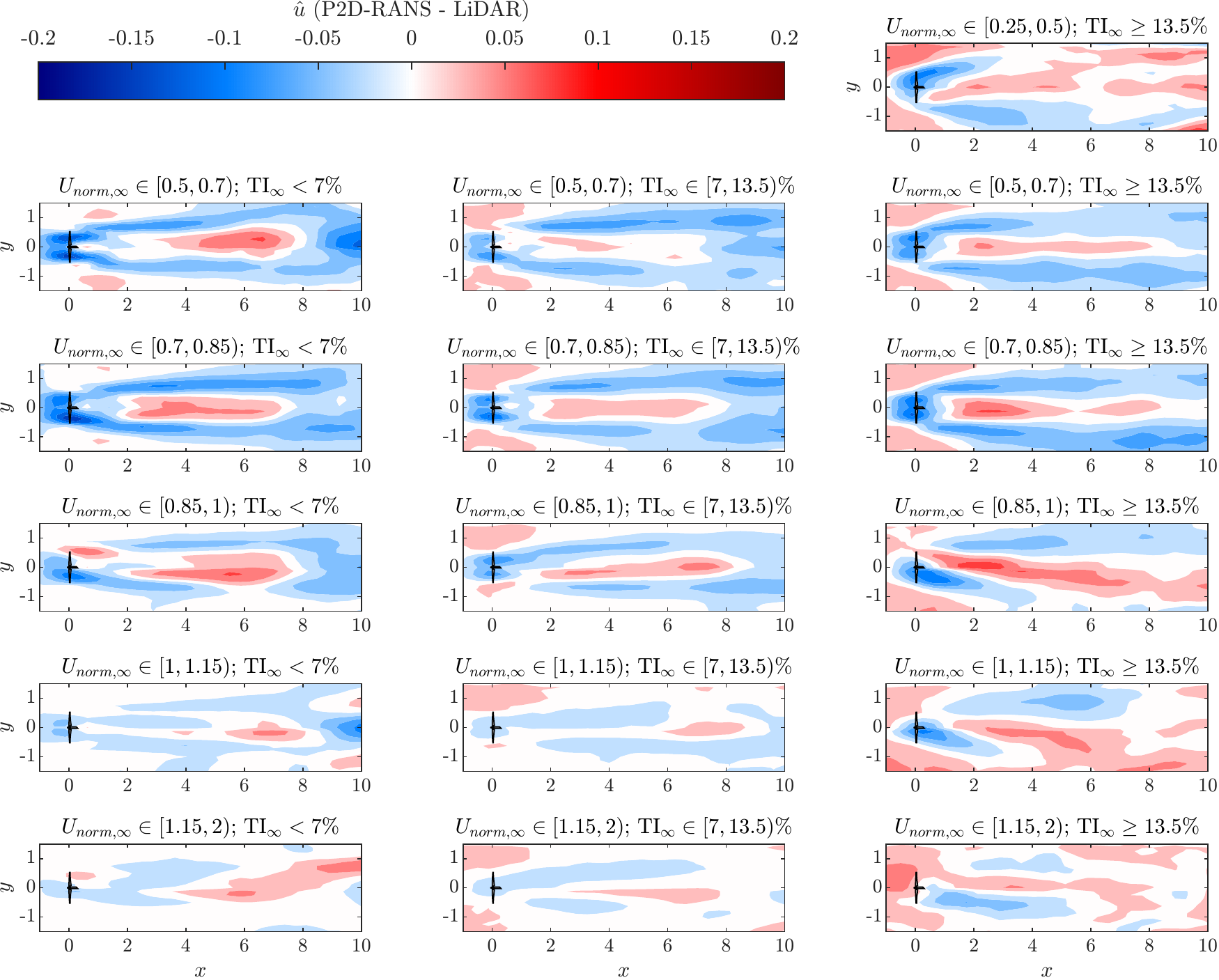}}
 \caption{Difference $\hat{u}$ velocity field between P2D-RANS and LiDAR after calibration.}\label{fig:Cal_u_diff}
 \end{figure*}

Fig. \ref{fig:Cal_EV_Ct}(c) shows a MAPE that lies below the 5\% threshold for all the wake after the calibration, which is considered satisfactory. To conclude, the calibrated numerical fields are shown in Fig. \ref{fig:Cal_u_P2D}, while the difference fields (P2D-RANS minus LiDAR) are reported in Fig. \ref{fig:Cal_u_diff}. Finally, Tables \ref{tab:EV} and \ref{tab:Ct} provide the $\nu_T$ and $c_t$ resulting from the calibration procedure.
 
\begin{table}
\caption{Optimal turbulent eddy viscosity of the P2D-RANS model. The value is made non-dimensional by $U_\infty D$.}\label{tab:EV}
\centering
\begin{tabular}{lccc} & TI$_\infty <7\%$ & TI$_\infty \in [7,13.5)\%$ & TI$_\infty\geq13.5\%$\\
\hline
$U_{\text{norm},\infty}\in[0,0.25)$ & N/A    & N/A  &       0.0575\\
$U_{\text{norm},\infty}\in[0.5,0.7)$ &0.0082 &0.0139 &0.0318 \\
$U_{\text{norm},\infty}\in[0.7,0.85)$ &0.0095 &0.0133 &0.0356 \\
$U_{\text{norm},\infty}\in[0.85,1)$ &0.0057 &0.0097 &0.0499 \\
$U_{\text{norm},\infty}\in[1,1.15)$ &0.003 &0.0051 &0.0663 \\
$U_{\text{norm},\infty}\in[1.15,2)$ &0.0081 &0.0176 &0.0323 \\
\end{tabular}
\end{table}

\begin{table}
\caption{Coefficient of thrust calibrated on LiDAR data.}\label{tab:Ct}
\centering
\begin{tabular}{lccc}& TI$_\infty <7\%$ & TI$_\infty \in [7,13.5)\%$ & TI$_\infty\geq13.5\%$\\
\hline
$U_{\text{norm},\infty}\in[0,0.25)$ & N/A    & N/A  &       1.83\\
$U_{\text{norm},\infty}\in[0.5,0.7)$ &1.21 &1.28 &1.51 \\
$U_{\text{norm},\infty}\in[0.7,0.85)$ &1.3 &1.24 &1.52 \\
$U_{\text{norm},\infty}\in[0.85,1)$ &0.81 &0.8 &1.11 \\
$U_{\text{norm},\infty}\in[1,1.15)$ &0.46 &0.48 &0.86 \\
$U_{\text{norm},\infty}\in[1.15,2)$ &0.23 &0.37 &0.64 \\
\end{tabular}
\end{table}
For the implementation of the P2D-RANS, the values of the calibrated rotor thrust coefficient and turbulent eddy-viscosity are assigned to the bin centroids displayed in Fig. \ref{fig:Cluster_stats}(a) and interrogated through a linear/nearest-neighbor Delauney interpolation to perform simulations with an arbitrary inflow. 

\section{Verification}\label{sec:Verification}
According to an AIAA technical report \cite{AIAA2002}, the process of determining if a model implementation accurately represents the conceptual description and the solution of a model is referred to as verification. The purpose of this section is indeed to verify the validity of the proposed shallow-water approach corrected for vertical fluxes and dispersive terms, the pressure correction algorithm of the P2D-RANS, and the numerical method. To this aim, the P2D-RANS solution of a single-wake velocity field is compared to the equivalent result calculated through an elliptic axisymmetric RANS solver. The latter uses the Newton algorithm to solve the full non-linear set of RANS equations in cylindrical coordinates discretized through a spectral collocation method based on Chebyshev polynomials (for more details, please refer to \citep{Viola2014,Iungo2018}, while the code is publicly available \cite{GRANS}. The simulations with the axisymmetric code are carried out using equivalent boundary conditions, turbine load, and turbulence model. The inflow conditions are arbitrarily chosen as $U_\infty=10$ m s$^{-1}$, TI$_\infty=10\%$, which in turn define the axial aerodynamic thrust and the turbulent eddy viscosity (Sec.  \ref{sec:Calibration}). 

\begin{figure*}
\centerline{\includegraphics[width=\textwidth]{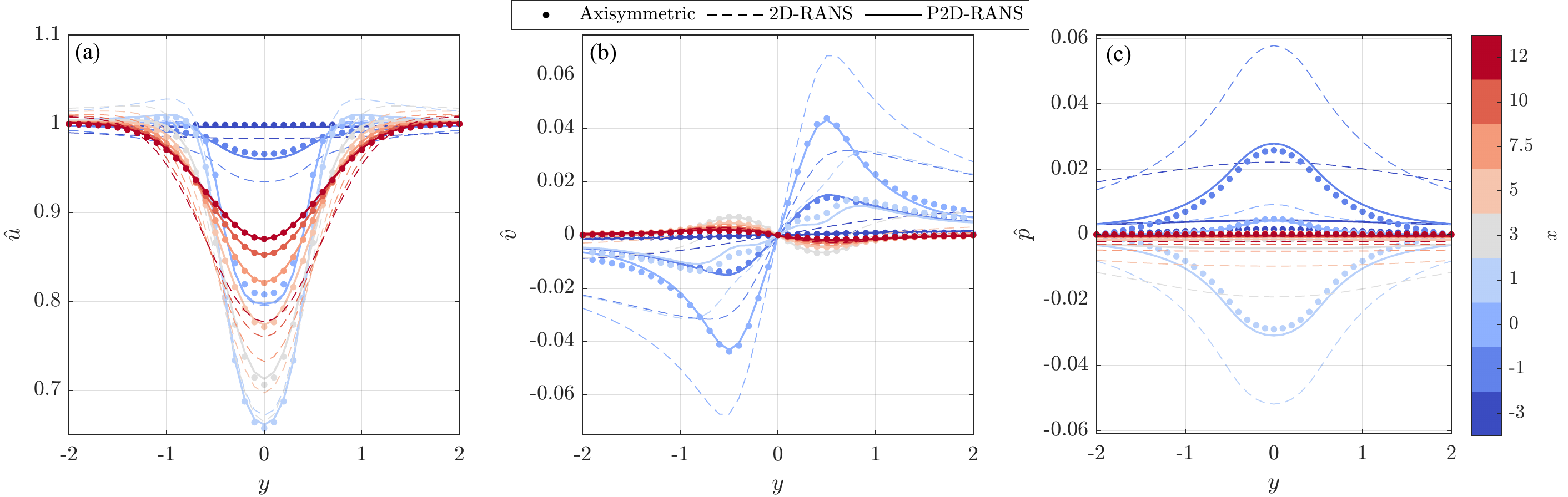}}
 \caption{Comparison between the depth-averaged fields of the elliptic axisymmetric RANS solver, a purely 2D-RANS code, and the P2D-RANS with 3D corrections: (a) streamwise velocity component; (b) spanwise velocity component; (c) pressure.}\label{fig:Verification_prof}
\end{figure*}

Fig. \ref{fig:Verification_prof} shows the spanwise profiles of $\hat{u}, \hat{v}, \hat{p}$ extracted at several downstream locations calculated by the axisymmetric solver (dots) and the P2D-RANS (solid lines), as well as the same quantities for a purely 2D solver, i.e. without 3D corrections (dashed lines). The agreement between the axisymmetric solver (i.e. the benchmark) and the P2D-RANS is excellent, with negligible discrepancies, which are due to the slightly different implementation of the boundary conditions for the two models \cite{Iungo2018}. The almost perfect overlapping between the pressure obtained by the axisymmetric RANS solving the fully elliptic Navier-Stokes equations and through the iterative marching scheme of the P2D-RANS described in Sec.  \ref{sec:Computational} serves as verification of the proposed pressure-velocity coupling scheme. Conversely, the profiles obtained with the merely 2D-RANS code exhibit significantly higher speedups, hampered wake recovery, excessive spanwise velocity magnitude, and pressure gradients compared to the benchmark simulations, which justifies the utilization of the proposed 3D corrections to relax the vertical confinement, which is a typical limitation of 2D wind farm flow simulators.

\begin{figure*}
\centerline{\includegraphics[width=\textwidth]{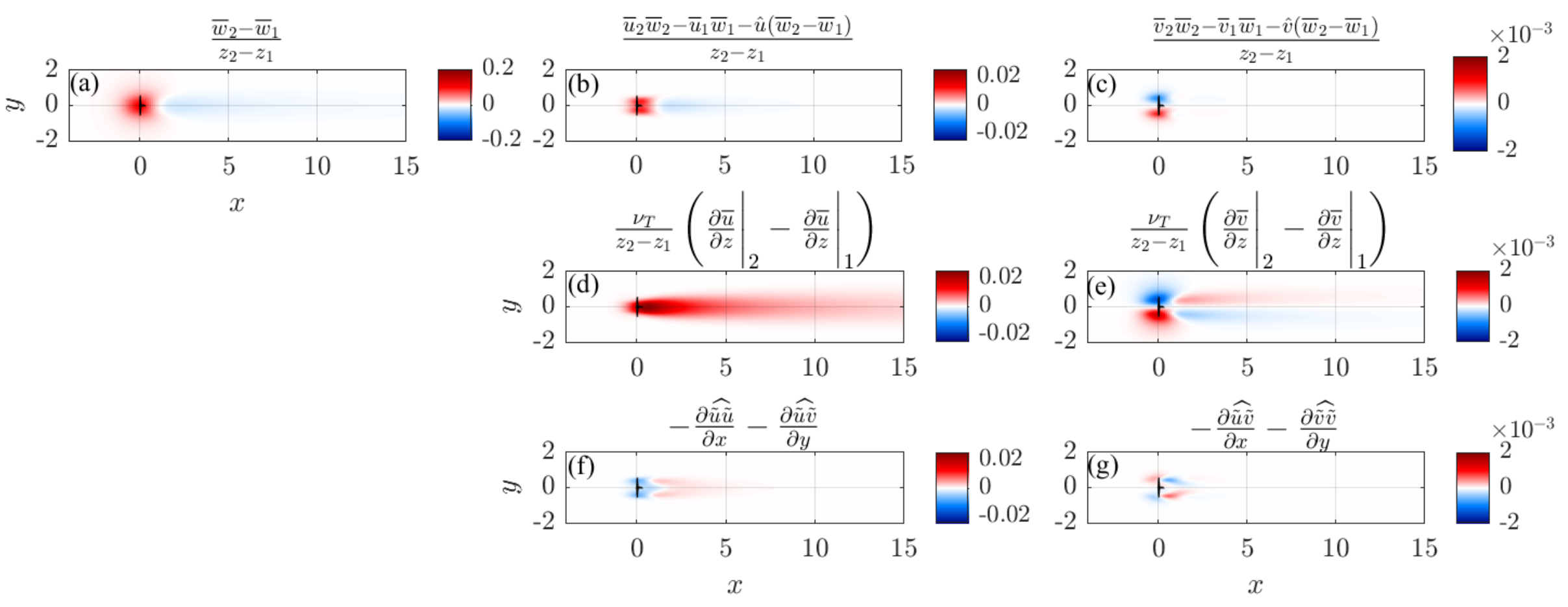}}
 \caption{3D corrections for the single turbine case: (a) vertical mass flux; (b) streamwise vertical advection; (c) spanwise vertical advection; (d) streamwise vertical turbulent flux; (e) spanwise vertical turbulent flux; (f) streamwise dispersive stresses; (g) spanwise dispersive stresses.}\label{fig:Verification_3Dcorr}
\end{figure*}

The 3D correction terms introduced in the P2D-RANS are assessed by plotting the individual contributions of the vertical fluxes and dispersive terms (overbarred and tilded terms, respectively, in Eqs. (\ref{eq:DepthAverage})) to the mass (Fig. \ref{fig:Verification_3Dcorr}(a)), $x$-momentum (Fig. \ref{fig:Verification_3Dcorr} (b, d, f)) and $y$-momentum (Fig. \ref{fig:Verification_3Dcorr} (c, e, g)) budgets. For an isolated turbine, these terms peak in the rotor area and drop sharply outside of the rotor span, decay further downstream as the wake recovers, and cross-stream components dissipate.
It is noteworthy that the results of the present verification hold valid for an arbitrary turbine thrust and eddy viscosity, namely for the full range of inflow conditions defined in the calibration phase (Sec.  \ref{sec:Calibration}).

For the simulation of a wind farm flow, the assumption of axially symmetrical and non-swirling flow underpinning the  method for the estimation of the 3D effects described in Appendix \ref{app:3D_corr} fails due to the merging of multiple wakes. Therefore, it is necessary to formulate an appropriate superposition method of the vertical fluxes and dispersive terms of the single turbines. To this aim, the flow resulting from the interaction of two in-tandem turbines with a streamwise spacing of $5D$ is calculated through the axisymmetric elliptic RANS and the P2D-RANS. The latter estimates the overall vertical mass and momentum fluxes through a superposition of those obtained for single-wake simulations and stored in a look-up table as a function of $U_{\text{norm},\infty}$ and TI$_\infty$ and interrogated through a nearest-neighbor method. Namely, the 3D corrections are evaluated for 15 values of $U_{\text{norm},\infty}$ from 0.2 to 2 and 12 values of TI$_\infty$ from 4\% to 50\% (i.e. 180 total single-wake cases), which cover the 98.5\% of the climatology of the Panhandle wind farm. The domain considered for single-wake fluxes extends from $-4D$ to $20D$ in the streamwise direction and $\pm 2D$ in the spanwise direction, since vertical fluxes and dispersive terms beyond such region are negligible for a wide range of operative conditions (see Fig. \ref{fig:Verification_3Dcorr}). Different methods have been considered for the estimate of the total fluxes resulting from wake overlapping, such as linear sum, root-squared signed sum, and the maximum absolute value of the single-wake contributions. The error with respect to the axisymmetric RANS, which solves directly the full flow equations without any simplifying assumption, is quantified to identify the best performing superposition principle. A total of 18 simulations of two in-tandem wind turbines with $5D$ streamwise spacing are performed for each method, each one having as input the centroid of the clusters in terms of $U_{\text{norm},\infty}$ and TI$_\infty$ previously defined. This analysis aims to characterize the flow-dependent error for various wake conditions.

\begin{figure*}
\centerline{\includegraphics[width=\textwidth]{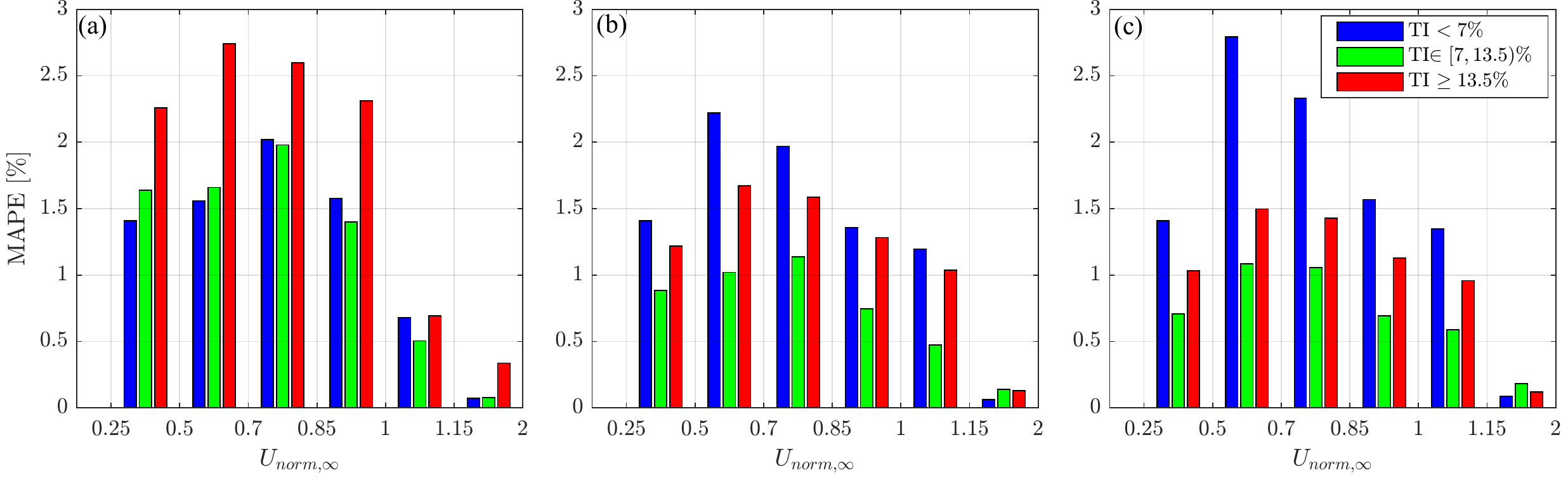}}
 \caption{Mean absolute relative error of $\hat{u}$ for between the axisymmetric RANS and the P2D-RANS using different superposition methods for the 3D corrections: (a) linear sum; (b) root-squared signed sum; (c) maximum absolute value.}\label{fig:Verification_VFS}
\end{figure*}

The MAPE of $\hat{u}$ for all the simulations is provided in Fig. \ref{fig:Verification_VFS} and indicates the root squared method with sign as that exhibiting the overall best agreement, with error exceeding 2\% for only one case (stable, region II).
It is noteworthy that the MAPE in $\hat{u}$ due to the superposition of the 3D corrections is always lower than 3\%, which is quite smaller than the errors reported for traditional engineering wake models, for which these types of superposition methods are applied directly on the velocity field \cite{Gunn2016,Machefaux2015}. 

\begin{figure*}
\centerline{\includegraphics[width=\textwidth]{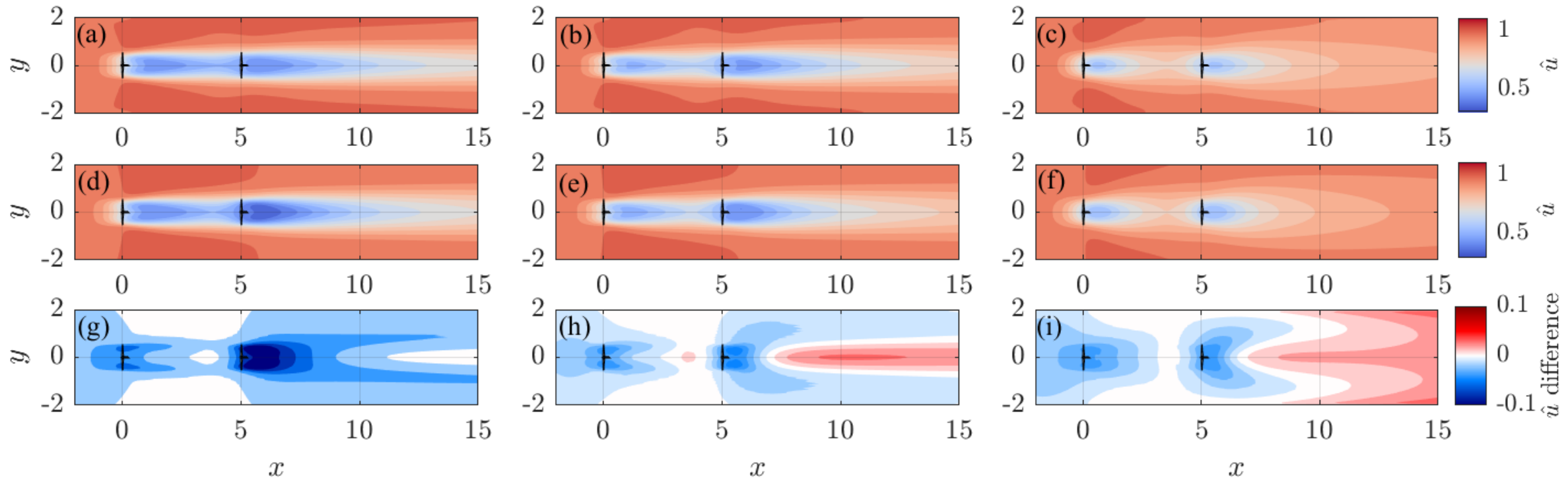}}
 \caption{Streamwise velocity simulated for two in-tandem wind turbines with a streamwise spacing of $5D$, $U_{\text{norm},\infty}=0.775$, and different turbulence intensity levels. By row: (a-c) axisymmetric RANS; (d-f) P2D-RANS using the root-squared superposition method; (g-i) difference (P2D-axisymmetric RANS). By columns: (a, d, g) TI$_\infty<7\%$; (b, e, h) TI$_\infty\in(7,13.5]\%$; (c, f, i) TI$_\infty \geq13.5\%$. }\label{fig:Verification_example}
\end{figure*}

Figure \ref{fig:Verification_example} shows three selected cases with two in-tandem turbines with $5D$ streamwise spacing for the centroid of the cluster $U_{\text{norm},\infty}\in (0.75, 0.8]$ (optimal $c_t$) and different TI$_\infty$ simulated through the reference elliptic RANS solver and the P2D-RANS with the root-squared signed sum method. The difference fields (Fig. \ref{fig:Verification_example} (g-i)) show that a systematic velocity underestimation takes place in the near wake region for the P2D-RANS, where vertical fluxes and dispersive terms peak, while the error decreases in magnitude and switches sign moving downstream. These features are observed in most of the simulations performed for this analysis and indicate that the empirical superposition technique of the 3D corrections may lead to substantial errors only in case of very close wake interaction (e.g. with streamwise spacing smaller than $3D$), which is however unlikely for well-designed wind farms. A more in-depth investigation on the accuracy of the 3D corrections superposition might be the object of future work. 

\section{Validation}\label{sec:Validation}
The validation of a computational model is formally defined as "the process of determining the degree to which a model is an accurate representation of the real world from the perspective of the intended uses of the model"\cite{AIAA2002}. In our case, real-world data are available in the form of 10-minute-averaged power from the SCADA data of the wind farm under investigation. The applications of the P2D-RANS model, considering its potential in accuracy and computational costs, can be broadly classified into two categories: $i)$ real-time diagnostic and control of wind turbines; $ii)$ wind farm performance analysis and optimal design. The former focuses on the estimation of the time variability of power production (typically in the form of 10-minute-averaged quantities) and accuracy of the P2D-RANS is assessed by comparing model predictions with the SCADA data for specific time series. The second application mainly entails the evaluation of wind turbine and wind farm performance in terms of power capture and quantification of wake losses as a function of atmospheric and turbine operational parameters. In this case, the following performance indicators are evaluated:
\begin{itemize}
\item percentage power losses \cite{El-Asha2017}, which are more relevant from the single-turbine modeling standpoint, and are defined as follows:
\begin{equation}\label{eq:Power_loss}
\Delta P_i(t_j)=1-\frac{P_i(t_j)}{P_{\infty,i}(t_j)},
\end{equation}
where $P_i(t_j)$ and $P_{\infty,i} (t_j)$ are the actual and unwaked power of the $i^\text{th}$ turbine at time $t_j$, respectively;
\item the wind farm efficiency, which is analogous to the previous parameter, yet for the entire turbine array \cite{Walker2016}:
\begin{equation}\label{eq:WF_eff}
    \eta=\frac{\sum_{i=1}^{N_t} P_i(t_j)}{\sum_i^{N_t} {P_{\infty,i}(t_j)}},
\end{equation}
with $N_t=25$ turbines for this study;
\item the annual (or cumulative) energy loss \cite{El-Asha2017}, which is more relevant from a financial standpoint since it considers the occurrence of each flow scenario within the local climatology, which can be recast as:
\begin{equation}\label{eq:Energy_loss}
   \Delta E=\frac{1}{N}\sum_{j=1}^N \sum_{i=1}^{N_t}(P_{\infty,i}(t_j)-P_i(t_j)) \cdot 8760, 
\end{equation}
where $N$ is the number of available quality controlled data for the flow conditions under consideration and the unit is kWh/y.
\end{itemize}
The unwaked power for each turbine, $P_{\infty,i}(t_j)$, is defined as the ideal power capture in absence of wake interactions, and it is estimated through a method similar to that proposed by Farrel et al. \cite{Farrell2021} to map a heterogeneous wind velocity field over a farm from pointwise sparse measurements. In this case, the same technique is applied to the power, using the power of unwaked turbines as input. Namely, the power from unwaked turbines (see Fig. \ref{fig:Panhandle_map_v2}(a)) is interpolated (extrapolated) at the location of waked turbines through a linear (nearest-neighbor) interpolation approach after a Delaunay triangulation. This ensures that the gross power takes into account possible heterogeneity in the wind field over the farm, whose relevance was already confirmed through the power curve analysis (section \ref{sec:Calibration}). 

It is noteworthy that the occurrence of overpower conditions, i.e. $P_i>P_{\infty,i}$, for a wind farm in flat terrain can be due either due to speedups caused by the induction pressure field of neighboring turbines or to an incorrect estimation of $P_{\infty,i}$, which might be associated with low availability of unwaked turbines for a certain wind direction. To exclude this last type of outliers yet retain cases with speedups, cases with overpower higher than 10\% are discarded only if the P2D-RANS model indicates that no pressure-induced speedup occurs at that turbine. In other words, the data point is rejected if SCADA data indicate $P_i/P_{\infty,i}>1.1$ and the P2D-RANS shows $P_i/P_{\infty,i}\leq1$ (1.3\% rejection rate).

Similarly to power, the undisturbed hub-height wind speed, $U_{\text{norm},\infty}$, turbulence intensity, TI$_\infty$, and wind direction, $\theta_w$, are estimated as averages of the respective quantities recorded by all the nacelle anemometers ($U_\text{norm}$ and TI$_\infty$) and yaw encoders ($\theta_w$) over the set of unwaked turbines at a specific time, $t_j$. Cases exhibiting error on the mean over the unwaked turbines at time $t_j$ based on the Student's $t$ distribution \cite{Wheeler2004} higher than 0.25 for $U_{\text{norm},\infty}$ (0.5\% of the overall datasets), 25\% of the average value for TI$_\infty$ (16\%), and 5$^\circ$ for $\theta_w$ (1.9\%) consequent to high flow variability within the park are discarded. Furthermore, periods when more than two turbines are either curtailed (based on the criteria described in Sec.  \ref{sec:Calibration}) or likely affected by wakes from neighboring turbine arrays that are not modeled are entirely excluded (35\%). This slightly relaxed criterion for the rejection of off-design farm operation is beneficial in terms of data availability and has minimal impact on the final statistics \cite{Nygaard2015}.

\begin{figure*}
\centerline{\includegraphics[width=\textwidth]{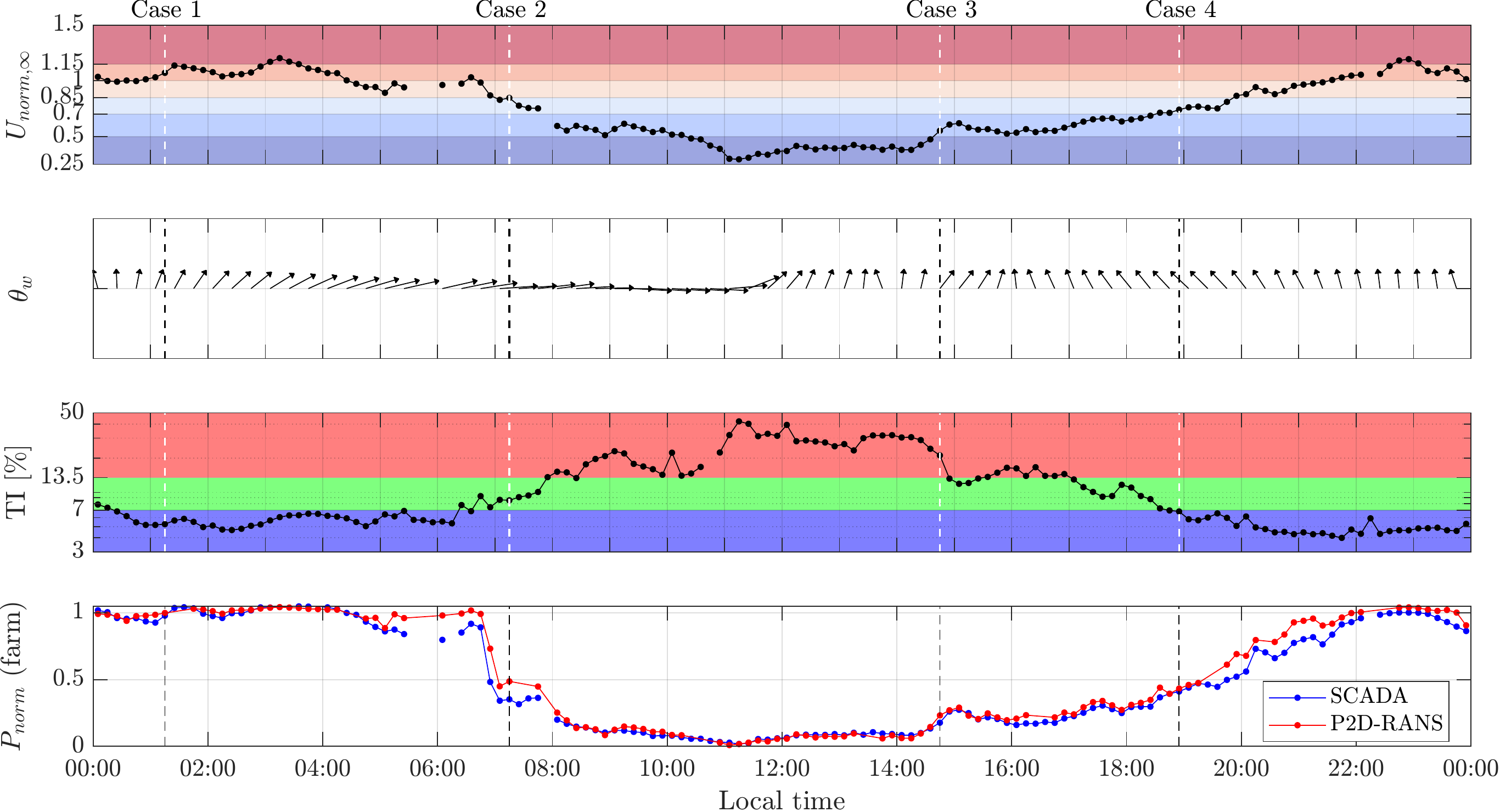}}
 \caption{Inflow characterization for August 26$^\text{th}$, 2016: (a) hub-height normalized wind speed; (b) wind direction; (c) turbulence intensity; (d) total normalized farm power capture. The background colors in (a) and (c) facilitate the identification of the $U_{\text{norm},\infty}$ and TI$_\infty$ clusters used in this work.}\label{fig:Time_series_Inflow}
 \end{figure*}

\begin{figure*}
\centerline{\includegraphics[width=\textwidth]{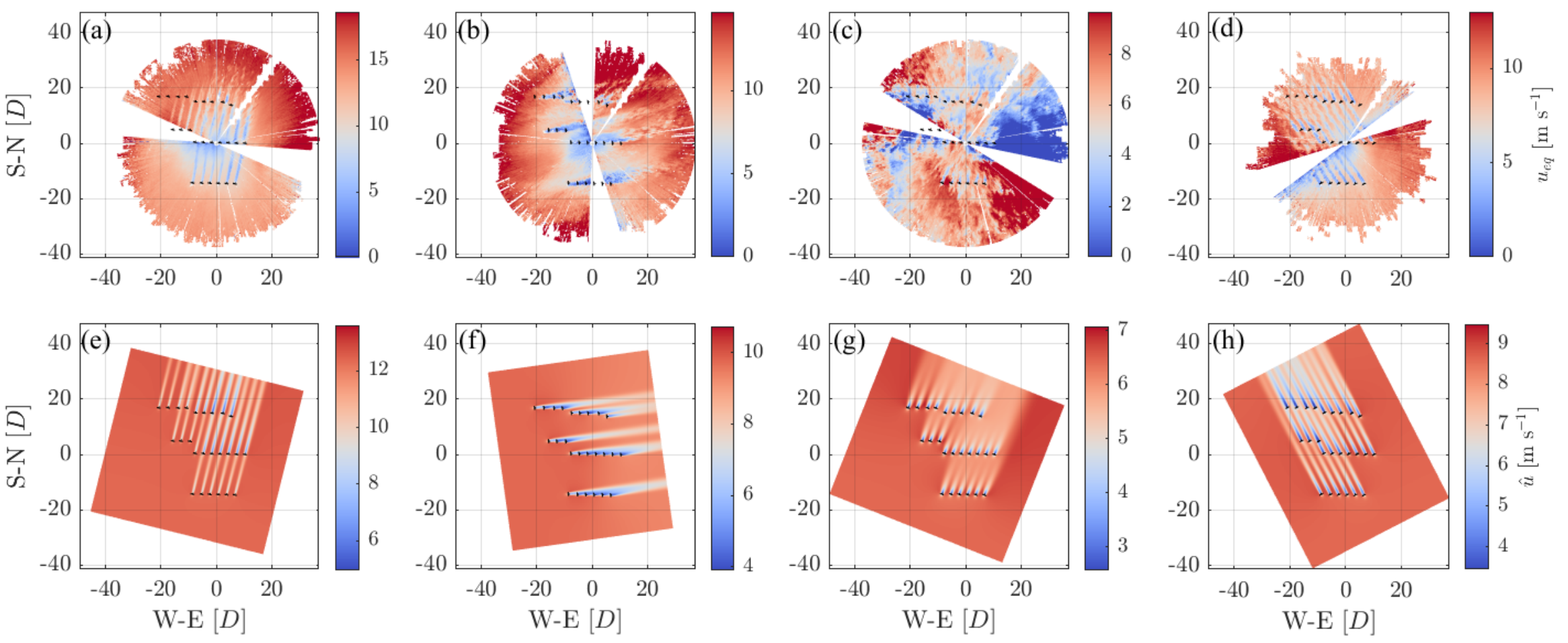}}
 \caption{Horizontal equivalent LiDAR (top row) and depth-averaged P2D-RANS (bottom rows) streamwise velocity for selected 10-minute periods on August 26$^\text{th}$, 2016: (a, e) Case 1, 1:10-1:20, $U_{\text{norm},\infty}=1.07$, $\theta_w=194^\circ$, TI$_\infty=5.3\%$;  (b, f) Case 2, 7:10-7:20, $U_{\text{norm},\infty}=0.85$, $\theta_w=262^\circ$, TI$_\infty=8.5\%$; (c, g) Case 3, 14:40-14:50, $U_{\text{norm},\infty}=0.55$, $\theta_w=202^\circ$, TI$_\infty=21.2\%$;  (d, h) Case 4, 18:50-19:00, $U_{\text{norm},\infty}=0.74$, $\theta_w=156^\circ$, TI$_\infty=6.8\%$;}\label{fig:Time_series_LiDAR}
 \end{figure*}

\begin{figure*}
\centerline{\includegraphics[width=\textwidth]{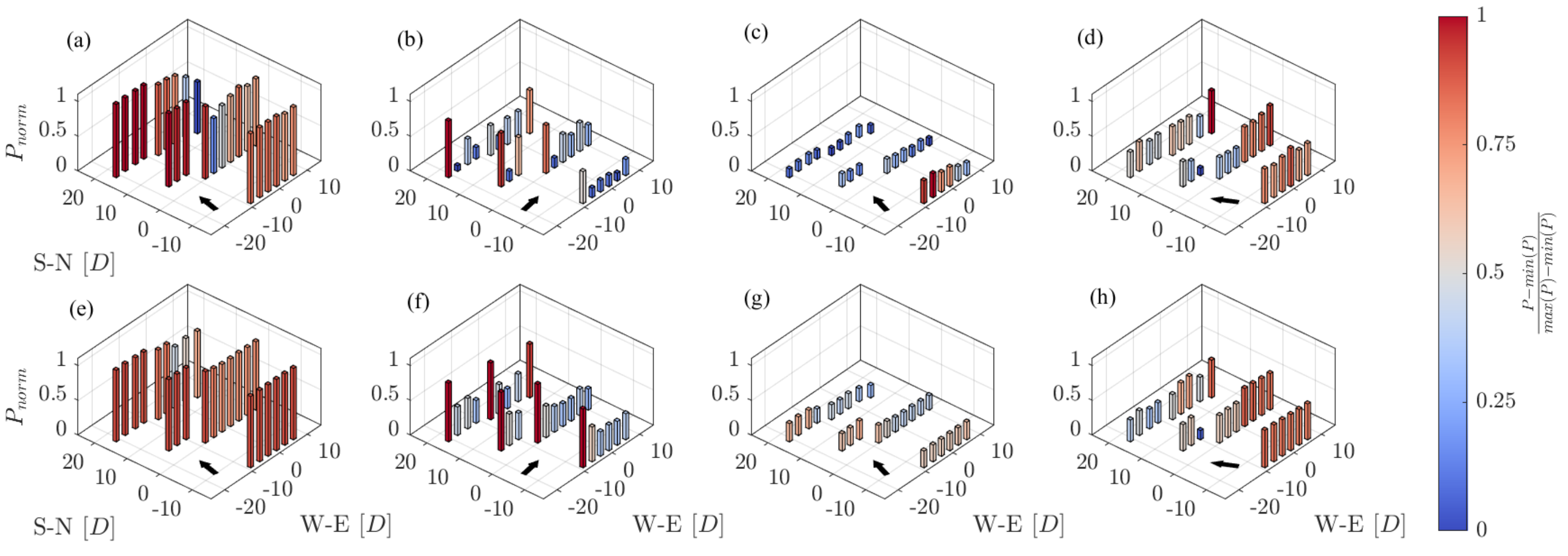}}
 \caption{As in Fig. \ref{fig:Time_series_LiDAR}, but for SCADA (top row) and P2D-RANS simulated (bottom row) power capture.}\label{fig:Time_series_Power}
 \end{figure*}

To showcase the capabilities of the P2D-RANS and highlight the challenges connected with the predictions of power production for individual onshore wind turbines, the simulations of a full day of turbine operation are first analyzed. The simulations are carried out for operations on August 26$^\text{th}$, 2016, and use the previously defined undisturbed hub-height wind speed, direction, and turbulence intensity as input with a 10-minute resolution. The selected day exhibits the typical stability-driven diurnal cycle of the wind resource \cite{El-Asha2017}, as shown in Fig. \ref{fig:Time_series_Inflow}. 
Specifically, the SCADA-recorded and P2D-RANS simulated power capture indicate a generally good agreement, except around 06:00 and 21:00 when the power is overestimated with the P2D-RANS, which is probably a consequence of the inherently complex transition in the atmospheric stability regime. Further insight is gained by selecting four illustrative cases indicated with vertical dashed lines in Fig. \ref{fig:Time_series_Inflow}. For these cases, experimental and numerical wind speed and power capture are reported in Figs. \ref{fig:Time_series_LiDAR} and \ref{fig:Time_series_Power}, respectively.

Case 1 represents a typical wind farm operation in region III of the power curve under stable atmospheric conditions and southerly wind direction.  Relatively long wakes are observed with a weak velocity deficit (Fig. \ref{fig:Time_series_LiDAR} (a, e)) leading to significant and localized power losses (Fig. \ref{fig:Time_series_Power} (a, e)). The main challenges in modeling this type of wind farm operations arise from the uncertainty in the quantification of the local wind direction throughout the farm, which is a major source of error in case of persistent wakes with limited lateral expansion. 

Case 2 shows the occurrence of strong wake interactions with relatively small streamwise spacing ($\sim 3D$) associated with moderate westerly winds for operations in region II of the power curve. The proximity of the waked rotors to the complex near-wake turbulent flow of multiple upstream turbines is arguably the most challenging (yet unlikely for a well-designed farm layout) case to simulate. Nevertheless, the P2D-RANS can reproduce satisfactorily the pattern of power losses across the farm (Fig. \ref{fig:Time_series_Power}(f)), albeit an evident overestimation in power capture is observed for specific turbines, which may be attributed to the detrimental effect on power capture of partial and/or intermittent wake interaction not reproduced by the steady RANS model.

Case 3 corresponds to highly convective daytime conditions, as portrayed by the utterly complex instantaneous flow field probed by the LiDAR (Fig. \ref{fig:Time_series_LiDAR}(c)). Thanks to the optimally calibrated $\nu_T$, the P2D-RANS model is capable in reproducing the enhanced turbulent mixing of the wakes, which results in moderate yet widespread power losses (Fig. \ref{fig:Time_series_Power}(g)), in agreement with the respective SCADA data (Fig. \ref{fig:Time_series_Power}(c)). The latter plots also show the occurrence of significant spanwise wind gradient at the southern-most array facing the incoming wind field, which is a quite common phenomenon at this site that justified the formulation of the turbine-specific unwaked power, $P_{\infty,i}$, adopted in Eqs. (\ref{eq:Power_loss}, \ref{eq:WF_eff}, \ref{eq:Energy_loss}).

Finally, case 4 represents an interesting south-east wind event, which results in $\sim 80\%$ power loss at turbine 07, which is accurately simulated by the P2D-RANS (Fig. \ref{fig:Time_series_Power} (d, h)).

This qualitative analysis is meant to remark the considerable complexity of a real wind farm flow and the challenges inherent to the simulation of the 10-minute-averaged power at the single turbine level, but also to stress the ability of the P2D-RANS model to reproduce the physical behavior of individual turbines experiencing a real-world turbulent ABL flow, yet keeping the required computational costs and resources very limited. 

More systematic quantification of the uncertainty of the P2D-RANS model is carried out through a statistical analysis encompassing all the available SCADA data. To this aim, a look-up table of simulated single-turbine power is built for a wide range of inflow conditions covering the overall climatology of the site. The input inflow matrix is the combination of several $U_{\text{norm},\infty}$ values between 0.2 and 2 (15 cases), TI$_\infty$ between 4\% and 50\% (12 cases), and wind directions from $0^\circ$ to $360^\circ$ with a resolution of $5^\circ$, resulting in 12960 simulations and a 98.5\% coverage of the climatic conditions of the wind farm. A further refinement of the input matrix did not produce any significant differences in the results. The simulated power at the generic time $t_j$ is obtained by interrogating independently for the $i^\text{th}$ turbine the power look-up table, $P_i(U_{\text{norm},\infty}, \text{TI}_\infty,\theta_w)$, through a tri-linear interpolation with the experimental values of $U_{\text{norm},\infty}(t_j)$, TI$_\infty(t_j)$ and $\theta_w(t_j)$ as inputs. 

\begin{figure*}
\centerline{\includegraphics[width=0.9\textwidth]{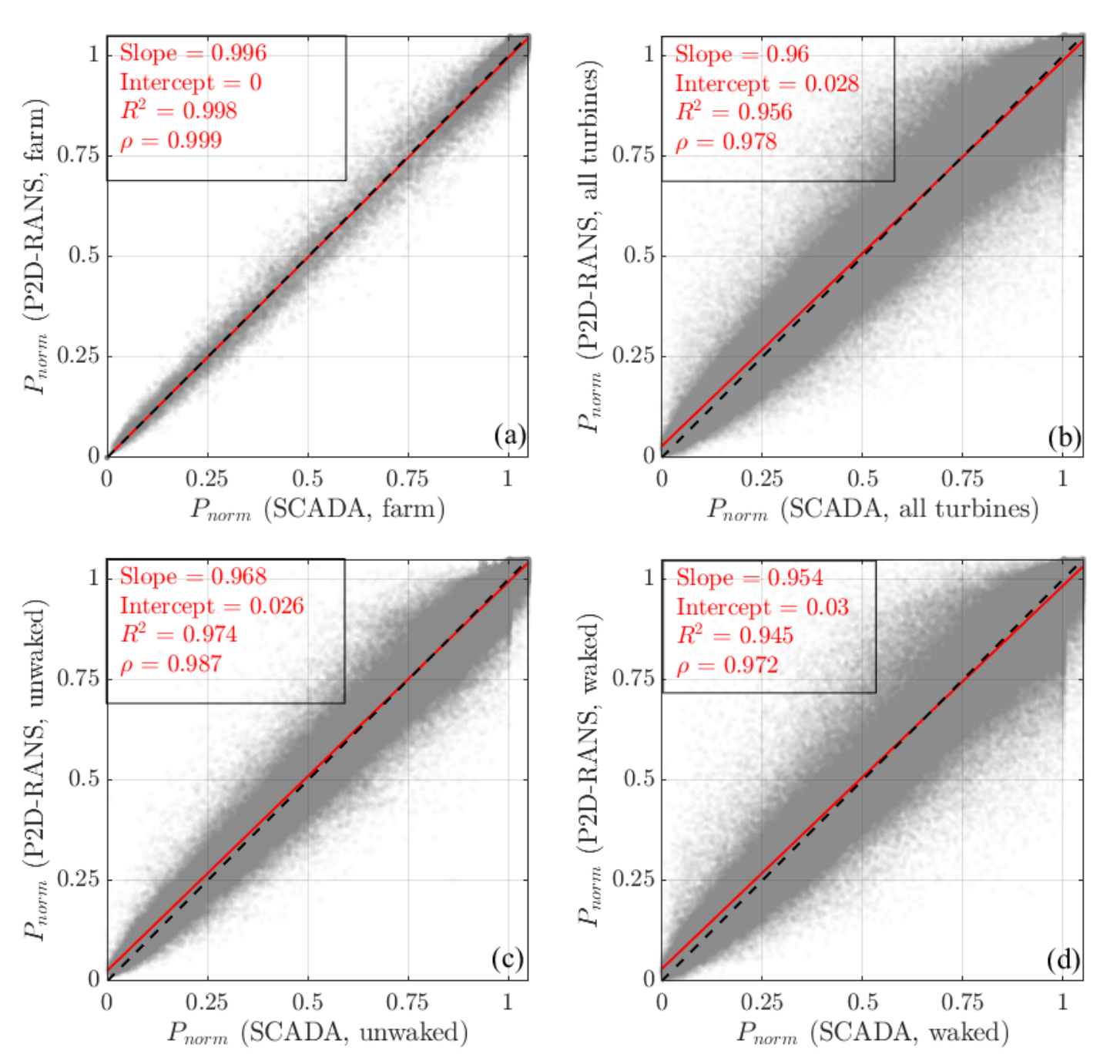}}
 \caption{Linear regression of SCADA vs. P2D-RANS 10-minute-averaged power: (a) normalized wind farm power; (b) normalized turbine power; (c) normalized power of unwaked turbines; (d) normalized power of waked turbines.}\label{fig:Validation_Power}
 \end{figure*}

The characterization of the accuracy of the P2D-RANS in simulating the time-resolved 10-minute-averaged power is carried out through a linear regression of the power, as reported in Fig. \ref{fig:Validation_Power}.
There is an outstanding agreement for the overall power of the entire farm (Fig. \ref{fig:Validation_Power}(a)) with a negligible bias and $R^2=0.998$, which drops when considering the regression between the power of individual turbines (Fig. \ref{fig:Validation_Power}(b), $R^2=0.956$). This feature indicates that a noticeable error cancellation occurs at the farm level, thus making the actual accuracy in power predictions for single turbine power a better indicator to assess the validity of the proposed flow model. For a better understanding of the error source, Figs. \ref{fig:Validation_Power} (c, d) report the linear regression of single-turbine power for unwaked and waked wind turbines, respectively. It is noteworthy that, although the prediction of the power of waked turbines exhibits larger discrepancies compared to that of unwaked turbines ($R^2=0.945$ vs. $R^2=0.974$), the latter still is encumbered with a non-negligible scattering. This mismatch between the predicted and measured power for unwaked turbines is due to spatial wind gradients, unsteady conditions within the 10-minute periods, power curve uncertainty, and undetected wakes, which represent a baseline error that is inevitably present regardless of the accuracy of the used wind farm model. The linear regression of power capture under waked conditions shows an almost unitary slope and nearly zero intercept, which indicates a negligible contribution of systematic biases in the prediction of the wake interactions to the overall uncertainty.

To provide a more focused statistical analysis of power and energy losses due solely to wake interactions, further validation is performed also including results obtained with the Jensen \cite{Jensen1983}, Multizone \cite{Gebraad2016}, and Gaussian \cite{Bastankhah2014,Niayifar2016} wake models implemented in the FLORIS package \cite{FLORIS_2021}. All these models use the nominal power curve (see Fig. \ref{fig:Power_curve}) and the optimally calibrated $c_t$ reported in Sec.  \ref{sec:Calibration}, while other wake parameters have default FLORIS settings (specific documentation and references are available at \cite{FLORIS_2021}).

\begin{figure*}[b!]
\centerline{\includegraphics[width=\textwidth]{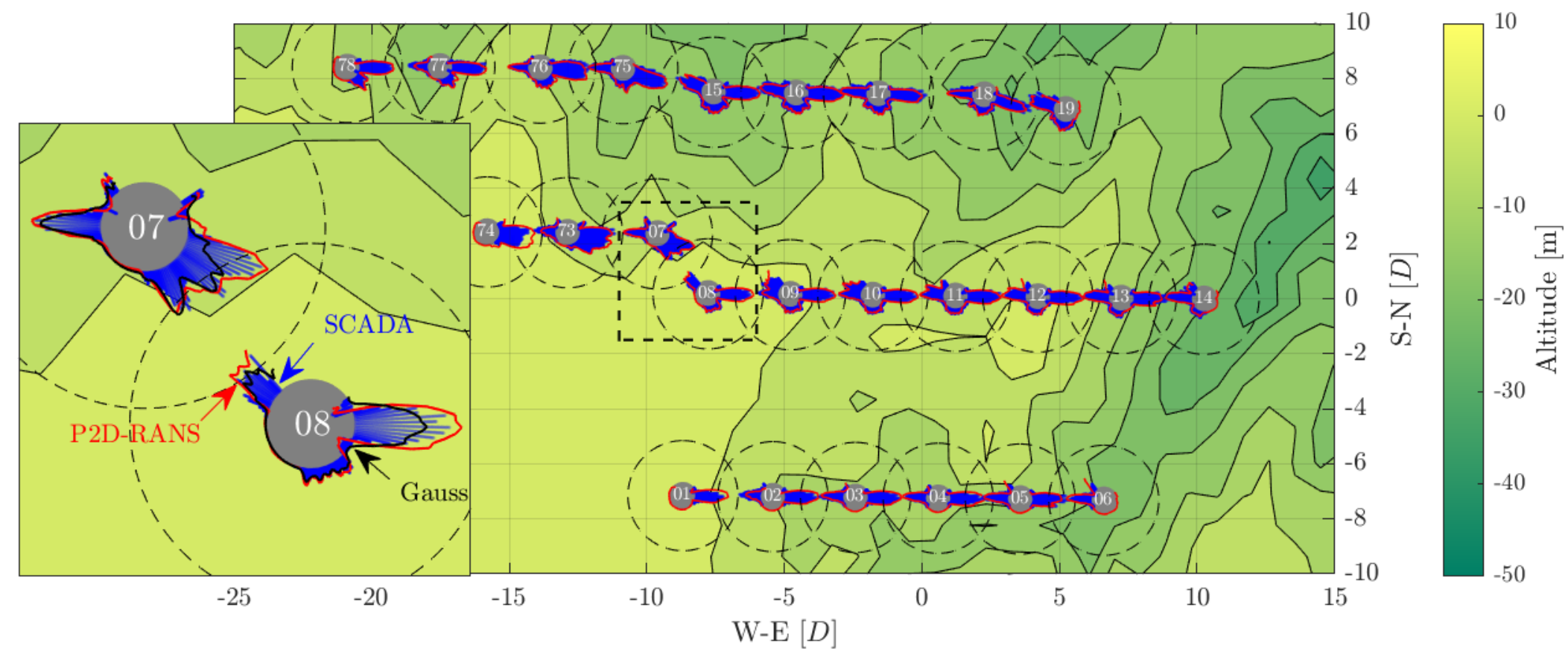}}
 \caption{Polar chart of the directional percentage power losses. The blue bars represent the SCADA data, while the red lines are the P2D-RANS results and the black (zoomed region only) the Gauss model of FLORIS. The grey dots represent the turbines, with the radius corresponding also the $\Delta P_i=0\%$ losses. The dashed circles indicate $\Delta P_i=100\%$.}\label{fig:Validation_PL}
 \end{figure*}

Firstly, the ability of the P2D-RANS model to predict the directional wake losses is tested by calculating the bin-average of the percentage power losses for each turbine (Eq. \ref{eq:Power_loss}) in $5^\circ$-wide sectors, as shown in Fig. \ref{fig:Validation_PL}. Statistics for bins having an error on the mean larger than $25\%$ with $95\%$ confidence are excluded.
The model can reproduce the directional pattern of power losses of each turbine satisfactorily with a NMAE $=30.5\%$ with respect to the SCADA data. In Fig. \ref{fig:Validation_PL}, the zoomed-in frame shows how the model can accurately reproduce the wake losses resulting from both near (between turbine 07 and 08, with spacing $4.7D$) and far (South direction for turbine 08, with spacing $\sim 15 D$) wake interactions, thus confirming the validity of the wake model for a wide range of streamwise distances, even exceeding the domain of LiDAR data used for the calibration. The zoomed region also indicates that the P2D-RANS provides accurate quantification of the physically consistent overpower occurring at turbine 08 for wind directions at the edge of a heavily-waked eastern sector (overpower is visualized as a negative power loss, i.e. the part of the polar plot that lies inside the grey circle corresponding to zero losses). For these cases the P2D over-performs other engineering wake models, which lack modeling of the pressure field, as the Gaussian wake model (black line in the same plot), and, thus, any effect on the incoming wind field induced by the thrust force of a turbine rotor.

\begin{figure*}[t!]
\centerline{\includegraphics[width=\textwidth]{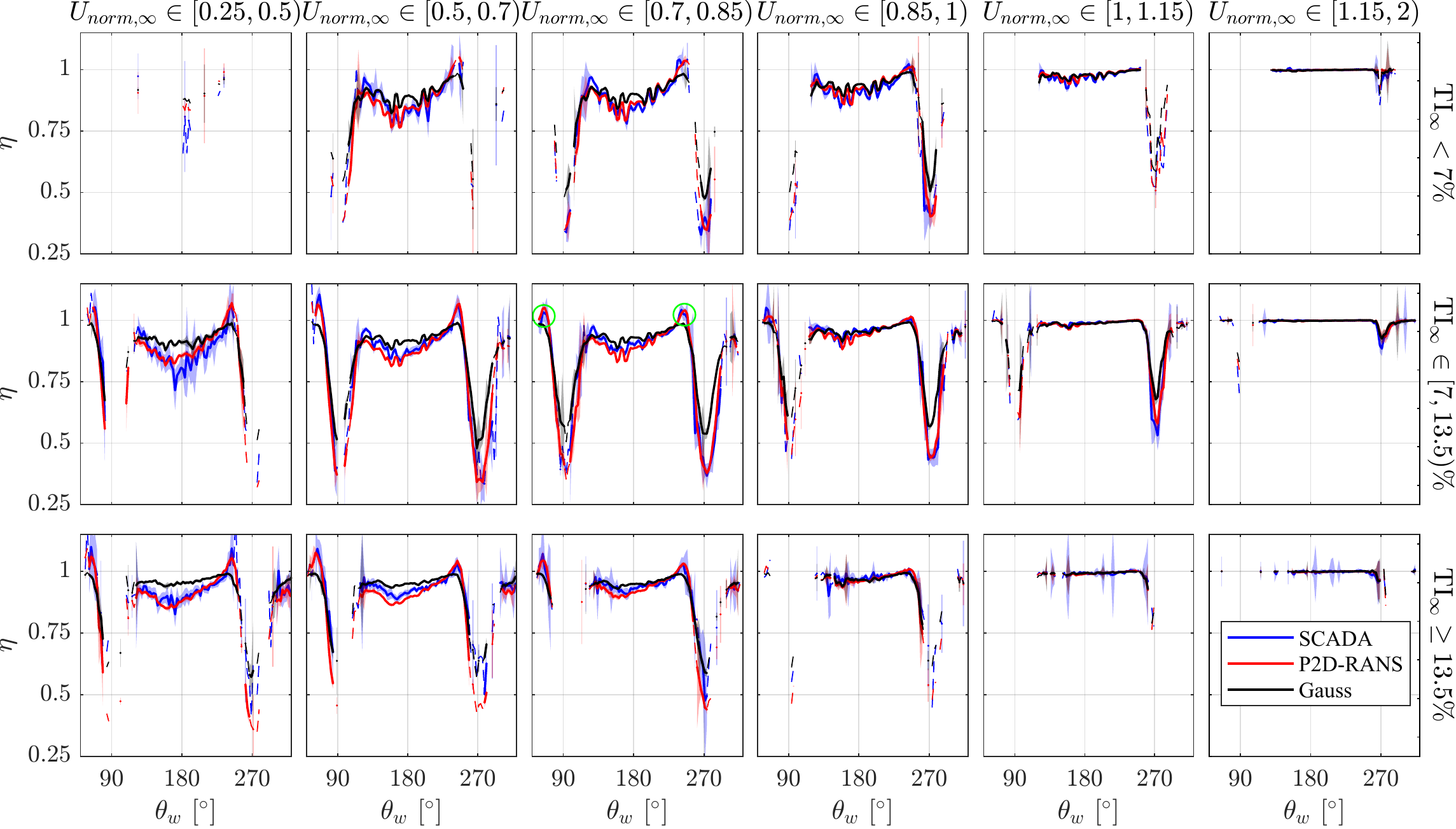}}
 \caption{Directional and clustered wind farm efficiency for all the $U_{\text{norm},\infty}$ and TI$_\infty$ clusters defined in Sec.  \ref{sec:Calibration}). The shaded area represents the error on the mean with 95\% confidence level. Dashed lines refer to points rejected due to an error on the mean that exceeds 0.25.}\label{fig:Validation_WF_eff}
 \end{figure*}

The analysis of the wind farm performance as a function of the incoming flow is further refined by bin-averaging the wind farm efficiency, $\eta$, for all the wind directions and separately for each $U_{\text{norm},\infty}$ and TI$_\infty$ cluster (Fig. \ref{fig:Validation_WF_eff}).
The resulting directional and clustered wind farm efficiency exhibits, as expected, significant variability with lower values occurring for low $U_{\text{norm},\infty}$ (i.e. high $c_t$), low TI$_\infty$ (i.e. slower wake recovery), and winds aligned with the turbine arrays (i.e. W-E direction). 
This analysis corroborates the high accuracy of the model in reproducing the wind farm flow at the turbine level. The Gauss model shows comparatively a poorer agreement with the experimental wind farm efficiency, especially for heavily waked sectors ($\theta_w \sim 90^\circ$ and $\theta_w \sim 270^\circ$), a general underestimation of losses for southerly winds, and a complete lack of wind farm overpower (i.e. $\eta>1$), which on the contrary is observed for the SCADA and the P2D-RANS for narrow wind sectors around $\theta_w=65^\circ$ and $\theta_w=245^\circ$.
 
\begin{figure*}[t!]
\centerline{\includegraphics[width=0.85\textwidth]{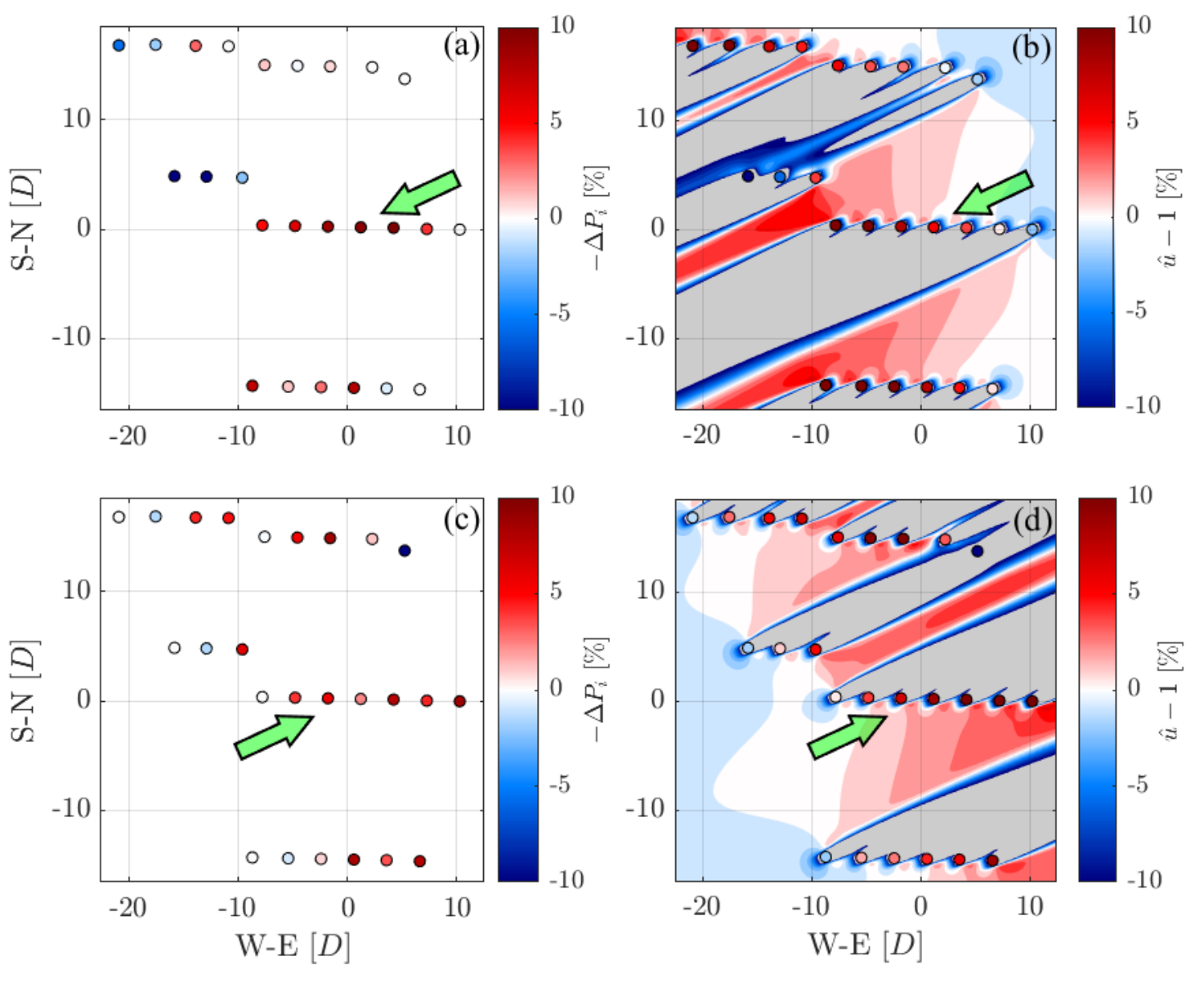}}
 \caption{Power gain for selected wind direction bins within the cluster $U_{\text{norm},\infty}\in [0.7, 0.85)$, TI$_\infty \in[7,13.5)\%$: (a) SCADA power gain for $\theta_w\in [62.5,67.5)^\circ$; (b) P2D-RANS power gain for $\theta_w\in (62.5,67.5)^\circ$; (c) SCADA power gain for $\theta_w\in [242.5,247.5)^\circ$; (d) P2D-RANS power gain for $\theta_w\in (242.5,247.5)^\circ$. The velocity fields in (b) and (d) corresponds to the P2D-RANS solution for the bin centroid.} \label{fig:Validation_Speedup}
 \end{figure*}

In this regard, two occurrences of wind farm efficiency greater than one, 
which are circled in green in Fig. \ref{fig:Validation_WF_eff}, are further investigated by extracting the individual turbine power gain (i.e. $-\Delta P_i$, which is more suitable to investigate speed-up conditions). Fig. \ref{fig:Validation_Speedup} displays $-\Delta P_i$ and the non-dimensional wind speed anomaly, $\hat{u}-1$, obtained from the P2D-RANS simulation corresponding to the bin centroids. The visualizations of the wind speed reveal the presence of channeling effects leading to power increases up to $\sim 10\%$, which are also observed in the SCADA data. This is further evidence of the importance of the rotor-induced pressure field on the performance of utility-scale wind turbines.
 
\begin{figure*}[t!]
\centerline{\includegraphics[width=\textwidth]{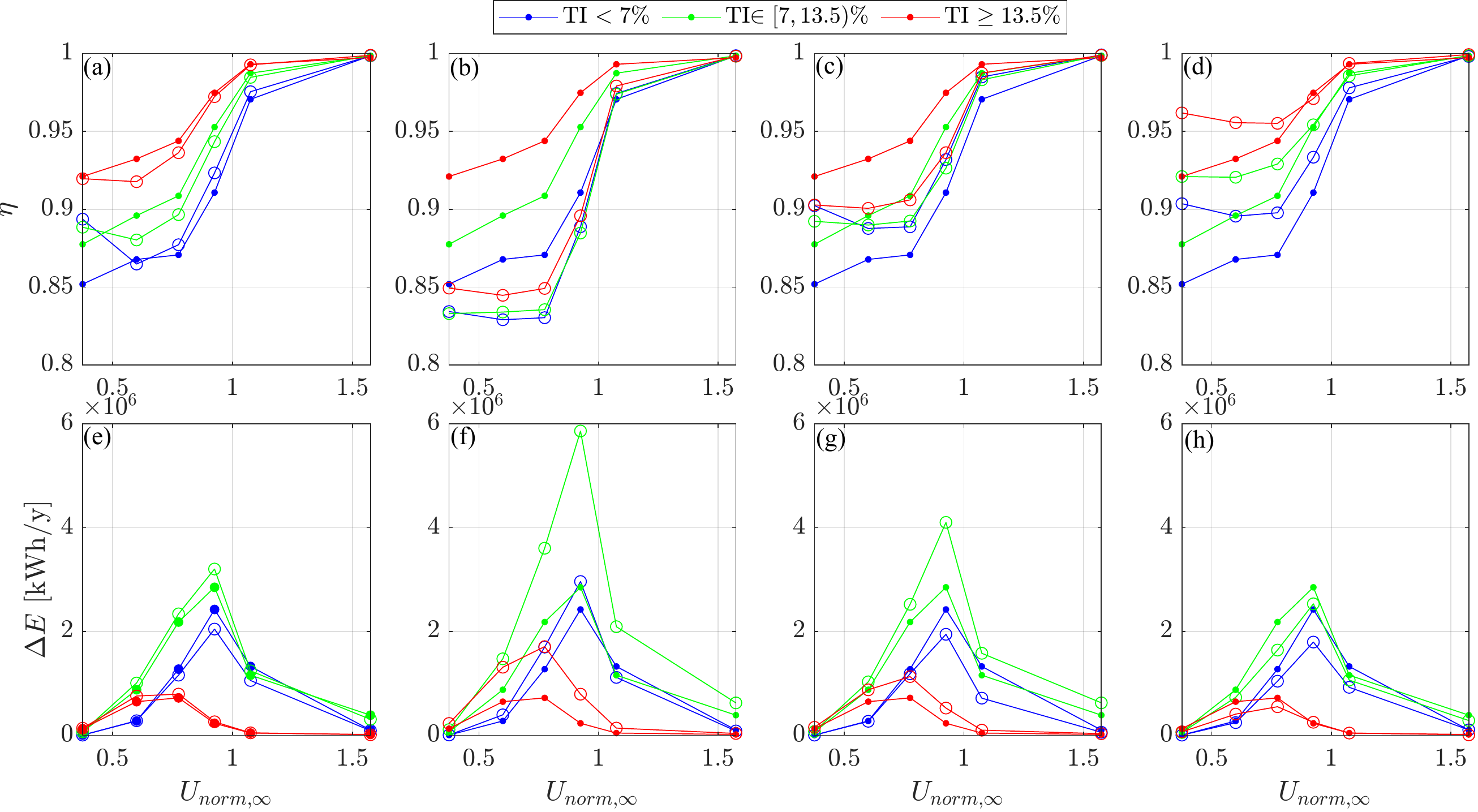}}
 \caption{Clustered wind farm efficiency (a-d) and annual energy losses (e-f) for difference models: (a) and (e) P2D-RANS; (b) and (f) Jensen; (c) and (g) Multizone; (d) and (h) Gauss. The dots indicate the SCADA statistics, the empty circles refer to the models. All points displayed satisfy the requirement of less than 25\% of error on the mean with a 95\% confidence.}\label{fig:Validation_Energy}
 \end{figure*}

The influence of the incoming flow on the wind farm performance is further characterized by disregarding power capture variability with the wind direction and calculating clustered wind farm efficiency and annual energy losses associated with the different $U_{\text{norm},\infty}$ and TI$_\infty$ bins (Fig. \ref{fig:Validation_Energy}). 
The omnidirectional wind farm efficiency exhibits a neat variability for the different TI$_\infty$ clusters, with higher efficiency achieved for convective conditions, and the expected trend as a function of $U_{\text{norm},\infty}$, as higher losses are observed in region II of the power curve. From the modeling standpoint, the P2D-RANS shows a general slight underestimation of wake losses in stable atmospheric conditions and overestimation for neutral and convective clusters. Nevertheless, the P2D-RANS achieves better agreement in clustered wind farm efficiency (Fig. \ref{fig:Validation_Energy}(a)) and annual energy losses (Fig. \ref{fig:Validation_Energy}(e)) than the other wake models considered for this study.
 
The Jensen model (Fig. \ref{fig:Validation_Energy} (b, f)) generally overestimates the wake losses and does not capture the change in wind farm efficiency due to atmospheric stability. The difference in wind farm efficiency among different TI$_\infty$ values is solely due to intra-cluster variability of wind direction, being the analytical wake-field insensitive to the turbulence intensity due to the constant wake expansion coefficient $k_e=0.05$. The Multizone model (Fig. \ref{fig:Validation_Energy} (c, g)) also fails to fully capture the effect of the atmospheric stability, since it still uses a fixed wake expansion coefficient $k_e=0.05$. The overestimation of wake losses is however assuaged thanks to the improved formulation of the velocity field, compared to Jensen. The Gaussian model with the TI-dependent wake expansion \cite{Niayifar2015} $k_e=0.004 + 0.38$TI$_\infty$ shows a significant improvement, although it generally underestimates the wake losses in region II of the power curve. The performance of the Gaussian model is remarkable, considering the extremely low computational time (0.1 s per simulation) and the lack of a site-specific calibration of the wake expansion coefficient.
 
\begin{figure*}[t!]
\centerline{\includegraphics[width=\textwidth]{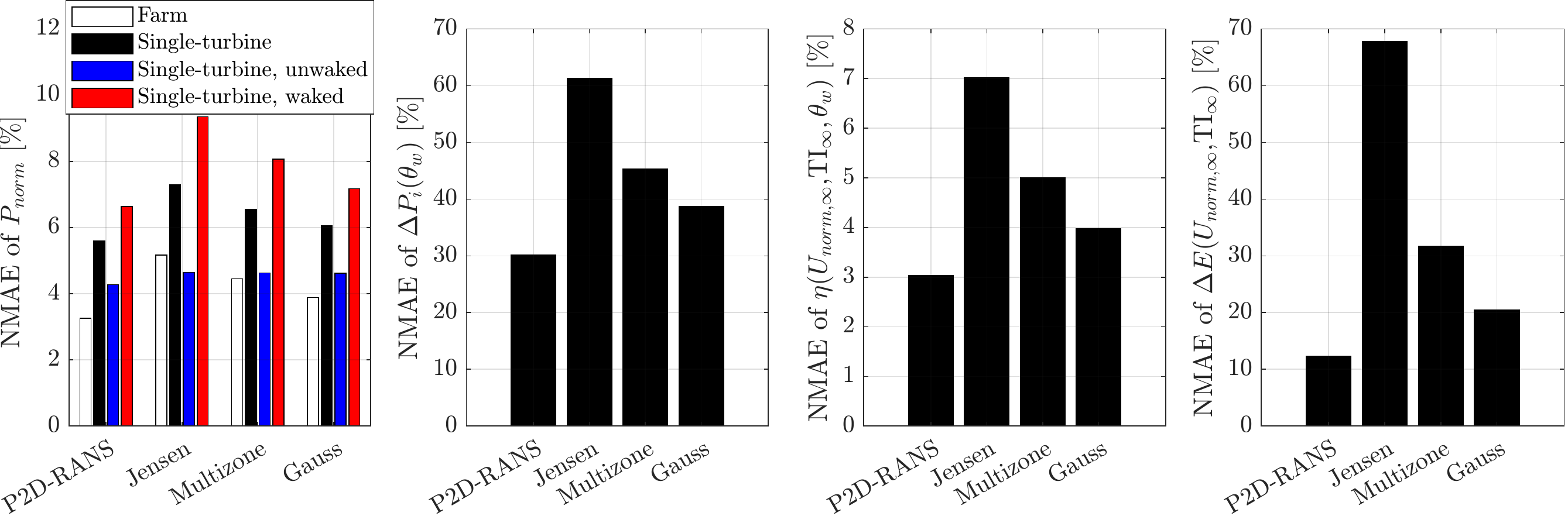}}
 \caption{Normalized mean absolute error (NMAE) for the different performance indicator of the P2D-RANS and FLORIS wake models: (a) power capture; (b) single-turbine directional percentage power losses; (c) directional and clustered wind farm efficiency; (d) clustered annual energy losses.}\label{fig:Validation_NMAE}
\end{figure*}
The NMAE for all the performance indicators presented in this section are summarized in Fig. \ref{fig:Validation_NMAE} and for all the models. Although the P2D-RANS shows the smallest error for all the statistics, there are important distinctions that it is worth pointing out. Table \ref{tab:DNMAE} provides the percentage improvement attained by the P2D-RANS compared to the other wake models to assist the interpretation of the results. The P2D-RANS achieves a significant reduction of error for the farm and single-turbine power (especially waked turbines) compared to Jensen and Multizone, while the improvement compared to the more accurate TI-dependent Gaussian model is more limited (8\% for the single turbine power). Better relative performances are observed for the directional single-turbine power losses, with a 22\% error reduction achieved by the P2D-RANS compared to the Gaussian model, a consequence of the more accurate description of the wake velocity field and the modeling of the rotor-induced pressure field and the associated speed-up conditions for specific wind directions. A further significant margin of improvement is obtained for the statistics taking into account atmospheric stability through the clustering in TI$_\infty$, which is a clear indication that the site-specific calibration of $\nu_T$ is beneficial. The NMAE reduction with respect to the Gaussian model is 24\% for the directional and clustered wind farm efficiency and 40\%  for the annual energy losses. This last parameter is particularly relevant for the quantification of the financial losses due to wake interactions at a specific site. Therefore, the significantly enhanced accuracy in the prediction of the energy losses achieved by the P2D-RANS compared to other models is actually required to justify the investment in such LiDAR-driven CFD approach, which is computationally and logistically more expensive than the other engineering wake models.
\begin{table}[]
\centering
\caption{Percentage improvement in NMAE of the P2D-RANS model compared to the FLORIS wake models.}\label{tab:DNMAE}
\begin{tabular}{lccc}
 & Jensen & Multizone & Gauss \\ \hline
$P_\text{norm}$ (farm)                       & 37     & 27        & 16    \\
$P_\text{norm}$ (single-turbine)             & 23     & 15        & 8     \\
$P_\text{norm}$ (single-turbine, unwaked)    & 8      & 8         & 7     \\
$P_\text{norm}$ (single-turbine, waked)      & 29     & 18        & 8     \\
$\Delta P_i(\theta_w)$                       & 51     & 33        & 22     \\
$\eta(U_{\text{norm},\infty}$,TI$_\infty,\theta_w)$ & 57     & 39        & 24    \\
$\Delta E(U_{\text{norm},\infty}$,TI$_\infty)$      & 82     & 61        & 40   \\
\end{tabular}
\end{table}

For the sake of completeness, the gross and net capacity factors and the overall wake losses for the full SCADA dataset are reported in Table \ref{tab:CapacityFactor}. The P2D-RANS is still the model giving the best prediction, nevertheless its capability in simulating the wake interaction under different atmospheric stability and operative conditions is not clearly identifiable using these global statistical parameters. 
\begin{table}[]
\centering
\caption{Gross capacity factor, net capacity factor, and wake loss factor estimated with the various wind farm models.}\label{tab:CapacityFactor}
\begin{tabular}{lccccc}
&SCADA  & P2D-RANS & Jensen & Multizone & Gauss  \\ \hline
Gross capacity factor & 0.801 & 0.800   & 0.799 & 0.799    & 0.800 \\
Net capacity factor   & 0.77 & 0.768   & 0.746 & 0.763    & 0.774 \\
Wake losses {[}\%{]}  & 3.87   & 4.00     & 6.63   & 4.51      & 3.25   \\ 
\end{tabular}
\end{table}

\section{Conclusions}\label{sec:Conclusions}
The theoretical fundamentals, numerical scheme, verification, and validation process of the Pseudo 2D-RANS (P2D-RANS) model have been discussed in this manuscript. The P2D-RANS uses a shallow-water approximation of the RANS equations to achieve an efficient solution of a wind farm flow vertically averaged over the rotor heights by inserting into the RANS solver appropriate corrective source terms, which take into account the contribution to the mass and momentum budgets of vertical fluxes and spatial dispersive stresses.

The set of depth-averaged and corrected flow equations is solved through an iterative marching scheme, which alternates sweeps in the streamwise directions for the solution of the momentum equations and global pressure corrections to ensure a solenoidal velocity field upon the convergence of the algorithm. The detailed derivation of the pressure correction, which may represent a generalizable and convenient numerical tool for the solution of partially-parabolized flows with source terms is also provided.

Clustered and statistically analyzed LiDAR measurements of wakes generated by utility-scale wind turbines have been leveraged to calibrate the turbulence closure and the actuator disk models implemented into the P2D-RANS. The optimally tuned turbulent eddy-viscosity exhibits physically reasonable trends as a function of the turbulence intensity of the incoming wind with a striking wake recovery enhancement detected for convective conditions, which might be connected with the occurrence of wake meandering. The thrust-force distribution over the rotor radial direction has been estimated for the various bins of incoming wind speed and turbulence intensity  at hub height. This data-driven approach has enabled the detection of a general reduction of the thrust force when turbine operations transition from region II to region III of the power curve.

The correctness of the conceptual and computational framework of the P2D-RANS and its numerical implementation has been verified versus the solution of an elliptic axisymmetric RANS solver for single and in-tandem wind turbines. The validation process of the P2D-RANS has been broken down into three phases. First, effects of the daily cycle of the atmospheric stability on operations of an onshore farm have been simulated with a 10-minute resolution, then assessed against the SCADA data at the single-turbine level. Subsequently, the 10-minute-averaged power has been simulated for the whole set of available SCADA data to perform statistically meaningful error analysis. The power of single turbines shows an $R^2=0.956$, a normalized mean absolute error of $5.6\%$, and negligible bias, which represents an 8\% error reduction compared to the Gaussian wake model. Finally, wake losses have been quantified through different indicators and extensively analyzed as a function of the inflow conditions. The P2D-RANS reproduces satisfactorily the single-turbine directional wake losses, including the effect of a local increase of the freestream wind speed, referred to as flow speed-up conditions, which are induced by the rotor pressure field for specific wind directions. Such speed-up conditions have also repercussions on the wind farm efficiency, which is predicted with a normalized mean absolute error of 3\%, 24\% smaller than for the Gaussian model.

The P2D-RANS reproduces convincingly the annual energy wake losses (12\% error) for different stability and operative conditions, which indicates the financial relevance of LiDAR-driven CFD modeling in the context of the optimal design of wind farms. It is also proven the necessity for correct modeling of the variability of wind farm efficiency with atmospheric stability.

The current implementation of the P2D-RANS can solve the steady flow over a wind farm with 25 turbines and extending $30D$ by $30D$ in an averaged time of $\sim80$ s on a commercial laptop equipped with a single i5 1.70GHz core, which can enable performing real-time wind farm monitoring and control. The required computational time has been observed to be roughly half when operating an Intel Xeon Platinum 8160 48-core 2.1GHz cluster, which implies that the full characterization of the performances of a mid-sized farm can be completed in $\sim170$ CPU-hours.




\section*{Acknowledgments}
This material is based upon work supported by the National Science Foundation (grant no. IIP-1362022; Collaborative Research – I/UCRC for Wind Energy, Science, Technology, and Research) and from the WindSTAR I/UCRC Members of Aquanis, Inc., EDP Renewables, Bachmann Electronic Corp., GE Energy, Huntsman, Hexion, Leeward Asset Management, LLC, Pattern Energy, EPRI, LMWind, Texas Wind Tower, and TPI Composites. This research has been funded by the National Science Foundation CBET Fluid Dynamics (grant no. 1705837). 

\section*{Conflict of interest declaration}
The authors have no conflicts to disclose.

\appendix

\section{Estimation of 3D correction for an axisymmetric flow\label{app:3D_corr}}
The 3D and dispersive terms in Eqs. \ref{eq:RANS_SW} are estimated assuming an axisymmetric, non-swirling flow. By setting the boundary of the domain to $z_1=-D/2$ and $z_2=D/2$, the unknown 3D terms can be conveniently recast as follows:
\begin{equation}\label{eq:SW_equations_terms}
    \begin{cases}
\frac{\overline{w}_2-\overline{w}_1}{z_2-z_1}=\frac{\overline{w}_{D/2}}{D/2} \\
\frac{\overline{u}_2\overline{w}_2-\overline{u}_1\overline{w}_1-\hat{u}(\overline{w}_2-\overline{w}_1)}{z_2-z_1}=\frac{(\overline{u}_{D/2}-\hat{u})\overline{w}_{D/2}}{D/2}\\
\frac{\nu_T}{z_2-z_1}\left(\frac{\partial{\overline{u}}}{\partial z}\bigg|_2-\frac{\partial{\overline{u}}}{\partial z}\bigg|_1 \right)=\frac{\nu_T}{D/2}\frac{\partial \overline{u}}{\partial z}\bigg|_{D/2} \\
\frac{\overline{v}_2\overline{w}_2-\overline{v}_1\overline{w}_1-\hat{v}(\overline{w}_2-\overline{w}_1)}{z_2-z_1}=\frac{(\overline{v}_{D/2}-\hat{v})\overline{w}_{D/2}}{D/2} \\
\frac{\nu_T}{z_2-z_1} \left(\frac{\partial{\overline{v}}}{\partial z}\bigg|_2-\frac{\partial{\overline{v}}}{\partial z}\bigg|_1 \right)=\frac{\nu_T}{D/2}\frac{\partial \overline{v}}{\partial z}\bigg|_{D/2}.\\
\end{cases}
\end{equation}
\begin{figure}[b!]
\centerline{\includegraphics[width=0.3\textwidth]{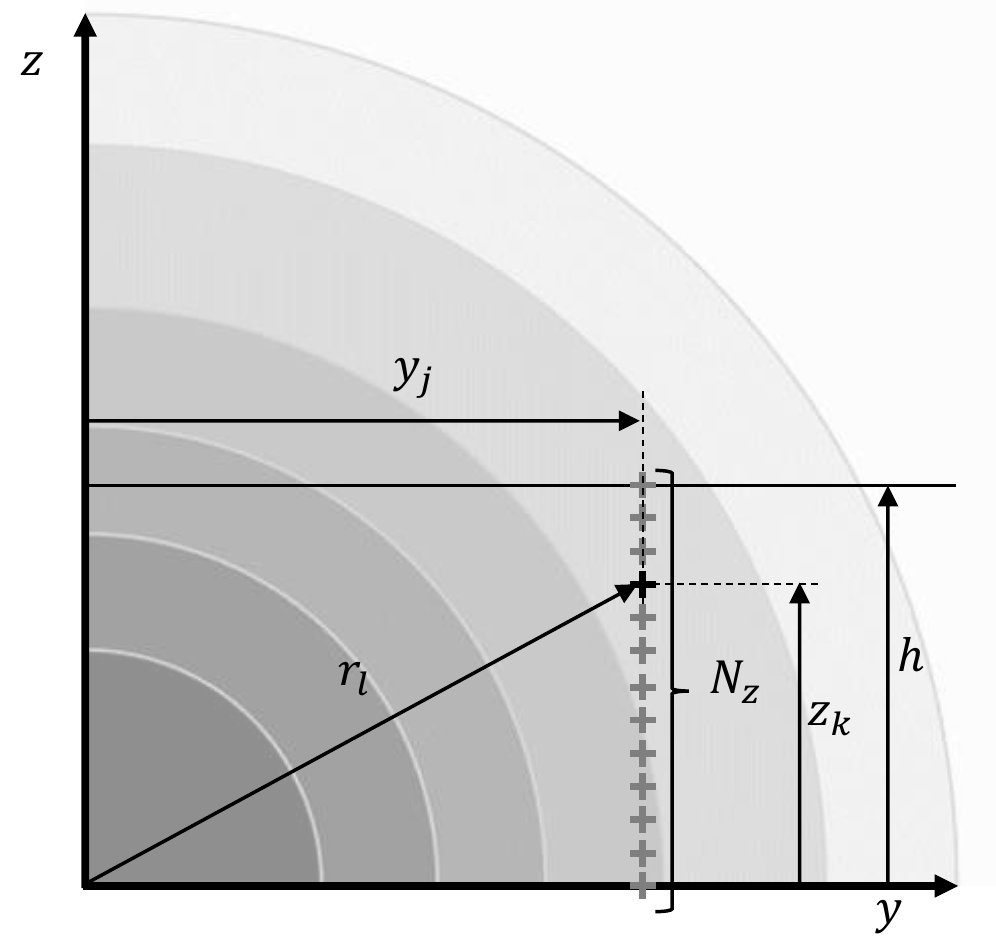}}
\caption{Sampling of the axisymmetric streamwise velocity for the calculation of the depth average. The iso-levels in the background indicate the magnitude of the axisymmetric velocity field, $\overline{u}_x$, which is sampled to calculate $\tilde{u}$.\label{fig:DepthAvg_sketch}}
\end{figure}
The depth-average $\hat{\cdot}$ is related to the 3D field by the operator defined in Eq. \ref{eq:DepthAverage}. 

Under the assumption of axisymmetric flow, the 3D field $\overline{u}(x_i,y_j,z_k)$ can be obtained by sampling the streamwise velocity component expressed in cylindrical coordinates, $\overline{u}_x(x,r)$ at $x = x_i$ and radial location $r_l =\sqrt{y_j^2+z_j^2}$ (see Fig. \ref{fig:DepthAvg_sketch}). Formalizing the sampling operation from the radial location $r_l$ to the Cartesian equivalent $(y_j,z_k)$  as a linear interpolation operator, $\Phi$, yields: 
\begin{eqnarray}
\label{eq:uhat_ux2}
\hat{u}_{ij}=&&\frac{1}{N_z}\sum_{k=1}^{N_z}\overline{u}(x_i,y_j,z_k)=\frac{1}{N_z}\sum_{k=1}^{N_z}\overline{u}_x\left(x_i,\overbrace{\sqrt{y_j^2+z_k^2}}^{r_{l}}\right)= \nonumber\\
&&=\frac{1}{N_z}\sum_{k=1}^{N_z} \sum_{l=1}^{N_r}\Phi_{k,l}\overline{u}_{x,il} \Rightarrow \hat{u}=M_u(\overline{u}_x)
\end{eqnarray}
Similarly, for the depth-averaged transverse velocity, one  can show that:
\begin{equation}
\label{eq:vhat_ur}
\hat{v}_{ij}=\frac{1}{N_z}\sum_{k=1}^{N_z} \sum_{l=1}^{N_r}\frac{y_j}{r_l}\Phi_{k,l}\overline{u}_{r,il}\Rightarrow \hat{v}=M_{v}(\overline{u}_r)
\end{equation}
It is noteworthy that the linear operators $M_u$ and $M_v$ of Eqs. \ref{eq:uhat_ux2} and \ref{eq:vhat_ur} are invertible, namely it always possible to calculate from a depth-averaged velocity (i.e. the P2D-RANS provisional solution) the equivalent axisymmetric velocity field without swirl. 

\section{Derivation of the pressure correction\label{app2}}
\begin{figure}[h!]
\centerline{\includegraphics[width=0.8\textwidth, trim=0 0 0 0,clip]{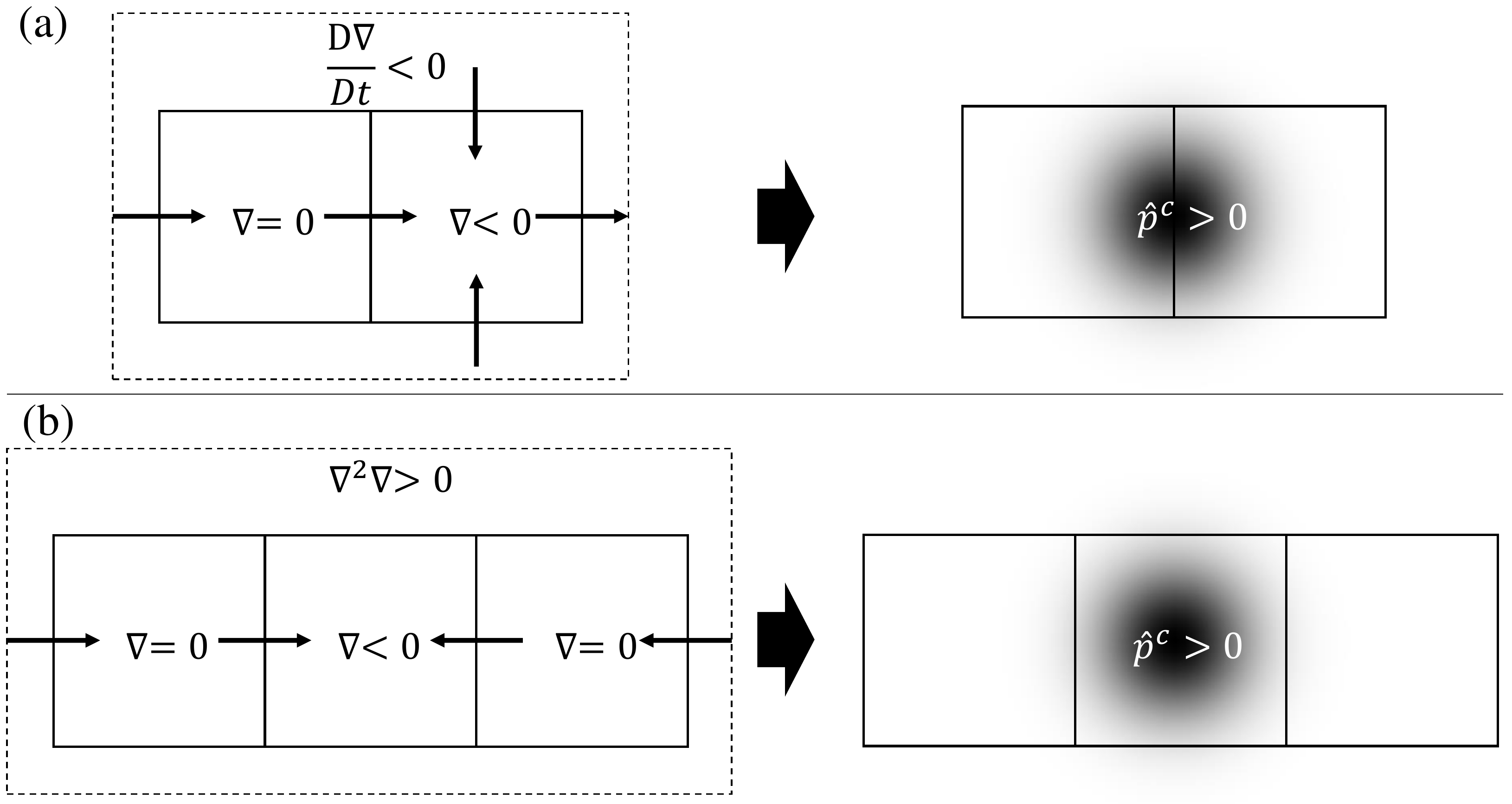}}
 \caption{Graphical representation of the effect of the LHS of Eq. (\ref{eq:Press_correction2}) on pressure correction: \textbf{(a)}: advection term; \textbf{(b)} diffusion term.}\label{fig:Press_correction_sketch}
\end{figure}

To derive the pressure correction, the velocity and pressure fields at the current iteration are separated into two components:
\begin{enumerate}
    \item the provisional field $(\hat{u},\hat{v},\hat{p})$, in which the velocity components satisfy the momentum equation with the currently estimated pressure $\hat{p}$;
    \item the correction field $(\hat{u}^c,\hat{v}^c,\hat{p}^c)$, that is added to the provisional field to satisfy the continuity equation.
\end{enumerate}
By assuming that the correction field is much smaller than the main field, i.e. $\hat{u}^c\ll\hat{u}$, $\hat{v}^c\ll\hat{v}$, $\hat{p}^c\ll\hat{p}$, and considering that the main field satisfies the RANS equations (Eq. \ref{eq:RANS_SW}), one can recast the correction field equations as:
 \begin{equation}\label{eq:NS_corrections}
 \begin{cases}
 \hat{u}\frac{\partial \hat{u}^c}{\partial x}+ \hat{v}\frac{\partial \hat{u}^c}{\partial y}+ \hat{u}^c\frac{\partial \hat{u}}{\partial x}+ \hat{v}^c\frac{\partial \hat{u}}{\partial y}=-\frac{\partial \hat{p}^c}{\partial x}+\nu_T\left( \frac{\partial^2 \hat{u}^c}{\partial x^2}+\frac{\partial^2 \hat{u}^c}{\partial y^2}\right)
 \\
  \hat{u}\frac{\partial \hat{v}^c}{\partial x}+ \hat{v}\frac{\partial \hat{v}^c}{\partial y}+ \hat{u}^c\frac{\partial \hat{v}}{\partial x}+ \hat{v}^c\frac{\partial \hat{v}}{\partial y}=-\frac{\partial \hat{p}^c}{\partial y}+\nu_T\left( \frac{\partial^2 \hat{v}^c}{\partial x^2}+\frac{\partial^2 \hat{v}^c}{\partial y^2}\right)
 \end{cases}
 \end{equation}
Taking the divergence of Eq. \ref{eq:NS_corrections} yields:
  \begin{equation}\label{eq:press_corr1}
 \begin{array}{l}
    \hat{u}^c\frac{\partial (\nabla \cdot \mathbf{\hat{u}})}{\partial x}+\hat{v}^c\frac{\partial (\nabla \cdot \mathbf{\hat{u}})}{\partial y}+\hat{u}\frac{\partial (\nabla \cdot \mathbf{\hat{u}^c})}{\partial x}+\hat{v}\frac{\partial (\nabla \cdot \mathbf{\hat{u}^c})}{\partial y}=\\+
    \gamma(\hat{u},\hat{v},\hat{u}^c,\hat{v}^c) -\frac{\partial^2 \hat{p}^c}{\partial x^2}-\frac{\partial^2 \hat{p}^c}{\partial y^2}\\
     +\nu_T\left( \frac{\partial^2 (\nabla\cdot \mathbf{\hat{u}^c})}{\partial x^2}+\frac{\partial^2 (\nabla\cdot \mathbf{\hat{u}^c})}{\partial y^2}\right)
 \end{array}
 \end{equation}
 where:
\begin{equation}\label{eq:spurious}
 \begin{array}{l}
     \gamma(\hat{u},\hat{v},\hat{u}^c,\hat{v}^c)    =2\left ( \frac{\partial \hat{u}}{\partial x}  \frac{\partial \hat{u}^c}{\partial x} +\frac{\partial \hat{v}}{\partial y}  \frac{\partial \hat{v}^c}{\partial y}        +\frac{\partial \hat{v}}{\partial x}  \frac{\partial \hat{u}^c}{\partial y}  +\frac{\partial \hat{v}^c}{\partial x}  \frac{\partial \hat{u}}{\partial y}  \right).
     \end{array}
\end{equation}
For the corrected fields to be solenoidal, the following equality must be satisfied:
\begin{equation}\label{eq:Divergence}
    \nabla \cdot \mathbf{\hat{u}^c}= -\nabla\cdot \mathbf{\hat{u}}-\frac{\overline{w}_2-\overline{w}_1}{z_2-z_1}=-\nabla.
\end{equation}
Eq. \ref{eq:Divergence} implies that even though $|\mathbf{\hat{u}^c}|<<|\mathbf{\hat{u}}|$, the divergence of both fields have the same magnitude. This allows neglecting the first two terms of the LHS of equation \ref{eq:press_corr1}:
 \begin{equation}\label{eq:press_corr2}
 \begin{array}{l}
     -\hat{u}\frac{\partial \nabla}{\partial x}-\hat{v}\frac{\partial \nabla}{\partial y} + \gamma\sim-\frac{\partial^2 \hat{p}^c}{\partial x^2}-\frac{\partial^2 \hat{p}^c}{\partial y^2}
     -\nu_T\left( \frac{\partial^2 \nabla}{\partial x^2}+\frac{\partial^2 \nabla}{\partial y^2}\right),
 \end{array}
 \end{equation}
The correction velocity components are unknown, so the term $\gamma$ defined in Eq. (\ref{eq:spurious}) is necessarily dropped \citep{Patankar1972,Moore1979}, which yields the final results reported as Eq. (\ref{eq:Press_correction1}). 
The error associated with this approximation is eventually reduced by applying the correction iteratively. The RHS of the pressure correction (Eq. (\ref{eq:Press_correction2})) contains an advection term and a diffusion term. The former creates a positive (negative) local pressure if a certain material volume of fluid is undergoing compression (expansion), while the latter creates a positive (negative) local pressure if a region of fluid is locally experiencing compression (expansion). The latter induces a pressure maximum (minimum) if the flow is locally experiencing a compression (expansion). Nonetheless, the pressure correction guides the solution towards a divergence-free velocity field. Figure \ref{fig:Press_correction_sketch} shows graphically the behavior of the pressure correction, $\hat{p}^c$, in response to a local compression. 

\section{Loads formulation}\label{app:Loads}
The non-dimensional streamwise momentum sink used to mimic the action of the generic $i^{\text{th}}$ rotor on the flow in the P2D-RANS is defined as:
\begin{equation}
\hat{f}_{x,i}=c s(x) \hat{\phi}(y) \hat{u}^2(x_{0,i},y),
\end{equation}
where $c$ is normalization constant, $s(x)$ is a Gaussian smoothing kernel with standard deviation $0.5D$, $\hat{\phi}(y)$ is a shape function containing the information on the radial distribution of the axial induction, and $x_{0,i}$ is the sampling location $2D$ upstream of the turbine where the freestream velocity is extracted. The introduction of the $\hat{u}^2(x_{0,i},y)$ modulates the local thrust in case of non-uniform inflow and enhances numerical stability by preventing reverse flow. The normalization constant is obtained by imposing the total thrust to equate the one dictated by the LiDAR-driven $c_t$ as follows:
\begin{equation}
    c \int_{-0.5}^{0.5} \hat{\phi}(y) \hat{u}^2(x_{0,i},y) \left[\int_{+\infty}^{-\infty} s(x) dx\right] dy = \frac{\frac{1}{2}\rho c_t \pi \frac{D^2}{4} U_{\infty,i}^2}{\rho U_\infty^2 D^2},
\end{equation}
where $U_{\infty,i}$ is the dimensional local incoming wind speed, defined as the average of the streamwise velocity over the rotor span at $x_{0,i}$. Since that $s(x)$ has unit integral, the former equality yields:
\begin{equation}
c=\frac{\pi c_t}{8 \int_{-0.5}^{0.5} \hat{\phi}(y) \hat{u}^2(x_{0,i},y)}\left( \frac{U_{\infty,i}}{U_{\infty}}\right)^2.
\end{equation}
In case of uniform inflow, the former equation reduces to:
\begin{equation}
c=\frac{\pi c_t}{8 \int_{-0.5}^{0.5} \hat{\phi}(y)}.
\end{equation}

\bibliographystyle{unsrt}
\bibliography{Bibliography}

\begin{thebibliography}{100}

\bibitem{SanzRodrigo2017}
Javier {Sanz Rodrigo}, Roberto~Aurelio {Ch{\'{a}}vez Arroyo}, Patrick Moriarty,
  Matthew Churchfield, Branko Kosovi{\'{c}}, Pierre~Elouan R{\'{e}}thor{\'{e}},
  Kurt~Schaldemose Hansen, Andrea Hahmann, Jeffrey~D. Mirocha, and Daran Rife.
\newblock {Mesoscale to microscale wind farm flow modeling and evaluation}.
\newblock {\em Wiley Interdisciplinary Reviews: Energy and Environment}, 6(2),
  2017.

\bibitem{Veers2019}
P~Veers, K~Dykes, E~Lantz, S~Barth, CL~Bottasso, O~Carlson, A~Clifton, J~Green,
  P~Green, H~Holttinen, D~Laird, V~Lehtom{\"{a}}ki, JK~Lundquist, J~Manwell,
  M~Marquis, C~Meneveau, P~Moriarty, X~Munduate, M~Muskulus, J~Naughton, L~Pao,
  J~Paquette, J~Peinke, A~Robertson, J.~S Rodrigo, AM~Sempreviva, JC~Smith,
  A~Tuohy, and R~Wiser.
\newblock {Grand challenges in the science of wind energy}.
\newblock {\em Science}, 366(6464), 2019.

\bibitem{Christiansen2006}
MB~Christiansen and CharlotteB Hasager.
\newblock {Using Airborne and Satellite SAR for Wake Mapping Offshore}.
\newblock {\em Wind Energy}, 9(March):437--455, 2006.

\bibitem{Platis2018}
A~Platis, SK~Siedersleben, J~Bange, A~Lampert, K~B{\"{a}}rfuss, R~Hankers,
  B~Ca{\~{n}}adillas, R~Foreman, J~Schulz-Stellenfleth, B~Djath, T~Neumann, and
  S~Emeis.
\newblock {First in situ evidence of wakes in the far field behind offshore
  wind farms}.
\newblock {\em Sci Rep}, 8(1):1--10, 2018.

\bibitem{Lundquist2019}
JK~Lundquist, KK~DuVivier, D~Kaffine, and JM~Tomaszewski.
\newblock {Costs and consequences of wind turbine wake effects arising from
  uncoordinated wind energy development}.
\newblock {\em Nat Energy}, 4(January):26--34, 2019.

\bibitem{Porte-agel2019}
Fernando Port{\'{e}}-agel, Majid Bastankhah, and Sina Shamsoddin.
\newblock {Wind-Turbine and Wind-Farm Flows : A Review}.
\newblock {\em Boundary-Layer Meteorology}, 2019.

\bibitem{Bailey2014}
Bruce~H Bailey.
\newblock {The Financial Implications of Resource Assessment}.
\newblock Technical report, TruePower, 2014.

\bibitem{El-Asha2017}
S~El-Asha, L~Zhan, and GV~Iungo.
\newblock {Quantification of power losses due to wind turbine wake interactions
  through SCADA , meteorological and wind LiDAR data}.
\newblock {\em Wind Energy}, 20(June):1823--1839, 2017.

\bibitem{Barthelmie2010}
RJB Barthelmie, SCP Pryor, STF Frandsen, KSH Hansen, JGS Schepers, K~Rados,
  W~Schelz, A~Neubert, LE~Jensen, and S~Neckelmann.
\newblock {Quantifying the Impact of Wind Turbine Wakes on Power Output at
  Offshore Wind Farms}.
\newblock {\em J Atmos Ocean Technol}, pages 1302--1317, 2010.

\bibitem{Sebastiani2020}
Alessandro Sebastiani, Antonio Segalini, Francesco Castellani, and Giorgio
  Crasto.
\newblock {Data analysis and simulation of the Lillgrund wind farm}.
\newblock {\em Wind Energy}, (November):1--15, 2020.

\bibitem{Magnusson1994}
M~Magnusson and AS~Smedman.
\newblock {Influence of Atmospheric Stability on Wind Turbine Wakes}.
\newblock {\em Wind Eng}, 18(3):139--152, 1994.

\bibitem{Iungo2014}
Giacomo~Valerio Iungo and Fernando Port{\'e}-Agel.
\newblock Volumetric lidar scanning of wind turbine wakes under convective and
  neutral atmospheric stability regimes.
\newblock {\em Journal of Atmospheric and Oceanic Technology},
  31(10):2035--2048, 2014.

\bibitem{CarbajoFuertes2018}
Fernando Carbajo~Fuertes, Corey~D Markfort, and Fernando Port{\'e}-Agel.
\newblock Wind turbine wake characterization with nacelle-mounted wind lidars
  for analytical wake model validation.
\newblock {\em Remote Sens}, 10(5):668, 2018.

\bibitem{Zhan2020}
Lu~Zhan, Stefano Letizia, and Giacomo Valerio~Iungo.
\newblock Lidar measurements for an onshore wind farm: Wake variability for
  different incoming wind speeds and atmospheric stability regimes.
\newblock {\em Wind Energy}, 23(3):501--527, 2020.

\bibitem{BarthelmieWE2010}
RJ~Barthelmie and LE~Jensen.
\newblock {Evaluation of wind farm efficiency and wind turbine wakes at the
  Nysted offshore wind farm}.
\newblock {\em Wind Energy}, 13:573/586, 2010.

\bibitem{Hansen2012}
KS~Hansen, RJ~Barthelmie, LE~Jensen, and A~Sommer.
\newblock {The impact of turbulence intensity and atmospheric stability on
  power deficits due to wind turbine wakes at Horns Rev wind farm}.
\newblock {\em Wind Energy}, 2012.

\bibitem{MeyerForsting2017}
Alexander~Raul {Meyer Forsting}, N.~Troldborg, and Mac Gaunaa.
\newblock {The flow upstream of a row of aligned wind turbine rotors and its
  effect on power production}.
\newblock {\em Wind Energy}, 20:63--77, 2017.

\bibitem{Wu2017}
Ka~Ling Wu and Fernando Port{\'{e}}-Agel.
\newblock {Flow adjustment inside and around large finite-size wind farms}.
\newblock {\em Energies}, 10(12):4--9, 2017.

\bibitem{Ebenhoch2017}
Raphael Ebenhoch, Blas Muro, Jan~{\AA}ke Dahlberg, Patrik {Berkesten
  H{\"{a}}gglund}, and Antonio Segalini.
\newblock {A linearized numerical model of wind-farm flows}.
\newblock {\em Wind Energy}, 20(5):859--875, 2017.

\bibitem{Bleeg2018}
James Bleeg, Mark Purcell, Renzo Ruisi, and Elizabeth Traiger.
\newblock {Wind farm blockage and the consequences of neglecting its impact on
  energy production}.
\newblock {\em Energies}, 11(6), 2018.

\bibitem{Branlard2020}
Emmanuel Branlard, Eliot Quon, Alexander~R. {Meyer Forsting}, Jennifer King,
  and Patrick Moriarty.
\newblock {Wind farm blockage effects: Comparison of different engineering
  models}.
\newblock {\em Journal of Physics: Conference Series}, 1618(6), 2020.

\bibitem{Schneemann2021}
J{\"{o}}rge Schneemann, Frauke Theuer, Andreas Rott, Martin
  D{\"{o}}renk{\"{a}}mper, and Martin K{\"{u}}hn.
\newblock {Offshore wind farm global blockage measured with scanning lidar}.
\newblock {\em Wind Energy Science}, 6(2):521--538, 2021.

\bibitem{Nishino2012}
Takafumi Nishino and Richard H~J Willden.
\newblock {The efficiency of an array of tidal turbines partially blocking a
  wide channel}.
\newblock {\em J Fluid Mech.}, 708:596--606, 2012.

\bibitem{McTavish2014}
S~McTavish, S~Rodrigue, and D~Feszty.
\newblock {An investigation of in-field blockage effects in closely spaced
  lateral wind farm configurations}.
\newblock {\em Wind Energy}, 18:1989--2011, 2014.

\bibitem{MeyerForsting2016}
Alexander~Raul {Meyer Forsting}, Niels Troldborg, and Mac Gaunaa.
\newblock {The flow upstream of a row of aligned wind turbine rotors and its
  effect on power production}.
\newblock {\em Wind Energy}, pages 1481--1498, 2016.

\bibitem{Jensen1983}
N~O Jensen.
\newblock {A note on wind generator interaction}.
\newblock Technical report, Ris{\o}, Roskilde, Denmark, 1983.

\bibitem{Frandsen2006}
Sten Frandsen, Rebecca Barthelmie, Sara Pryor, Ole Rathmann, S{\o}ren Larsen,
  J{\o}rgen H{\o}jstrup, and Morten Th{\o}gersen.
\newblock {Analytical modelling of wind speed deficit in large offshore wind
  farms}.
\newblock {\em Wind Energy}, 9(1-2):39--53, 2006.

\bibitem{Schlichting1979}
H.~Schlichting.
\newblock {\em {Boundary Layer Theory, 9th Ed.}}
\newblock McGraw-Hill, 1979.

\bibitem{Abramovich1963}
G.~N. Abramovich.
\newblock {\em {The Theory of Turbulent Jets}}.
\newblock MIT Press, 1963.

\bibitem{Lissaman1979}
PBS Lissaman.
\newblock Energy effectiveness of arbitrary arrays of wind turbines.
\newblock {\em J Energy}, 3(6):323--328, 1979.

\bibitem{Voutsinas1990}
S.~Voutsinas.
\newblock {On the analysis of wake effects in wind parks}.
\newblock {\em Wind Engineering}, 14(4):204--219, 1990.

\bibitem{Larsen1988}
G.C. Larsen.
\newblock {A Simple Wake Calculation Procedure}.
\newblock Technical report, Technical University of Denmark, Roskild, Denmark,
  1988.

\bibitem{Bastankhah2014}
Majid Bastankhah and Fernando Port{\'{e}}-Agel.
\newblock {A new analytical model for wind-turbine wakes}.
\newblock {\em Renewable Energy}, 70:116--123, 2014.

\bibitem{Ishihara2018}
Takeshi Ishihara and Guo~Wei Qian.
\newblock {A new Gaussian-based analytical wake model for wind turbines
  considering ambient turbulence intensities and thrust coefficient effects}.
\newblock {\em Journal of Wind Engineering and Industrial Aerodynamics},
  177(May):275--292, 2018.

\bibitem{Xie2015}
Shengbai Xie and Cristina Archer.
\newblock {Self-similarity and turbulence characteristics of wind turbine wakes
  via large-eddy simulation}.
\newblock {\em Wind Energy}, 18(10):1815--1838, 2015.

\bibitem{Cheng2018}
Wai~Chi Cheng and Fernando Port{\'{e}}-Agel.
\newblock {A Simple Physically-Based Model for Wind-Turbine Wake Growth in a
  Turbulent Boundary Layer}.
\newblock {\em Boundary-Layer Meteorology}, 169(1):1--10, 2018.

\bibitem{Schreiber2020}
Johannes Schreiber, Amr Balbaa, and Carlo~L. Bottasso.
\newblock {Brief communication: A double-Gaussian wake model}.
\newblock {\em Wind Energy Science}, 5(1):237--244, 2020.

\bibitem{Shapiro2019}
Carl~R. Shapiro, Genevieve~M. Starke, Charles Meneveau, and Dennice~F. Gayme.
\newblock {A wake modeling paradigm for wind farm design and control}.
\newblock {\em Energies}, 12(15), 2019.

\bibitem{Blondel2020}
Fr{\'{e}}d{\'{e}}ric Blondel and Marie Cathelain.
\newblock {An alternative form of the super-Gaussian wind turbine wake model}.
\newblock {\em Wind Energy Science}, 5:1225--1236, 2020.

\bibitem{Pope2000}
S.~B. Pope.
\newblock {\em Turbulent Flows}.
\newblock Cambridge University Press, 2000.

\bibitem{Crasto2012}
G.~Crasto, A.~R. Gravdahl, F.~Castellani, and E.~Piccioni.
\newblock {Wake modeling with the actuator Disc concept}.
\newblock {\em Energy Procedia}, 24(January):385--392, 2012.

\bibitem{Mortensen2016}
Niels~G Mortensen.
\newblock {Wind resource assessment using the WAsP software}.
\newblock Technical Report December, DTU Wind Energy E-0135, 2016.

\bibitem{Gebraad2016}
Pieter~MO Gebraad, FW~Teeuwisse, JW~Van~Wingerden, Paul~A Fleming, SD~Ruben,
  JR~Marden, and LY~Pao.
\newblock Wind plant power optimization through yaw control using a parametric
  model for wake effects—a cfd simulation study.
\newblock {\em Wind Energy}, 19(1):95--114, 2016.

\bibitem{Zhan2020WES}
Lu~Zhan, Stefano Letizia, and Giacomo~Valerio Iungo.
\newblock Optimal tuning of engineering wake models through lidar measurements.
\newblock {\em Wind Energy Sci}, 5(4):1601--1622, 2020.

\bibitem{Gunn2016}
K.~Gunn, C.~Stock-Williams, M.~Burke, R.~Willden, C.~Vogel, W.~Hunter,
  T.~Stallard, N.~Robinson, and S.~R. Schmidt.
\newblock {Limitations to the validity of single wake superposition in wind
  farm yield assessment}.
\newblock {\em J Phys: Conf Ser}, 749(1), 2016.

\bibitem{Zong2020}
Haohua Zong and Fernando Port{\'e}-Agel.
\newblock A momentum-conserving wake superposition method for wind farm power
  prediction.
\newblock {\em Journal of Fluid Mechanics}, 889, 2020.

\bibitem{Barthelmie2006}
R.J. Barthelmie, L:~Folkerts, Gunner~C. Larsen, K.~Rados, S.~C. Pryor, S.T.
  Frandsen, B.~Lange, and G.~Schepers.
\newblock {Comparison of Wake Model Simulations with Offshore Wind Turbine Wake
  Profiles Measured by Sodar}.
\newblock {\em Journal of Atmospheric and Oceanic Technology}, 23:888--901,
  2006.

\bibitem{Barthelmie2009}
R~J Barthelmie, K~Hansen, S~T Frandsen, O~Rathmann, J~G Schepers, W~Schlez,
  J~Phillips, K~Rados, A~Zervos, E~S Politis, and P~K Chaviaropoulos.
\newblock {Modelling and Measuring Flow and Wind Turbine Wakes in Large Wind
  Farms Offshore}.
\newblock {\em Wind Energy}, 12(June):431--444, 2009.

\bibitem{Moriarty2014}
Patrick Moriarty, Javier~Sanz Rodrigo, Pawel Gancarski, Matthew Chuchfield,
  Jonathan~W. Naughton, Kurt~S. Hansen, Ewan MacHefaux, Eoghan Maguire,
  Francesco Castellani, Ludovico Terzi, Simon~Philippe Breton, and Yuko Ueda.
\newblock {IEA-task 31 WAKEBENCH: Towards a protocol for wind farm flow model
  evaluation. Part 2: Wind farm wake models}.
\newblock {\em Journal of Physics: Conference Series}, 524(1), 2014.

\bibitem{Archer2018}
Cristina~L. Archer, Ahmadreza Vasel-Be-Hagh, Chi Yan, Sicheng Wu, Yang Pan,
  Joseph~F. Brodie, and A.~Eoghan Maguire.
\newblock {Review and evaluation of wake loss models for wind energy
  applications}.
\newblock {\em Applied Energy}, 226(February 2018):1187--1207, 2018.

\bibitem{Santoni2018}
Christian Santoni, Edgardo~J Garcia-Cartgena, Umberto Ciri, Giacomo~Valerio
  Iungo, and Stefano Leonardi.
\newblock {Coupling of mesoscale Weather Research and Forecasting model to a
  high fidelity Large Eddy Simulation}.
\newblock {\em J. Phys.: Conf. Ser.}, 1037:062010, 2018.

\bibitem{Santoni2020}
Christian Santoni, Edgardo~J Garcia-Cartgena, Umberto Ciri, Lu~Zhan,
  Giacomo~Valerio Iungo, and Stefano Leonardi.
\newblock {One-way mesoscale-microscale coupling for simulating a wind farm in
  North Texas: Assessment against SCADA and LiDAR data}.
\newblock {\em Wind Energy}, 23:691--710, 2020.

\bibitem{Mehta2014}
D~Mehta, A~H~Van Zuijlen, B~Koren, J~G Holierhoek, and H~Bijl.
\newblock {Journal of Wind Engineering Large Eddy Simulation of wind farm
  aerodynamics: A review}.
\newblock {\em J. of Wind Engineering and Industrial Aerodynamics}, 133:1--17,
  2014.

\bibitem{Breton2017}
S~Breton, J~Sumner, JN~S{\o}rensen, KS~Hansen, S~Sarmast, and S~Ivanell.
\newblock {A survey of modelling methods for high-fidelity wind farm
  simulations using large eddy simulation}.
\newblock {\em Phil Trans R Soc A}, 375(20160097), 2017.

\bibitem{Sanderse2011}
B~Sanderse, SP~Van Der~Pijl, and B~Koren.
\newblock {Review of computational fluid dynamics for wind turbine wake
  aerodynamics}.
\newblock {\em Wind Energy}, 14:799--819, 2011.

\bibitem{Bou-Zeid2005}
Elie Bou-Zeid, Charles Meneveau, and Marc Parlange.
\newblock {A scale-dependent Lagrangian dynamic model for large eddy simulation
  of complex turbulent flows}.
\newblock {\em Physics of Fluids}, 17(2):1--18, 2005.

\bibitem{Wan2011}
Feng Wan and Fernando Port{\'{e}}-Agel.
\newblock {Large-Eddy Simulation of Stably-Stratified Flow Over a Steep Hill}.
\newblock {\em Boundary-Layer Meteorology}, pages 367--384, 2011.

\bibitem{Rethore2009}
Pierre-Elouan R{\'{e}}thor{\'{e}}.
\newblock {\em {Wind Turbine Wake in Atmospheric Turbulence}}.
\newblock PhD thesis, Technical University of Denmark, 2009.

\bibitem{VanDerLaan2015}
M~Paul Van Der~Laan, Niels~N S{\o}rensen, Pierre-Elouan R{\'e}thor{\'e}, Jakob
  Mann, Mark~C Kelly, Niels Troldborg, J~Gerard Schepers, and Ewan Machefaux.
\newblock An improved $\kappa-\epsilon$ model applied to a wind turbine wake in
  atmospheric turbulence.
\newblock {\em Wind Energy}, 18(5):889--907, 2015.

\bibitem{Taylor1980}
P.~A. {Taylor}.
\newblock {On wake decay and row spacing for WECS farms}.
\newblock In {\em 3rd International Symposium on Wind Energy Systems}, pages
  451--468, January 1980.

\bibitem{Sforza1981}
P.~M. Sforza, P.~Sheerin, and M.~Smorto.
\newblock {Three-dimensional wakes of simulated wind turbines}.
\newblock {\em AIAA Journal}, 19(9):1101--1107, 1981.

\bibitem{Crespo1985}
A~Crespo, F.~Manuel, D.~Moreno, E.~Fraga, and J.~Hernandez.
\newblock {Numerical Analysis of Wind Turbine Wakes}.
\newblock In {\em Proceedings of an International Workshop held at the European
  Culture Center of Delphi, Greece}, pages 15--25, 1985.

\bibitem{Ott2011}
S{\o}ren Ott, Jacob Berg, and Morten Nielsen.
\newblock {Linearised CFD Models for Wakes}.
\newblock Technical report, Ris{\o}, Roskilde, Denmark, 2011.

\bibitem{Martinez-Tossas2019}
Luis~A. Mart{\'{i}}nez-Tossas, Jennifer Annoni, Paul~A. Fleming, and Matthew~J.
  Churchfield.
\newblock {The aerodynamics of the curled wake: a simplified model in view of
  flow control}.
\newblock {\em Wind Energy Science}, 4(1):127--138, 2019.

\bibitem{Liu1983}
Mei~Kao Liu, Mark~A. Yocke, and Thomas~C. Myers.
\newblock {Mathematical Model for the Analysis of Wind-Turbine Wakes.}
\newblock {\em Journal of energy}, 7(1):73--78, 1983.

\bibitem{Ainslie1988}
J.F. Ainslie.
\newblock {Calculating the flow field in the wake of wind turbines}.
\newblock {\em Journal of Wind Engineering and Industrial Aerodynamics,},
  27:213--224, 1988.

\bibitem{Iungo2015}
Giacomo~Valerio Iungo, Francesco Viola, Umberto Ciri, Mario~A Rotea, and
  Stefano Leonardi.
\newblock {Data-driven RANS for simulations of large wind farms}.
\newblock {\em J. Phys.: Conf. Ser.}, 625:012025, 2015.

\bibitem{Iungo2017}
Giacomo~Valerio Iungo, Vignesh Santhanagopalan, Umberto Ciri, Francesco Viola,
  Lu~Zhan, Mario~A Rotea, and Stefano Leonardi.
\newblock {Parabolic RANS solver for low-computational-cost simulations of wind
  turbine wakes}.
\newblock {\em Wind Energy}, (October):1--14, 2017.

\bibitem{Santhanagopalan2018}
V.~Santhanagopalan, M.~A. Rotea, and G.~V. Iungo.
\newblock {Performance optimization of a wind turbine column for different
  incoming wind turbulence}.
\newblock {\em Renewable Energy}, 116:232--243, 2018.

\bibitem{Bradstock2020}
Philip Bradstock and Wolfgang Schlez.
\newblock {Theory and Verification of a new 3D RANS Wake Model}.
\newblock {\em Wind Energy Science}, 5:1425--1434, 2020.

\bibitem{Martinez-Tossas2020}
Luis Mart{\'{i}}nez-Tossas, Jennifer King, Eliot Quon, Christopher Bay, Rafael
  Mudafort, Nicholas Hamilton, and Paul Fleming.
\newblock {The curled wake model: A three-dimensional and extremely fast
  steady-state wake solver for wind plant flows}.
\newblock {\em Wind Energy Science}, 6(2021):555--570, 2021.

\bibitem{Crespo1991}
A.~Crespo and J.~Hend\'andez.
\newblock {Parabolic and elliptic models of wind turbine wakes. application to
  the interaction between different wakes and turbines}.
\newblock {\em PHOENICS J. Comp. Fluid Dyn.}, pages 104--127, 1991.

\bibitem{Masson1997}
Christian Masson, Idriss Ammara, and Ion Paraschivoiu.
\newblock {An aerodynamic method for the analysis of isolated horizontal-axis
  wind turbines}.
\newblock {\em International Journal of Rotating Machinery}, 3(1):21--32, 1997.

\bibitem{Leclerc1999}
Christophe Leclerc, Christian Masson, and Idriss Ammara.
\newblock {Turbulence Modeling of the Flow Around Horizontal Axis Wind
  Turbines}.
\newblock {\em Wind Engineering}, 23(5):279--294, 1999.

\bibitem{Ammara2002}
Idriss Ammara, Christophe Leclerc, and Christian Masson.
\newblock {A Viscous Three-Dimensional Differential/Actuator-Disk Method for
  the Aerodynamic Analysis of Wind Farms}.
\newblock {\em Journal of Solar Energy Engineering}, 124(4):345, 2002.

\bibitem{Soleimanzadeh2014}
Maryam Soleimanzadeh, Rafael Wisniewski, and Arno Brand.
\newblock State-space representation of the wind flow model in wind farms.
\newblock {\em Wind Energy}, 17(4):627--639, 2014.

\bibitem{Annoni2015}
Jennifer Annoni and Peter Seiler.
\newblock {A low-order model for wind farm control}.
\newblock {\em Proceedings of the American Control Conference},
  2015-July(1):1721--1727, 2015.

\bibitem{King2016}
Ryan King, Peter Hamlington, Katherine Dykes, and Peter Graf.
\newblock Adjoint optimization of wind farm layouts for systems engineering
  analysis.
\newblock In {\em 34th Wind Energy Symposium}, page 2199, 2016.

\bibitem{Adcock2018}
C~Adcock and RN~King.
\newblock {Data-Driven Wind Farm Optimization Incorporating Effects of
  Turbulence Intensity}.
\newblock In {\em 2018 Annual American Control Conference}, pages 1--6,
  Wisconsin Center, Milwaukee, USA, 2018. {Annual American Control Conference}.

\bibitem{King2017}
RN~King, K~Dykes, P~Graf, and PE~Hamlington.
\newblock {Optimization of wind plant layouts using an adjoint approach}.
\newblock {\em Wind Energy Science}, 2(1):115--131, 2017.

\bibitem{Larsen2008}
G.C. Larsen.
\newblock {Wake meandering: A pragmatic approach}.
\newblock {\em Wind Energy}, 11(4):377--395, 2008.

\bibitem{Keck2012}
Rolf~Erik Keck, Dick Veldkamp, Helge~Aagaard Madsen, and Gunner Larsen.
\newblock {Implementation of a mixing length turbulence formulation into the
  dynamic wake meandering model}.
\newblock {\em Journal of Solar Energy Engineering, Transactions of the ASME},
  134(2):1--13, 2012.

\bibitem{Boussinesq1897}
Joseph Boussinesq.
\newblock {\em Th{\'e}orie de l'{\'e}coulement tourbillonnant et tumultueux des
  liquides dans les lits rectilignes a grande section}, volume~1.
\newblock Gauthier-Villars, 1897.

\bibitem{Gomez-Elvira2005}
Rafael G{\'{o}}mez-Elvira, Antonio Crespo, Emilio Migoya, Fernando Manuel, and
  Julio Hern{\'{a}}ndez.
\newblock {Anisotropy of turbulence in wind turbine wakes}.
\newblock {\em Journal of Wind Engineering and Industrial Aerodynamics},
  93(10):797--814, 2005.

\bibitem{Cabezon2011}
D~Cabez{\'o}n, E~Migoya, and A~Crespo.
\newblock Comparison of turbulence models for the computational fluid dynamics
  simulation of wind turbine wakes in the atmospheric boundary layer.
\newblock {\em Wind Energy}, 14(7):909--921, 2011.

\bibitem{VanderLaan2013}
Paul van~der Laan, N.~N. Sorensen, P.-E. Rethore, J.~Mann, M.C. Kelly, and J.G.
  Schepers.
\newblock {Nonlinear eddy viscosity models applied to wind turbine wakes}.
\newblock In {\em ICOWES2013 Conference}, number March 2016, pages 514--525,
  2013.

\bibitem{Lange2003}
Bernhard Lange, Hans~Peter Waldl, Algert~Gil Guerrero, Detlev Heinemann, and
  Rebecca~J. Barthelmie.
\newblock {Modelling of offshore wind turbine wakes with the wind farm program
  FLaP}.
\newblock {\em Wind Energy}, 6(1):87--104, 2003.

\bibitem{Keck2014a}
RE~Keck, R~Mikkelsen, N~Troldborg, M~de~Mar{\'{e}}, and K.~S. Hansen.
\newblock {Synthetic atmospheric turbulence and wind shear in large eddy
  simulations of wind turbine wakes}.
\newblock {\em Wind Energy}, 17(April 2014):1247--1267, 2014.

\bibitem{Keck2015}
RE~Keck, M~de~Mar{\'{e}}, MJ~Churchfield, S~Lee, G~Larsen, and HA~Madsen.
\newblock {Two improvements to the dynamic wake meandering model: including the
  effects of atmospheric shear on wake turbulence and incorporating turbulence
  build-up in a row of wind turbines}.
\newblock {\em Wind Energy}, 18(April 2013):111--132, 2015.

\bibitem{Iungo2018}
Giacomo~Valerio Iungo, Stefano Letizia, and Lu~Zhan.
\newblock {Quantification of the axial induction exerted by utility-scale wind
  turbines by coupling LiDAR measurements and RANS simulations}.
\newblock In {\em Journal of Physics: Conference Series}, volume 1037, 2018.

\bibitem{Prospathopoulos2011}
J.~M. Prospathopoulos, E.~S. Politis, K.~G. Rados, and P.~K. Chaviaropoulos.
\newblock {Evaluation of the effects of turbulence model enhancements on wind
  turbine wake predictions}.
\newblock {\em Wind Energy}, 14:285--300, 2011.

\bibitem{ElKasmi2008}
Amina {El Kasmi} and Christian Masson.
\newblock {An extended k - $\epsilon$ model for turbulent flow through
  horizontal-axis wind turbines}.
\newblock {\em Journal of Wind Engineering and Industrial Aerodynamics},
  96(1):103--122, 2008.

\bibitem{VanDerLaan2015b}
M.~P. Van Der~Laan, K.~S. Hansen, N.~N. S{\o}rensen, and P.~E.
  R{\'{e}}thor{\'{e}}.
\newblock {Predicting wind farm wake interaction with RANS: An investigation of
  the Coriolis force}.
\newblock {\em Journal of Physics: Conference Series}, 625(1), 2015.

\bibitem{Hennen2017}
Joep Hennen and Sa{\v{s}}a Kenjere{\v{s}}.
\newblock {Contribution to improved eddy-viscosity modeling of the wind turbine
  - to - wake interactions}.
\newblock {\em International Journal of Heat and Fluid Flow},
  68(September):319--336, 2017.

\bibitem{Cea2007}
L.~Cea, J.~Puertas, and V{\'{a}}zquez-Cend{\'{o}}n~M. E.
\newblock {Depth averaged modelling of turbulent shallow water flow with
  wet-dry fronts}.
\newblock {\em Archives of Computational Methods in Engineering},
  14(3):303--341, 2007.

\bibitem{Iungo2013}
G.~V. Iungo, F.~Viola, S.~Camarri, F.~Port{\'{e}}-Agel, and F.~Gallaire.
\newblock {Linear stability analysis of wind turbine wakes performed on wind
  tunnel measurements}.
\newblock {\em Journal of Fluid Mechanics}, 737:499--526, 2013.

\bibitem{Meyers2010}
Johan Meyers and Charles Meneveau.
\newblock {Large Eddy Simulations of large wind-turbine arrays in the
  atmospheric boundary layer}.
\newblock In {\em 48th AIAA Aerospace Sciences Meeting Including the New
  Horizons Forum and Aerospace Exposition 4 - 7 January 2010, Orlando,
  Florida}, number January, 2010.

\bibitem{Medici2006}
D.~Medici and P.~H. Alfredsson.
\newblock {Measurements on a wind turbine wake: 3D effects and bluff body
  vortex shedding}.
\newblock {\em Wind Energy}, 9(3):219--236, 2006.

\bibitem{Chamorro2009}
Leonardo~P. Chamorro and Fernando Port{\'{e}}-Agel.
\newblock {A wind-tunnel investigation of wind-turbine wakes: Boundary-Layer
  turbulence effects}.
\newblock {\em Boundary-Layer Meteorology}, 132(1):129--149, 2009.

\bibitem{Boersma2018}
Sjoerd Boersma, Bart Doekemeijer, Mehdi Vali, Johan Meyers, and Jan-Willem~van
  Wingerden.
\newblock A control-oriented dynamic wind farm model: Wfsim.
\newblock {\em Wind Energy Sci}, 3(1):75--95, 2018.

\bibitem{Pletcher2012}
Richard~H Pletcher, John~C Tannehill, and Dale Anderson.
\newblock {\em Computational fluid mechanics and heat transfer}.
\newblock CRC press, 2012.

\bibitem{Pratap1975}
V.~S. Pratap and D.~B. Spalding.
\newblock {Numerical Computations of the Flow in Curved Ducts}.
\newblock {\em Aeronautical Quarterly}, 3(July 1975):219--228, 1975.

\bibitem{Briley1974}
W.~R. Briley.
\newblock {Numerical method for predicting three-dimensional steady viscous
  flow in ducts}.
\newblock {\em Journal of Computational Physics}, 28:8--28, 1974.

\bibitem{Moore1979}
J.~Moore and J.~G. Moore.
\newblock {Calculation Procedure for Three-Dimensional, Viscous, Compressible
  Duct Flow - 1. Inviscid Flow Considerations.}
\newblock {\em ASTM Special Technical Publication}, 101(December 1979):75--82,
  1979.

\bibitem{Dodge1977}
P.~R. Dodge.
\newblock {Numerical method for 2D and 3D viscous flows}.
\newblock {\em AIAA Journal}, 15(7):961--965, 1977.

\bibitem{Chilukuri1980}
R.~Chilukuri and R.~H. Pletcher.
\newblock {Numerical Solutions to the Partially Parabolized Navier-stokes
  Equations for Developing Flow in a Channel}.
\newblock {\em Numerical Heat Transfer}, 3(2):169--188, 1980.

\bibitem{HaglundElGaidi2018}
S.~{Haglund El Gaidi}.
\newblock {Partially Parabolic Wind Turbine Flow Modelling}.
\newblock Technical Report April, Royal Institute of Technology, 2018.

\bibitem{Murthy1990}
Jayathi~Y. Murthy and Suhas~V. Patankar.
\newblock {A partially parabolic calculation procedure for duct flows in
  irregular geometries: Part L: Formulation}.
\newblock {\em Numerical Heat Transfer, Part B: Fundamentals}, 16(1):1--15,
  1990.

\bibitem{Maulik2021}
romit Maulik, Vishwas Rao, Ashwin Renganathan, Stefano Letizia, and
  Giacomo~Valerio Iungo.
\newblock Cluster analysis of wind turbine wakes measured through a scanning
  doppler wind lidar.
\newblock In {\em AIAA Scitech 2021 Forum}, 2021.

\bibitem{Beck2017}
Hauke Beck and Martin K{\"{u}}hn.
\newblock {Dynamic data filtering of long-range doppler LiDAR wind speed
  measurements}.
\newblock {\em Remote Sensing}, 9(6), 2017.

\bibitem{IEC61400_12_1}
{International Electrotechnical Commission 61400-12-1}.
\newblock Wind energy generation systems – {P}art 12-1: {P}ower performance
  measurements of electricity producing wind turbines.
\newblock International Standard 61400-12-2, International Electrotechnical
  Commission (IEC), Geneva, Switzerland, 2013., 2017.

\bibitem{IEC61400_12_2}
{International Electrotechnical Commission, 61400-12-2}.
\newblock Wind turbine generator systems - {P}art 12-2: {P}ower performance of
  electricity-producing wind turbines based on nacelle anemometry.
\newblock International Standard 61400-12-2, International Electrotechnical
  Commission (IEC), Geneva, Switzerland, 2013., 2013.

\bibitem{Hamilton2020}
Nicholas Hamilton, Christopher~J. Bay, Paul Fleming, Jennifer King, and Luis~A.
  Mart{\'{i}}nez-Tossas.
\newblock {Comparison of modular analytical wake models to the Lillgrund wind
  plant}.
\newblock {\em Journal of Renewable and Sustainable Energy}, 12(5), 2020.

\bibitem{Teng2020}
Jian Teng and Corey~D. Markfort.
\newblock {A calibration procedure for an analytical wake model using wind farm
  operational data}.
\newblock {\em Energies}, 13(14):1--19, 2020.

\bibitem{Letizia2021}
Stefano Letizia, Lu~Zhan, and Giacomo {Valerio Iungo}.
\newblock {LiSBOA: LiDAR Statistical Barnes Objective Analysis for optimal
  design of LiDAR scans and retrieval of wind statistics. Part I: Theoretical
  framework}.
\newblock {\em Atmospheric Measurement Techniques}, 14:2065--2093, 2021.

\bibitem{Letizia2021b}
Stefano Letizia, Lu~Zhan, and Giacomo {Valerio Iungo}.
\newblock {LiSBOA: LiDAR Statistical Barnes Objective Analysis for optimal
  design of LiDAR scans and retrieval of wind statistics. Part II: Applications
  lidar measurements of wind turbine wakes}.
\newblock {\em Atmospheric Measurement Techniques}, 14:2095--2113, 2021.

\bibitem{inpaintnans}
{Matlab file exchange community website}.
\newblock {inpaint\_nans.m}.
\newblock
  \url{https://www.mathworks.com/matlabcentral/fileexchange/4551-inpaint_nans},
  2021.
\newblock Accesed: 2021-07-30.

\bibitem{Vermeulen1980}
P.~E.~J. Vermeulen.
\newblock An experimental analysis of wind turbine wake.
\newblock In {\em 3rd Int. Symp. Wind Energy Syst., Copenhagen, August}, 1980.

\bibitem{Chamorro2012}
L.~P. Chamorro, M.~Guala, R.~E.A. Arndt, and F.~Sotiropoulos.
\newblock {On the evolution of turbulent scales in the wake of a wind turbine
  model}.
\newblock {\em Journal of Turbulence}, 13(December):1--13, 2012.

\bibitem{AIAA2002}
AIAA.
\newblock {Guide for the Verification and Validation of Computational}.
\newblock Technical Report G-077, American Institute of Aeronautics and
  Astronautics, 2002.

\bibitem{Viola2014}
F.~Viola, G.~V. Iungo, S.~Camarri, F.~Port{\'{e}}-Agel, and F.~Gallaire.
\newblock {Prediction of the hub vortex instability in a wind turbine wake:
  Stability analysis with eddy-viscosity models calibrated on wind tunnel
  data}.
\newblock {\em Journal of Fluid Mechanics}, 750:R1, 2014.

\bibitem{GRANS}
{UTD WindFluX Global RANS solver for simulations of axisymmetric wind turbine
  wakes}.
\newblock \url{https://github.com/UTD-WindFluX/G-RANS}.
\newblock Accessed: 2021-10-27.

\bibitem{Machefaux2015}
E.~Machefaux, G.~C. Larsen, and J.~P.Murcia Leon.
\newblock {Engineering models for merging wakes in wind farm optimization
  applications}.
\newblock {\em J Phys: Conf Ser}, 625(1), 2015.

\bibitem{Walker2016}
Keith Walker, Neil Adams, Brian Gribben, Breanne Gellatly, Nicolai~Gayle
  Nygaard, Andrew Henderson, Miriam {Marchante Jim{\'{e}}mez}, Sarah~Ruth
  Schmidt, Javier {Rodriguez Ruiz}, Daniel Paredes, Gemma Harrington, Niall
  Connell, Oliver Peronne, Miguel Cordoba, Paul Housley, Robert Cussons,
  M{\aa}ns H{\aa}kansson, Andreas Knauer, and Eoghan Maguire.
\newblock {An evaluation of the predictive accuracy of wake effects models for
  offshore wind farms}.
\newblock {\em Wind Energy}, 19(5):979--996, 2016.

\bibitem{Farrell2021}
Alayna Farrell, Jennifer King, Caroline Draxl, Rafael Mudafort, Nicholas
  Hamilton, Christopher~J. Bay, Paul Fleming, and Eric Simley.
\newblock {Design and analysis of a wake model for spatially heterogeneous
  flow}.
\newblock {\em Wind Energy Sci.}, 6(3):737--758, 2021.

\bibitem{Wheeler2004}
A.~J. Wheeler and R.~J. Ahmad.
\newblock {\em Introduction to engineering experimentation}.
\newblock Pearson Higher Education, {1 Lake St., Upper Saddle River, New
  Jersey, 07458.}, 2004.

\bibitem{Nygaard2015}
Nicolai~Gayle Nygaard.
\newblock {Systematic quantification of wake model uncertainty}.
\newblock {\em EWEA Offshore}, pages 1--10, 2015.

\bibitem{Niayifar2016}
Amin Niayifar and Fernando Port{\'{e}}-Agel.
\newblock {Analytical modeling of wind farms: A new approach for power
  prediction}.
\newblock {\em Energies}, 9(9):1--13, 2016.

\bibitem{FLORIS_2021}
NREL.
\newblock Floris. version 2.4.
\newblock
  \url{https://population.un.org/wpp/Graphs/DemographicProfiles/Line/900},
  2021.
\newblock Accessed: 2021-09-01.

\bibitem{Niayifar2015}
Amin Niayifar and Fernando Port{\'{e}}-Agel.
\newblock {A new analytical model for wind farm power prediction}.
\newblock {\em Journal of Physics: Conference Series}, 625(1), 2015.

\bibitem{Patankar1972}
S.~V. Patankar and D.~B. Spalding.
\newblock {A calculation procedure for heat, mass and momentum transfer in
  three-dimensional parabolic flows}.
\newblock {\em International Journal of Heat and Mass Transfer},
  15(10):1787--1806, 1972.

\end{thebibliography}

\end{document}